\documentclass[a4paper,11pt]{article}

\usepackage[margin=2.5cm]{geometry}
\usepackage{booktabs}
\usepackage{xcolor}
\usepackage[most]{tcolorbox}

\usepackage{mathrsfs}

\usepackage{authblk}

\usepackage{subcaption}
\usepackage{multicol}

\usepackage{caption}
\usepackage{amsmath,amsthm}
\usepackage{amssymb}
\usepackage{comment}
\usepackage{amsfonts}
\makeatletter  
\newif\if@restonecol  
\makeatother

\usepackage[linesnumbered,ruled,vlined]{algorithm2e}
\usepackage{algpseudocode}  
  
\usepackage{tikz}
\usetikzlibrary{positioning,backgrounds,patterns,calc}
\tikzstyle{vert}=[circle,draw=black,minimum size=8pt,inner sep=1pt]
\tikzstyle{vertex2}=[circle,draw=black,minimum size=15pt,inner sep=2pt]
\tikzstyle{edge}=[]
\tikzstyle{ypath}=[ultra thick]
\tikzstyle{dottedEdge}=[dotted,thick]
\tikzstyle{small-vertex}=[circle,draw=black,minimum size=6pt,inner sep=0pt,fill=white]
\tikzstyle{thinedges}=[draw=gray!30]
 \tikzstyle{boxes}=[draw,thick, rounded corners=3mm,text width=2.7cm,align=center,text opacity=1,fill opacity=1,fill=white]
\tikzstyle{unk}=[fill=gray!25!white]

\definecolor{ACMRed}{RGB}{255,0,0}
\definecolor{ACMDarkBlue}{RGB}{0,0,255}

\usepackage{bbding}
\usepackage{placeins}

{\bfseries}{\normalfont}
\newtheorem{obs}{Observation}{\bfseries}{\normalfont}

\DeclareMathOperator{\opUPQR}{\textrm{UPQR}}
\DeclareMathOperator{\HD}{HD}

\newcommand{\decprob}[3]{%
\medskip
  {\centering
    \begin{minipage}{1\linewidth}%
      {#1} \vspace{.2cm} \\
      \textbf{Input:} #2 \vspace{.1cm}\\ 
      \textbf{Question:} #3
    \end{minipage}%
  } 
\medskip
}

\newcommand{\decprobQ}[2]{%
\medskip
  {\centering
    \begin{minipage}{1\linewidth}%
      {#1} \\
      \textbf{Question:} #2
    \end{minipage}%
  } 
}
\newcommand{\decprobI}[2]{%
\medskip
  {\centering
    \begin{minipage}{1\linewidth}%
      {#1}\\
      \textbf{Input:} #2
    \end{minipage}%
  } 
\medskip
}

\newenvironment{probox}{
    \vspace{-.2cm}
    \begin{center}
      \begin{tcolorbox}[colback=white,
      colframe=black,
      width=.98\textwidth,
      left=1mm,
      right=1mm,
      top=0mm,
      bottom=0mm,
      arc=3mm, auto outer arc,
      boxrule=.8pt]}
    {\end{tcolorbox}
    \end{center}
    \vspace{-.2cm}}

\newenvironment{itemize+}
  { 
    \vspace*{-.2\baselineskip}
    \begin{itemize}
    \addtolength{\itemsep}{-.2\baselineskip}
    \addtolength{\baselineskip}{-.2\baselineskip}
  }
  { \end{itemize}
    \vspace*{-.2\baselineskip}
  }

\newenvironment{itemize++}
  { 
    \vspace*{-.1\baselineskip}
    \begin{itemize}
    \addtolength{\itemsep}{-.4\baselineskip}
    \addtolength{\baselineskip}{-.4\baselineskip}
  }
  { \end{itemize}
    \vspace*{-.1\baselineskip}
  }

\newenvironment{enumerate+}
  { 
    \vspace*{-.4\baselineskip}
    \begin{enumerate}
    \addtolength{\itemsep}{-.1\baselineskip}
    \addtolength{\baselineskip}{-.1\baselineskip}
  }
  { \end{enumerate}
    \vspace*{-.4\baselineskip}
  }
\newcommand{\myparagraph}[1]{\medskip\noindent\textbf{{#1}}}

\newcommand{\appsymb}{$\star$}
\newcommand{\appref}[1]{{\hyperref[proof:#1]{\appsymb}}}
\newcommand{\appBref}[1]{{\hyperref[#1]{\appsymb}}}
\usepackage{thm-restate}
\usepackage{hyperref}

\usepackage{cleveref}
\usepackage{appendix}

\usepackage{bm}

\usepackage{cases} 
\usepackage{microtype,ellipsis}
\usepackage{color}

\usepackage{units}
\usepackage[linecolor=green!70!black, backgroundcolor=green!10, bordercolor=black, textsize=scriptsize]{todonotes}
\usepackage{marginnote}

\usepackage[numbers,sort]{natbib}

\makeatletter
\def\NAT@spacechar{~}
\makeatother

\usepackage{paralist}

\usepackage{enumitem}

\newlist{propertylist}{enumerate}{10}
\setlist[propertylist]{label=\roman*),leftmargin=*}

\newcommand{\OL}{\textsc{Optimal-Lobbying}\xspace}

\newcommand{\WMDS}{\textsc{Weighted Maximum Density Subgraph}\xspace}
\newcommand{\WMDSshort}{\textsc{WMDS}\xspace}
\newcommand{\MC}{\textsc{Minimum Cut}\xspace}

\newcommand{\PVC}{\textsc{Positive Vertex Cover}\xspace}

\newcommand{\NP}{\ensuremath{\textsf{NP}}\xspace}

\newcommand{\W}[1]{\ensuremath{\textsf{W[#1]}}\xspace}

\crefname{rrule}{Rule}{Rules}

\graphicspath{{images/}}

\usepackage{etoolbox}
\usepackage{multirow}

\newcommand{\PrU}{\textrm{U}}
\newcommand{\PrTR}{\textrm{TR}}
\newcommand{\PrCR}{\textrm{CR}}

\newcommand{\CT}{\textrm{standard-form clause}\xspace}

\newcommand{\PvsNPequiv}{equivalent under polynomial-time Turing reductions}

\newcommand{\qquota}{{\ensuremath{q}}}
\newcommand{\nqquota}{{\ensuremath{\bar{q}}}}

\newcommand{\Robustness}{\textsc{UPQR-Robustness-Manipulation}\xspace}
\newcommand{\Possible}{\textsc{UPQR-Possible-Manipu\-lation}\xspace}
\newcommand{\Necessary}{\textsc{UPQR-Necessary-Manipu\-lation}\xspace}
\newcommand{\Exact}{\textsc{UPQR-Exact-Manipulation}\xspace}
\newcommand{\Hamming}{\textsc{UPQR-HD-Manipulation}\xspace}
\newcommand{\Bribery}{\textsc{UPQR-Bribery}\xspace}
\newcommand{\Microbribery}{\textsc{UPQR-Microbribery}\xspace}

\crefname{section}{Section}{Sections}
\crefname{lemma}{Lemma}{Lemmas}
\crefname{corollary}{Corollary}{Corollaries}
\crefname{proposition}{Proposition}{Propositions}
\crefname{observation}{Observation}{Observations}
\crefname{definition}{Definition}{Definitions}
\crefname{table}{Table}{Tables}
\crefname{figure}{Figure}{Figures}

\crefname{theorem}{Theorem}{Theorems}
\crefname{obs}{Observation}{Observations}

\newtheorem{theorem}{Theorem}
\newtheorem{definition}{Definition}
\newtheorem{lemma}{Lemma}
\newtheorem{corollary}{Corollary}

\newtheorem{proposition}{Proposition}

\title{Complexity of Manipulation and Bribery in Premise-Based\\ Judgment Aggregation with Simple Formulas}

\date{}

\author[1,2,3]{Robert Bredereck}
\author[4]{Junjie Luo}

\affil[1]{Institut f\"ur Softwaretechnik und Theoretische Informatik,
  Technische Universität Berlin, Berlin, Germany}
\affil[2]{Institut f\"ur Informatik, Humboldt-Universität zu Berlin, Berlin, Germany}
\affil[3]{Institut f\"ur Informatik, TU Clausthal, Clausthal-Zellerfeld, Germany
  \texttt{robert.bredereck}@tu-clausthal.de}
\affil[4]{School of Mathematics and Statistics, Beijing Jiaotong University, Beijing, China
  \texttt{jjluo1}@bjtu.edu.cn}

\providecommand{\keywords}[1]{\textbf{\textit{Key words---}} #1}

\begin{document}
\maketitle

\begin{abstract}  
Judgment aggregation is a framework to aggregate individual opinions on multiple, logically connected issues
into a collective outcome.
These opinions are cast by judges, which can be for example referees, experts, advisors or jurors, depending on the application and context.
It is open to manipulative attacks such as \textsc{Manipulation}
where judges cast their judgments strategically.
Previous works have shown that most computational problems corresponding to these manipulative attacks
are \NP-hard.
This desired computational barrier, however, often relies on formulas that are either of unbounded size or of complex structure.

We revisit the computational complexity for various \textsc{Manipulation} and \textsc{Bribery}
problems in premise-based judgment aggregation,
now focusing on simple and realistic formulas.
We restrict all formulas to be clauses that are (positive) monotone, Horn-clauses, or have bounded length.
For basic variants of \textsc{Manipulation}, we show that these restrictions make several variants,
which were in general known to be \NP-hard, polynomial-time solvable.
Moreover, we provide a P vs.\ NP dichotomy for a large class of clause restrictions (generalizing monotone and Horn clauses)
by showing a close relationship between variants of \textsc{Manipulation}  and variants of \textsc{Satisfiability}.
For Hamming distance based \textsc{Manipulation},
we show that \NP-hardness even holds for positive monotone clauses of length three, but the problem becomes polynomial-time solvable for positive monotone clauses of length two.
For \textsc{Bribery}, we show that \NP-hardness even holds for positive monotone clauses of length two, but it becomes polynomial-time solvable for the same clause set if there is a constant budget.
\end{abstract}

%


\maketitle

\newcommand{\nameofc}{cult potential}
\newcommand{\nameofm}{marketability}
\newcommand{\nameofh}{high profitability}
\newcommand{\nameofs}{strong competitors' existence}
\newcommand{\nameofe}{market entering potential}
\newcommand{\nameofr}{short-term risk}
\newcommand{\nameofcInit}{\textbf{c}ult potential}
\newcommand{\nameofmInit}{\textbf{m}arketability}
\newcommand{\nameofhInit}{\textbf{h}igh profitability}
\newcommand{\nameofsInit}{\textbf{s}trong competitors' existence}
\newcommand{\nameofeInit}{market \textbf{e}ntering potential}
\newcommand{\nameofrInit}{short-term \textbf{r}isk}
\newcommand{\nameofcLong}{\nameofc~($c$)}
\newcommand{\nameofmLong}{\nameofm~($m$)}
\newcommand{\nameofhLong}{\nameofh~($h$)}
\newcommand{\nameofsLong}{\nameofs~($s$)}
\newcommand{\nameofeLong}{\nameofe~($e:=\neg s \vee c$)}
\newcommand{\nameofrLong}{\nameofr~($r:=\neg m \vee \neg h$)}
\newcommand{\nameofcLongInit}{\nameofcInit~($c$)}
\newcommand{\nameofmLongInit}{\nameofmInit~($m$)}
\newcommand{\nameofhLongInit}{\nameofhInit~($h$)}
\newcommand{\nameofsLongInit}{\nameofsInit~($s$)}
\newcommand{\nameofeLongInit}{\nameofeInit~($e:=\neg s \vee c$)}
\newcommand{\nameofrLongInit}{\nameofrInit~($r:=\neg m \vee \neg h$)}

\section{Introduction}
Justine is the head of a committee deciding on financial support for new startup companies.
For her decisions, she uses publicly available evaluations of experts (judges) with respect to a set of
basic features such as \nameofcLongInit{}, \nameofmLongInit{}, \nameofhLongInit{}, and \nameofsLongInit{}.
As a brilliant mathematician and economist, Justine developed a model that can reliably predict
the success of the startup by putting the features into logical relation.
For example, she defined two further composed features ``\nameofeInit{}'' as $e:=\neg s \vee c$ and
``\nameofrInit{}'' as $r:=\neg m \vee \neg h$.
Since the expert's evaluations are different, she needs to aggregate them to feed her model.
Her first idea was to take the majority on each feature,
but she recognizes that she may obtain the following evaluations from three experts:
\begin{center}
\begin{tabular}{l r r r r r r r r r r r}
 expert 1: & $s$ & $\wedge$ & $\neg c$ & $\wedge$ & $m$ & $\wedge$ & $h$ & $\wedge$ & $\neg e$ & $\wedge$ & $\neg r$ \\
 expert 2: & $s$ & $\wedge$ & $c$ & $\wedge$ & $\neg m$ & $\wedge$ & $h$ & $\wedge$ & $e$ & $\wedge$ & $r$ \\
 expert 3: & $\neg s$ & $\wedge$ & $\neg c$ & $\wedge$ & $m$ & $\wedge$ & $\neg h$ & $\wedge$ & $e$ & $\wedge$ & $r$ 
\end{tabular}
\end{center}
where the majority opinions claim \nameofsLong{}, \nameofmLong{}, and \nameofhLong{}, and disclaim \nameofcLong{}.
However, the majority opinions also claim \nameofeLong{} and \nameofrLong{};  
an obviously paradoxical situation.
Justine does a quick literature review and identifies her aggregation problem as ``judgment aggregation''
and the observed paradox as a variant of the well-known doctrinal paradox~\cite{KS86}.
To avoid this paradox and since the experts are anyway better in evaluating basic features than evaluating composed features,
she decides to adapt the concept of premise-based judgment aggregation rules~\cite{dietrich2007judgment}
for her decision process:
Basic features form the \emph{premises}, and composed features are \emph{conclusions} (which logically connect premises).
The aggregation process is performed only on the premises and conclusions are deduced from them.
That is, the outcome in the above example is $s$, $\neg c$, $m$, $h$, and, hence, also~$\neg e$ and~$\neg r$.

Justine is happy with the aggregation process, but she is worried about the reliability of the results.
(1) For example, what if an expert made a mistake?
Can she compute efficiently whether a set of important features remains stable
even if some expert provided a wrong evaluation?
(2) What if an expert evaluated strategically or untruthfully due to bribery or lobbyism?
Is it difficult for an expert to compute a successful strategy?
A quick literature review identifies all questions posed by Justine as variants of \textsc{Manipulation} and \textsc{Bribery},
which are computationally intractable.

Although the intractability of strategic evaluation and bribery seems to be good news,
Justine is skeptical about the relevance of these results for her application.
In her model, all formulas are length-two \emph{Horn clauses}.
All intractability results found, however, use rather complex or long formulas as conclusions, and, hence, do not apply in her situation. 
So Justine is stuck with the literature's state of the art and cannot decide whether her model
(with simple formulas) 
 is vulnerable towards manipulation or bribery.

In this paper, we help Justine (and all others in similar situations) and provide a fine-grained computational complexity analysis
of \textsc{Manipulation} and \textsc{Bribery} for judgment aggregation with simple formulas.
Our results are essentially good news for Justine:
Concerning question~(1), we show that checking the robustness of an outcome (i.e.\ whether a single judge/mistake can change it)
turns out to become polynomial-time solvable for simple formulas.
Concerning question~(2), we show that most manipulative actions (such as strategic  evaluation or bribery) remain computationally hard
even for structurally simple formulas.

\subsection{Related Work}
For a detailed introduction to the general topic of judgment aggregation we refer to excellent surveys in the field~\cite{End16,EandC6judgement,GP14,Lis12,LP09}.
We focus here on a description of work related to computational complexity of strategic behavior for premise-based judgment aggregation rules,
which is most relevant to our work.
We will point the interested reader to specific surveys for broader information.

\citet{DL07str} introduced strategic behavior to judgment aggregation.
They introduced the concept of strategy-proofness to analyze whether certain strategic behavior can influence the outcome
(ignoring the computational complexity of manipulative attacks).
Following \citeauthor{BTT89a}~\cite{BTT89a,BTT92}, intractability of manipulative attacks is usually seen as
``barrier against manipulation'' and, hence, a desired property.
\citet{endriss2012complexity} were the first who analyzed the computational complexity
of strategic behavior in judgment aggregation and showed that it is \NP-hard for a judge
to decide whether she can cast a judgment set that influences the collective outcome in a beneficial way
(assuming Hamming distance based preferences over judgment sets),
even for the simple premise-based majority rule, where a premise is accepted if it is accepted by more than half of judges.
We also consider, among others, Hamming distance based preferences, but for uniform premise-base quota rules,
which is a family of rules that contains their premise-based majority rule as prominent special case.
Moreover, \citet{endriss2012complexity} allow arbitrarily complex formulas to obtain their results.
\citet{BEER13,baumeister2015complexity,DBLP:journals/jcss/BaumeisterEERS20} continued this line of research and
where the first who extended the results to the more general uniform premise-base quota rules.
They also initiated the analysis of further variants of strategic behavior for judgment aggregation,
including further variants of \textsc{Manipulation} 
or cases where an external agent influences the structure (\textsc{Control})
or the opinions of the judges (\textsc{Bribery}), showing \NP-hardness for most considered problems. 
However, these \NP-hardness results usually rely on complex or size-unbounded formulas, leaving open the complexity of cases with simple and realistic formulas, which is the focus of our paper.
Similar to \citet{endriss2012complexity}, they also did not impose any systematic restriction on the
complexity of the formula.
We continue this line of research but we focus on practically meaningful simple formula and analyze
to which extent desired computational complexity transfers.
Besides premise-based rules, \citet{DBLP:conf/atal/Haan17} studied the complexity of \textsc{Manipulation}, \textsc{Bribery} and \textsc{Control} for the Kemeny judgment aggregation rule,
and \citet{DBLP:conf/atal/BaumeisterBW21} investigated the complexity of manipulation by changing the processing order in sequential rules.
For a broader overview on strategic behavior in judgment aggregation, we refer to a recent survey~\cite{COMSOCtrends8judgment}.

Our work also fits well into the line of research initiated by the seminal paper of \citet{FHHR11} showing that
the barrier against manipulative attacks sometimes disappears in context of restricted domains.
In context of voting one usually considers restricted preference domains (see \citet{ELP22surv} for a recent survey) whereas we focus on restricted formulas.

More generally, fine-grained analysis has also been considered for outcome determination in judgment aggregation, which is usually computationally intractable for many judgment aggregation rules (e.g., Kemeny rule)~\cite{DBLP:journals/jair/EndrissHLS20}. 
\citet{DBLP:conf/ecai/Haan16} investigated the parameterized complexity of outcome determination of the Kemeny rule by considering structural parameters, including the maximum size of formulas, which is also one restriction considered in this paper.
\citet{DBLP:conf/kr/Haan18} studied the influence of different formula restrictions (e.g., Horn and Krom formulas) on the complexity of outcome determination in judgment aggregation.
Finally, the problem of agenda safety in judgment aggregation, which determines for a given set of formulas and a given aggregation procedure whether the procedure's outcome is always consistent for any combination of judgment sets, has also been considered from the perspective of parameterized complexity~\cite{DBLP:conf/atal/EndrissHS15}.

\subsection{Contributions and Organization}
We analyze the computational complexity of variants of \textsc{Manipulation} and \textsc{Bribery} in  premise-based judgment aggregation with simple formulas.
We restrict all formulas to clauses and systematically study the effect of the number of positive (negative) literals in a clause and clause length on the computational complexity.
In particular, we consider Horn clauses (implication-like
conclusions which for instance are fundamental in logic programming~\cite{lloyd87,ceri2012logic}), (positive) monotone clauses, and clauses of bounded length.
In \cref{sec:prelim} we describe the formal model and introduce our notation.

In \cref{sec:manip} we revisit the computational complexity for basic variants of \textsc{Manipulation},
showing that the restriction to clauses makes several variants, which were in general known to be \NP-hard~\cite{baumeister2015complexity}, polynomial-time solvable.
Our main result in this section is a P vs.\ NP dichotomy for a large class of clause restrictions (generalizing monotone and Horn clauses)
by showing a close relationship between variants of \textsc{Manipulation}  and variants of \textsc{Satisfiability}.
For details, we refer to \cref{tabel: results of manipulation} in our conclusion (\cref{sec: conclusion}).

In \cref{sec:HDmanip} we revisit Hamming distance based \textsc{Manipulation}.
Our main result in this section is that for positive monotone clauses the problem becomes polynomial-time solvable for clauses of length~$\ell=2$
but remains \NP-hard when~$\ell=3$. 
This is particularly surprising since \textsc{Satisfiability} is trivial for positive monotone clauses even for unbounded length.
The latter result is reached by showing \NP-hardness of a natural variant of \textsc{Vertex Cover} which we believe to be interesting on its own.
The \NP-hardness also holds for monotone or Horn clauses of length~$\ell=2$. 

In \cref{sec:brib} we revisit the computational complexity of two variants of \textsc{Bribery}, where we show that the \NP-hardness even holds for positive monotone clauses of length~$\ell=2$, probably the most basic case.
We then consider the restricted case with a fixed budget, and show that in this case the problem becomes polynomial-time solvable for positive monotone clauses of length~$\ell=2$.
For general clause sets, we show an interesting relation between Hamming distance based \textsc{Manipulation} and \textsc{Bribery} with the same clause set: the \NP-hardness of Hamming distance based \textsc{Manipulation} implies the \NP-hardness of the corresponding \textsc{Bribery}.

\section{Model and Preliminaries}
\label{sec:prelim}

\subsection{Premise-Based Judgment Aggregation}
\hyphenation{Bau-mei-ster}
We adopt the judgment aggregation framework described  by \citet{baumeister2015complexity}
and \citet{endriss2012complexity} and slightly simplify the notation for premise-based rules.

The topics to be evaluated are collected in the \emph{agenda}~$\Phi = \Phi_p \uplus \Phi_c$ that
consists of a finite set of premises~$\Phi_p$ (propositional variables)
as well as a finite set of conclusions~$\Phi_c$ (propositional formulas built from the premises using
standard logical connectives $\neg$, $\lor$, and~$\land$).\footnote{This
implies that the agenda is closed under propositional variables, that is, if $\phi$ is a formula in the
agenda, then so is every propositional variable occurring within $\phi$.}
In this paper we only study disjunctive agendas, i.e., there is no formulas containing~$\land$ (see more details in Section~\ref{sec:clause-restrict}).
The agenda does not contain any doubly negated formulas and is closed under complementation, that is, $\neg \alpha \in \Phi $ if and only if $\alpha \in \Phi$.
An evaluation on the agenda is expressed as a \emph{judgment set} $J \subseteq \Phi$.
A judgment set is \emph{complete} if each premise and conclusion is contained either in the negated or non-negated form and
\emph{consistent} if there is an assignment that satisfies all premises and conclusions in the judgment set simultaneously. 
The set of all complete and consistent subsets of~$\Phi$ is denoted by $\mathcal{J}(\Phi)$.

Let $N = \{1, . . . , n\}$ be a set of $n > 1$ \emph{judges}.
A profile is a vector of judgment sets $\bm{J} = (J_1 , \dots, J_n ) \in {\mathcal{J}(\Phi)}^n$.
We denote by $(\bm{J}_{-i}, {J_i}')$ the profile that is like $\bm{J}$, except that $J_i$ has been replaced
by ${J_i}'$.
A \emph{judgment aggregation procedure} for agenda $\Phi$ and judges $N = \{1, . . . , n\}$ is
a function $F : {\mathcal{J}(\Phi)}^n \rightarrow 2^{\Phi}$ that maps a profile $\bm{J}$ to
a single judgment set, which is called a \emph{collective judgment set}.
A procedure is called \emph{complete} (\emph{consistent}) if the collective judgment set under the procedure is always \emph{complete} (\emph{consistent}).

The most natural procedure is probably the \emph{majority rule}, 
which accepts a formula if and only if it is accepted by more than half of the individual judges.
\citet{dietrich2007judgment} introduced the \emph{quota rule} as a generalization of the majority rule, where each formula has an given acceptance threshold.
As shown in the introductory example, the majority rule (and the quota rule) does not satisfy consistency.
In this paper, we consider the \emph{uniform premise-based quota rule}~\cite{baumeister2015complexity}, which first applies the quota rule to the premises and then accepts all conclusions that are satisfied by these collectively accepted premises.
The formal definition follows.

\begin{definition}[Uniform Premise-based Quota Rule for $q \in [0,1)$]\label{def:UPQR}
A uniform premise-based quota rule~$\opUPQR_q : {\mathcal{J}(\Phi)}^n \rightarrow 2^{\Phi}$
divides the premises~$\Phi_p$ into two disjoint subsets $\Phi_\qquota$ and $\Phi_\nqquota$, 
where $\Phi_\qquota$ consists of all premises in the non-negated form and $\Phi_\nqquota$ consists of all premises in the negated form.
For each~$\bm{J} \in \mathcal{J}(\Phi)^n$ the outcome $\opUPQR_q(\bm{J})$
is the collective judgment set that contains every premise from~$\Phi_\qquota$ 
that appears at least~$\lfloor q n + 1 \rfloor$ times in the profile~$\bm{J}$, every premise from~$\Phi_\nqquota$ that appears at least~$\lceil n-qn \rceil$
times in the profile, as well as all conclusions that are satisfied by these premises.
\end{definition}

Notice that since $\lfloor q n + 1 \rfloor + \lceil n-qn \rceil=n+1$, it is guaranteed that for each premise~$\alpha \in \Phi_p$, we have either $\alpha \in \opUPQR_q(\bm{J})$ or $\neg \alpha \in \opUPQR_q(\bm{J})$.
Then by the way conclusions are selected, the outcome $\opUPQR_q(\bm{J})$ is always complete and consistent.

The following example formally restates the introductory example.

\myparagraph{Example.}
The premise set~$\Phi_p$ contains two parts $\Phi_\qquota=\{s,c,m,h\}$ and~$\Phi_\nqquota=\{\neg s,\neg c,\neg m,\neg h\}$.
The conclusion set~$\Phi_c$ contains~$\neg s \lor c$, $\neg m \lor \neg h$ and their negations.
The profile is given as follows:

\begin{center}
\begin{tabular}
{p{2.5cm} p{0.3cm} p{0.3cm} p{0.3cm} p{0.3cm} p{0.4cm} p{1cm} p{1.6cm}}
\toprule[1pt]
Judgment Set & $s$ & $c$ & $m$ & $h$  & & $\neg s \lor c$ & $\neg m \lor \neg h$
 \\ \midrule
$J_1$ & 1 & 0 & 1 & 1 & & 0 & 0  \\
$J_2$ & 1 & 1 & 0 & 1 & & 1 & 1  \\
$J_3$ & 0 & 0 & 1 & 0 & & 1 & 1  \\
$\opUPQR_{1/2}$ & 1 & 0 & 1 & 1 & $\Rightarrow$ & 0 & 0 \\
\bottomrule[1pt]
\end{tabular}
\end{center}
In the table we use 1 or 0 to represent whether the formula is contained in the judgment set or not.
As an example, $J_1=\{s, \neg c,m,h,\neg (\neg s \lor c),\neg (\neg m \lor \neg h)\}$.
The collective judgment set is obtained by applying $\opUPQR_q$ with~$q=1/2$.
Thus, to be included in the outcome, every positive premise needs to be accepted by $\lfloor \frac{3}{2} + 1 \rfloor=2$ judges, and every negative premise needs to be accepted by $\lceil 3-\frac{3}{2} \rceil=2$ judges.
We first have that~$s$, $\neg c$ , $m$ and~$h$ are included in the outcome.
Or we just say~$s=1$, $c=0$, $m=1$ and~$h=1$.
Then we get that~$\neg s \lor c=0$ and~$\neg m \lor \neg h=0$,
which means~$\neg (\neg s \lor c)$ and~$\neg (\neg m \lor \neg h)$ are included in the outcome.

\subsection{Decision Variables}
In order to analyze the influence of judges on the outcome~$\opUPQR_q(\bm{J})$, we call a variable~$x$ \emph{decided} by judge~$i$ 
if the outcome with respect to~$x$ is decided by the judgment set of judge $i$, 
i.e., for any judgment set $J^* \in \mathcal{J}(\Phi)$, it holds that
$x \in \opUPQR_q \left(\bm J_{-i}, J^* \right)$ if and only if $x \in J^*$.
In the above example, variables~$c$ and~$m$ are decided by the third judge while~$s$ and~$h$ are not.

According to the definition of $\opUPQR_q$, we get the following characterization for variables decided by a judge.

\begin{obs}
\label{obs:decide-condition}
A variable $x \in \Phi_\qquota$ (resp.\ $x \in \Phi_\nqquota$) is decided by judge $i$ if and only if except for judge $i$ there are exactly $\lfloor q n\rfloor$ (resp.\ $\lceil n-qn-1 \rceil$) judges that accept $x$.
\end{obs}

It follows that if variable $x$ is not decided by judge $i$, then except for judge $i$ the number of judges that accept $x$ is either at most $\lfloor q n\rfloor-1$ or at least $\lfloor q n\rfloor +1$
(resp.\ at most $\lceil n-qn-2 \rceil$ or at least $\lceil n-qn \rceil$).
In both cases the outcome of $x$ is independent of the judgment set of judge $i$.

\begin{obs}
\label{obs:not-decide}
If variable $x$ is not decided by judge $i$, then judge $i$ cannot change the outcome of $x$, i.e., for any judgment set $J^* \in \mathcal{J}(\Phi)$, it holds that
$x \in \opUPQR_q \left(\bm J_{-i}, J^* \right)$ if and only if $x \in \opUPQR_q \left(\bm J\right)$.
\end{obs}



We will use these observations in the analysis of many variants of \textsc{Manipulation}. We call a variable a \emph{decision} variable if the variable is decided by the manipulator.
We do not consider the corresponding definition for conclusions since for our problems only those conclusions whose outcome can be changed by the judge need to be considered and the rest can simply be deleted from the problem input.


\subsection{Clause Restrictions}\label{sec:clause-restrict}

We restrict the conclusions to be clauses and their negations, 
where a clause is defined as a disjunction of literals.
In particular, we consider \emph{positive monotone clauses} which are clauses with no negative literals, \emph{monotone clauses} which are clauses with only positive literals or with only negative literals, \emph{Horn clauses} which are clauses with at most one positive literal, and clauses of bounded length, where the length of a clause is the number of literals contained in it.
Moreover, we generalize these restrictions and define classes of clause restrictions based on a classification
with respect to the number of positive and negative literals in a clause.

\begin{definition}
\label{def: CT}
A clause set $\mathcal{C}$ is called a \textbf{\CT set} if $\mathcal{C}$ is a union of some $\mathcal{S}_i^j$, where $\mathcal{S}_i^j$ is the set of clauses
which contain exactly $i$~literals and exactly~$j$ of them are negative.
Denote $\mathcal{S}_{k}^{0}$ and $\mathcal{S}_{k}^{k}$ as $\mathcal{M}_k^+$ and $\mathcal{M}_k^-$.
\end{definition} 

This classification is useful as most clause classes we care about can be defined as the union of some $\mathcal{S}_i^j$.
For example, positive monotone clauses can be denoted by $\bigcup_{i=1}^{\infty} \mathcal{S}_i^0=\bigcup_{i=1}^{\infty} \mathcal{M}_i^+$,
Horn clauses can be denoted by $\bigcup_{i=1}^{\infty} (\mathcal{M}_i^- \cup \mathcal{S}_i^{i-1})$, and
clauses of length 3 can be denoted by $\bigcup_{j=0}^{3} \mathcal{S}_3^j$.

We remark that our results in this paper can be directly translated to the case where we restrict the conclusions to be conjunctions of literals, and their negations, since the negation of a clause is a conjunction of literals and the agenda is closed under complementation.
In other words, for each pair of clause~$\alpha$ of its negation~$\neg \alpha$ in the agenda, we can rewrite $\neg \alpha$ as a conjunction of literals and $\alpha$ as the negation of a conjunction of literals, such that all conclusions in the agenda are conjunctions of literals or their negations.

\subsection{\textsc{Satisfiability}}
\label{sec:pre-sat}

We will build connections between variants of \textsc{Manipulation} and variants of \textsc{Satisfiability}.
A formula is said to be \emph{satisfiable} if there exists a value assignment that assigns every variable appearing in the formula 1 or 0 such that the formula is valued as 1.
\textsc{Satisfiability} is the problem of deciding whether a given formula is satisfiable.
\textsc{3-Sat} is a restricted variant of \textsc{Satisfiability} where each formula is a conjunction of clauses of length 3.
We define more general variants of \textsc{Satisfiability} based on the clause set. 

\begin{definition}
Let $\mathcal{C}$ be a clause set. \textsc{$\mathcal{C}$-Sat}
is the problem of deciding whether a given formula $C_1 \land \dots \land C_m$ with $C_i \in \mathcal{C}$ is satisfiable or not.
\end{definition}
 
Now \textsc{3-Sat} corresponds to \textsc{$\bigcup_{j=0}^{3} \mathcal{S}_3^j$-Sat}.

\section{Basic Manipulation Problems}
\label{sec:manip}

In this section, we analyze the computational complexity of problems modeling simple variants of strategic behavior
of some judge.
The core idea is that a judge might cast an untruthful judgment set in order to influence
the collective judgment set towards some desired judgment set.
Note that we provide alternative, simpler (yet equivalent) problem definitions compared to those known from the
literature~\cite{baumeister2015complexity}.
In contrast to~\citet{baumeister2015complexity} who focus on the assumption on the preferences of the manipulator over all possible outcomes, which requires rather technical concepts of preference relations between judgment sets,
we take a different approach and directly model the requirements on the preferred outcome.
For example, the simplest variant of manipulation from~\citet{baumeister2015complexity}, \textsc{UPQR-U-Possible-Manipulation},
actually models the question whether the collective outcome is ``robust against one judge providing a faulty
judgment set'' as asked by Justine in the introduction.
Formally, we consider the following problems.

\begin{probox}
\decprobI{\textsc{UPQR Manipulation} basic variants \hfill{\scriptsize (Problem names from\cite{baumeister2015complexity} listed below.)}}
{An agenda $\Phi$, a profile $\bm{J} = ( J_1 , \dots, J_n ) \in  {\mathcal{J}(\Phi)}^n$, the manipulator's desired consistent
(possibly incomplete) set $J \subseteq J_n$, and a uniform rational threshold~$q \in [0,1)$.}\\
\decprobQ{\Robustness \hfill{\scriptsize (=\textsc{UPQR-U-Possible-Manipulation}\cite{baumeister2015complexity})}}
{$\exists J^* \in \mathcal{J}(\Phi): \opUPQR_q(\bm{J}) \cap J \not= \opUPQR_q(\bm{J}_{-n},J^*) \cap J$?}
\decprobQ{\Possible \hfill{\scriptsize (=\textsc{UPQR-CR-Possible-Manipulation}\cite{baumeister2015complexity})}}
{$\exists J^* \in \mathcal{J}(\Phi): (\opUPQR_q(\bm{J}_{-n},J^*) \cap J)  \setminus (\opUPQR_q(\bm{J}) \cap J) \not= \emptyset$?}
\decprobQ{\Necessary \hfill{\scriptsize (=\textsc{UPQR-CR-Necessary-Manipulation}\cite{baumeister2015complexity})}}
{$\exists J^* \in \mathcal{J}(\Phi): \opUPQR_q(\bm{J}) \cap J \subsetneq \opUPQR_q(\bm{J}_{-n},J^*) \cap J$?}
\decprobQ{\Exact \hfill{\scriptsize (=\textsc{UPQR-TR-Necessary-Manipulation}\cite{baumeister2015complexity})}}
{$\exists J^* \in \mathcal{J}(\Phi): J \subseteq \opUPQR_q( \bm{J}_{-n},J^* )$?}
\end{probox}

Intuitively, the manipulator only cares about the formulas in the desired set~$J$, which is a subset of the manipulator's judgment set.
\Robustness asks whether the manipulator can achieve a different outcome with respect to~$J$.
\Possible asks whether the manipulator can achieve an outcome that contains a formula from~$J$
which is not contained in the truthful outcome.
\Necessary asks whether the manipulator can achieve an outcome that contains a formula from~$J$ which is not contained in the truthful outcome, and meanwhile contains all formulas that are in both~$J$ and the truthful outcome.
\Exact asks whether the manipulator can achieve an outcome that contains all formulas from~$J$.

\citet{baumeister2015complexity} showed that all the four variants of \textsc{UPQR Manipulation} with
the desired set being incomplete are \NP-com\-plete.
However, a complex formula in conjunctive normal form is needed in the conclusion set in these reductions.
In this section, we give a more refined analysis by considering how restricting the conclusions to different \CT{} sets influences the computational complexity
of \textsc{UPQR Manipulation} for all four basic variants.
For completeness, before presenting the tractable cases in Section \ref{sec:P-case-basic-M} and intractable cases in Section \ref{sec:hard-case-basic-M}, we first provide the original problem definitions in~\citet{baumeister2015complexity} and argue that our problem definitions are equivalent in Section \ref{sec:original-problems}.

\subsection{Relation to \citet{baumeister2015complexity}}
\label{sec:original-problems}

Before we start the actual analysis our our problems, we show the equivalence between our problem definitions and those in \citet{baumeister2015complexity}.

\subsubsection{Preferences over judgment sets}
In order to define their manipulation problems, \citet{baumeister2015complexity}
used the following the notations for their concepts of preferences over judgment sets.
Let $\PrU$ be the set of all weak orders over $\mathcal{J}(\Phi)$.
For a weak order $\succeq$ over $\mathcal{J}(\Phi)$ and for all $X,Y \in \mathcal{J}(\Phi)$, define $X \succ Y$ by $X \succeq Y$ and $Y \not\succeq X$, and define $X \sim Y$ by $X \succeq Y$ and $Y \succeq X$.
We say $X$ is weakly preferred to $Y$ whenever $X \succeq Y$,
and we say $X$ is preferred to $Y$ whenever $X \succ Y$.
Given some (possibly incomplete) judgment set~$J$, define
\begin{enumerate}
\item the set of \emph{unrestricted $J$-induced (weak) preferences} as the set $\PrU_J$ of weak orders $\succeq$ in $\PrU$
      such that for all $X , Y \in \mathcal{J}(\Phi)$, $X \sim Y$ whenever $X \cap J = Y \cap J$;
\item the set of \emph{top-respecting $J$-induced (weak) preferences} as $\PrTR_J \subseteq \PrU_J$ such that $\succeq$ $\in \PrTR_J$
      if and only if for all $X \in \mathcal{J}(\Phi)$ with $X \cap J \not= J$, it holds that $J \succ X$ ;
\item the set of \emph{closeness-respecting $J$-induced (weak) preferences} as $\PrCR_J \subseteq \PrU_J$ such that $\succeq$ $\in \PrCR_J$
      if and only if for all $X , Y \in \mathcal{J}(\Phi)$ with $Y \cap J \subseteq X \cap J$, we have $X \succeq Y$.
\end{enumerate}

Let $J$, $X$, and $Y$ be three judgment sets for the same agenda $\Phi$, where $J$ is possibly incomplete 
and let $T_J \in \{ \PrU_J , \PrTR_J , \PrCR_J \}$ be a type of $J$-induced preferences.
We say judgment set~$X$ is \emph{necessarily} preferred to judgment set~$Y$ for type $T_J$ if $X \succ Y$ for all $\succeq$ $\in T_J$
and judgment set~$X$ is \emph{possibly} preferred to judgment set~$Y$ for type $T_J$ if there is some $\succeq$ $\in T_J$ with $X \succ Y$.

\subsubsection{Original Problem Definitions}

For each $T \in \{ \PrU, \PrTR, \PrCR\}$ and any rational quota~$q \in [0,1)$, \citet{baumeister2015complexity} defined their \textsc{Manipulation} problems as follows.

\begin{probox}
\decprob{\textsc{$\opUPQR_q$-$T$-Necessary-Manipulation} (resp. \textsc{$\opUPQR_q$-$T$-Possible-Manipulation})}
{An agenda $\Phi$, a profile $\bm{J} = ( J_1 , \dots, J_n ) \in  {\mathcal{J}(\Phi)}^n$, the manipulator's desired consistent
(possibly incomplete) set $J \subseteq J_n$.}
{Does there exist a judgment set $J^* \in \mathcal{J}(\Phi)$ such that
$\opUPQR_q( \bm{J}_{-n},J^* ) \succ \opUPQR_q ( \bm{J} )$
for all $\succeq \in T_J$ (resp.\ for some $\succeq \in T_J$)?} 
\end{probox}

\subsubsection{Alternative Characterizations}

Since the problem definitions in its various variants together with the respective $J$-induces preference models are rather
complicated to parse, we provided alternative, more direct questions that are helpful to analyze and understand the
considered problems at the beginning of this section.
In the following we show that our problem definitions are equivalent to the ones in \citet{baumeister2015complexity} (see also Table \ref{table:def-equal}).
Note that some variants have trivial solutions and can thus be ignored in our analysis.
Moreover, provide a ``motivation''-paragraph for each relevant manipulation variant,
where we discuss potential application settings.

\begin{table*}
\centering
\caption{Relation between original problem definitions in \cite{baumeister2015complexity} and our definitions. We ignore \textsc{$\opUPQR_q$-$\PrU$-Necessary-Manipulation} as it is known to be possibly strategy proof~\cite{baumeister2015complexity}.}
\label{table:def-equal}
\renewcommand{\arraystretch}{1.2}
\begin{tabular}
{lll}
\toprule
Definitions in \cite{baumeister2015complexity} & & Our definitions \\ \midrule
\textsc{$\opUPQR_q$-$\PrU$-Possible-Manipulation} & $\leftrightarrow$ & \Robustness \\
\textsc{$\opUPQR_q$-$\PrTR$-Necessary-Manipulation} & $\leftrightarrow$ & \Exact \\
\textsc{$\opUPQR_q$-$\PrTR$-Possible-Manipulation} & $\leftrightarrow$ & \Robustness \\
\textsc{$\opUPQR_q$-$\PrCR$-Necessary-Manipulation} & $\leftrightarrow$ & \Necessary \\
\textsc{$\opUPQR_q$-$\PrCR$-Possible-Manipulation} & $\leftrightarrow$ & \Possible \\
\bottomrule
\end{tabular}
\end{table*}

\begin{description}
\item[\textsc{$\opUPQR_q$-$\PrU$-Necessary-Manipulation}:] 
  This case is known to be possibly strategy proof, that is, no
  manipulated outcome would be necessarily preferred to truthful outcome~\cite{baumeister2015complexity}.
  Consequently, we ignore this problem variant.
\item[\textsc{$\opUPQR_q$-$\PrU$-Possible-Manipulation}:]
  Recall that unrestricted $J$-induced (weak) preferences only require that two outcomes~$X$ and~$Y$ must be equally good,
  when they are identical with respect to~$J$.
  That is, when~$X$ and~$Y$ not identical with respect to~$J$, then both $X \succ Y$ and $Y \succ X$ are possible.
  Therefore, a manipulator that wants to obtain a possibly better outcome only needs to archive an outcome that is
  different to the original outcome with respect to the desired set.
  
  Thus, \textsc{$\opUPQR_q$-$\PrU$-Possible-Manipulation} asks whether the manipulating judge can achieve a different outcome
  (that contains at least one more or one less formula from the desired set compared to the truthful outcome)
  with respect to the desired set.

  Formally, we have:
  $\exists J^* \in \mathcal{J}(\Phi): \opUPQR_q( \bm{J}_{-n},J^* ) \succ \opUPQR_q (\bm{J})$ for some $\succeq \in \PrU_J$? $\Leftrightarrow$ \\
  $\exists J^* \in \mathcal{J}(\Phi): \opUPQR_q(\bm{J_{-n}},J^*) \cap J \neq \opUPQR_q(\bm{J}) \cap J$?
  
  Thus, $\opUPQR_q$-$\PrU$-Possible-Manipulation is equivalent to \Robustness.
  
  \textbf{Motivation:} This problem can be seen as a question for \textbf{stability} or \textbf{robustness}~\cite{BFKNST2017}.
  Interpreting the desired set as ``set of relevant propositions'', it asks whether a particular judge can change
  the collective outcome with respect to  relevant propositions e.g.\ by changing her mind or making a mistake.
  Possible applications are when we know that some specific judge (formally the manipulator) is not certain or perfectly
  qualified for a reliable evaluation and we want to know whether we can trust the aggregated outcome with respect to specific
  important formulas (formally the desired set).

\item[\textsc{$\opUPQR_q$-$\PrTR$-Necessary-Manipulation}:]
  Recall that top-respecting $J$-induced (weak) preferences only require that~$J$ is preferred to every outcome
  that does not already contain~$J$.
  That is, an outcome~$X$ is \emph{necessarily }better than an outcome~$Y$ when~$Y$ did not contain~$J$ but~$X$ does.

  Thus, $\opUPQR_q$-$\PrTR$-Necessary-Manipulation asks whether the manipulator can achieve an outcome that contains
  the desired set, and in addition the truthful outcome does not contain the whole desired set.
  
  Formally, we have:
  $\exists J^* \in \mathcal{J}(\Phi): \opUPQR_q( \bm{J}_{-n},J^* ) \succ \opUPQR_q (\bm{J})$ for all $\succeq \in \PrTR_J$? $\Leftrightarrow$ \\
  $\exists J^* \in \mathcal{J}(\Phi): J \subseteq \opUPQR_q( \bm{J_{-n}},J^* )$ and $J \not \subseteq \opUPQR_q (\bm{J})$?
  
  Since it is trivial to check the second condition $J \not \subseteq \opUPQR_q (\bm{J})$,
  $\opUPQR_q$-$\PrTR$-Necessary-Manipulation is essentially equivalent to \Exact.
  
  \textbf{Motivation:} The motivation for this manipulation variant is straightforward.
  Whenever one is aware of the precise goal of a potential manipulator, one can ask whether the
  manipulator can fully reach that goal.
\item[\textsc{$\opUPQR_q$-$\PrTR$-Possible-Manipulation}:]
  Recall that top-respecting $J$-induced (weak) preferences only require that~$J$ is preferred to every outcome
  that does not already contain~$J$.
  That is, an outcome~$X$ is only required to be better than an outcome~$Y$ when~$Y$ did not contain~$J$ but~$X$ does.
  Thus, when~$Y$ does not contain~$J$, then $X$~is \emph{possibly} better than~$Y$ whenever it is different from~$Y$ with respect to~$J$.

  Thus, $\opUPQR_q$-$\PrTR$-Possible-Manipulation asks whether the manipulator can can achieve a different outcome with respect to the desired set,
  and in addition the truthful outcome does not contain the desired set.

  Formally, we have:
  $\exists J^* \in \mathcal{J}(\Phi): \opUPQR_q( \bm{J}_{-n},J^* ) \succ \opUPQR_q (\bm{J})$ for some $\succeq \in \PrTR_J$? $\Leftrightarrow$ \\
  $\exists J^* \in \mathcal{J}(\Phi): \opUPQR_q(\bm{J_{-n}},J^*) \cap J  \not= \opUPQR_q(\bm{J}) \cap J$ and $J \not \subseteq \opUPQR_q (\bm{J})$? \\
  Since it is trivial to check the second condition $J \not \subseteq \opUPQR_q (\bm{J})$,
  $\opUPQR_q$-$\PrTR$-Possible-Manipulation is essentially the same as \Robustness.
  
\item[{$\opUPQR_q$-$\PrCR$-Necessary-Manipulation}:]
  Recall that closeness-respecting $J$-induced (weak) preferences require that an outcome~$X$ must be better than an outcome~$Y$
  when the set of formulas from~$J$ included in~$X$ is a superset of the set of formulas from~$J$ included in~$Y$.
  That is, an outcome~$X$ is \emph{necessarily} better than an outcome~$Y$ when~$X$ contain all formulas from~$J$ also included in $Y$ plus some more.

  Thus, $\opUPQR_q$-$\PrCR$-Necessary-Manipulation asks whether the manipulator can achieve an outcome that contains
  at least one conclusion from the desired set that was not in the truthful outcome, and at the same time
  contains all formulas from the desired set that were part of the truthful outcome.
  
  Formally, we have:
  $\exists J^* \in \mathcal{J}(\Phi): \opUPQR_q( \bm{J}_{-n},J^* ) \succ \opUPQR_q (\bm{J})$ for all $\succeq \in \PrCR_J$? $\Leftrightarrow$ \\
  $\exists J^* \in \mathcal{J}(\Phi): \opUPQR_q(\bm{J}) \cap J \subsetneq \opUPQR_q(\bm{J_{-n}},J^*) \cap J?$ \\
  This is the same as \Necessary.
  
  \textbf{Motivation:} This variant is relevant when the goal set (desired set) of the manipulator is known and
  the valuation of the manipulator is strictly monotone (that is, obtaining one more desired formula is an improvement).
  Strict monotonicity is useful to model a risk-averse or conservative manipulator who would only manipulate
  if no formula from the desired set has to be given up.

\item[{$\opUPQR_q$-$\PrCR$-Possible-Manipulation}:]
  Recall that closeness-respecting $J$-induced (weak) preferences only requires that an outcome~$Y$ must be better than an outcome~$X$
  when the set of formulas from~$J$ included in~$Y$ is a superset of the set of formulas from~$J$ included in~$X$.
  That is, an outcome~$X$ is \emph{possibly} better than~$Y$ whenever it contains some formula from~$J$ which is not already included in~$Y$.
  
  Thus, $\opUPQR_q$-$\PrCR$-Possible-Manipulation asks whether the manipulator can achieve an outcome
  that contains at least one formula from the desired set that was not in the truthful outcome.

  Formally, we have:
  $\exists J^* \in \mathcal{J}(\Phi): \opUPQR_q( \bm{J}_{-n},J^* ) \succ \opUPQR_q (\bm{J})$ for some $\succeq \in \PrCR_J$? $\Leftrightarrow$ \\
  $\exists J^* \in \mathcal{J}(\Phi): (\opUPQR_q(\bm{J_{-n}},J^*) \cap J)  \setminus (\opUPQR_q(\bm{J}) \cap J) \not= \emptyset$? \\
  This is the same as \Possible.
  
  \textbf{Motivation:} This variant can be very useful if we have limited information about the manipulator.
  In case we know the absolute goal (formally the desired set of formulas) but not the respective priorities,
  we should consider this manipulator model since it allows to detect potentially dangerous situations where
  a manipulator may ``exchange'' one formula of the desired set from the truthful outcome by another,
  more prioritized formula.

\end{description}

\subsubsection{Minor Definition Details}

While we have seen that the manipulation variants we consider resemble those known in the literature, 
our actual definition of the decision problems still slightly differ from those in the literature~\cite{baumeister2015complexity}
as we put the threshold value~$q$ as part of the input.
For our polynomial-time algorithms, the quota is only interesting for computing which variables can be decided by
the manipulator. Thus, it has no influence on the computational complexity.
Our hardness reductions usually assume that $q=1/2$ and $n=3$, but they can all be adapted to work for any rational quota~$q$, due to the following lemma.

\begin{lemma}
\label{lem:any-q}
For any input instance~$I$ consisting of an agenda $\Phi$, a profile $\bm{J}= ( J_1 , J_2, J_3 ) \in  {\mathcal{J}(\Phi)}^3$, the manipulator's desired set $J \subseteq J_3$, and a threshold~$\frac{1}{2}$, for any $q \in[0,1)$, we can compute in polynomial time an ``equivalent'' instance~$I'$ consisting of the same agenda $\Phi$, a profile $\bm{J}'=( J'_1 , \dots, J'_n )  \in  {\mathcal{J}(\Phi)}^n$ with $J'_n=J_3$ and $n=\max\{\lceil\frac{1}{q}\rceil,\lceil\frac{1}{1-q}\rceil\}+1$, the manipulator's desired set $J'=J$, and the threshold~$q$, such that $\opUPQR_\frac{1}{2}(\bm{J}_{-3},J^*)=\opUPQR_q(\bm{J}'_{-n},J^*)$ for any $J^* \in \mathcal{J}(\Phi)$.
\end{lemma} 

\begin{proof}
We first make some observations about the desired instance $I'$.
Denote $\tau=\lfloor q n + 1 \rfloor$.
From $n=\max\{\lceil\frac{1}{q}\rceil,\lceil\frac{1}{1-q}\rceil\}+1$
we get $qn>q\lceil\frac{1}{q}\rceil \ge 1$ and hence $qn+1>2$.
Similarly, from $n=\max\{\lceil\frac{1}{q}\rceil,\lceil\frac{1}{1-q}\rceil\}+1$ we get $(1-q)n>(1-q)\lceil\frac{1}{1-q}\rceil \ge 1$ and hence $qn+1<n$.
Thus we have~$qn+1 \in (2,n)$, and then $\tau=\lfloor q n + 1 \rfloor \in [2,n-1]$.
For any variable $x\in \Phi$, if $x \in J'_i$ for $1\le i\le n-1$, then $x \in \opUPQR_q(\bm{J}'_{-n},J^*)$ for any $J^* \in \mathcal{J}(\Phi)$, and hence $x$ is not decided by the manipulator.
Similarly, if $\neg x \in J'_i$ for $1\le i\le n-1$, then $\neg x \in \opUPQR_q(\bm{J}'_{-n},J^*)$ for any $J^* \in \mathcal{J}(\Phi)$, and hence $x$ is not decided by the manipulator.
On the other hand, if except for the manipulator there are exactly $\tau-1$ judges that accept $x$, then $x$ is decided by the manipulator.

Based on these observations, we construct the profile~$\bm{J}'\in  {\mathcal{J}(\Phi)}^n$ with $J'_n=J_3$ as follows. 
For each variable~$x \in \Phi$ that is not decided by the manipulator in $I$, if $x \in \opUPQR_\frac{1}{2}(\bm{J})$, then we make $x \in J'_i$ for $1\le i\le n-1$ such that $x \in \opUPQR_q(\bm{J}')$ and $x$ is not decided by the manipulator; otherwise we make $\neg x \in J'_i$ for $1\le i\le n-1$ such that $\neg x \in \opUPQR_q(\bm{J}')$ and $x$ is not decided by the manipulator.
For each variable~$x \in \Phi$ that is decided by the manipulator in $I$, from the remaining $n-1$ judges we make exactly $\tau-1$ judges accept $x$ such that $x$ is decided by the manipulator.
This finishes the premise part.
The conclusion part follows trivially since all judgment sets should be complete and consistent.

Finally, we show $\opUPQR_\frac{1}{2}(\bm{J}_{-3},J^*)=\opUPQR_q(\bm{J}'_{-n},J^*)$ for any $J^* \in \mathcal{J}(\Phi)$.
It suffices to show that $\opUPQR_\frac{1}{2}(\bm{J}_{-3},J^*) \cap \Phi_p=\opUPQR_q(\bm{J}'_{-n},J^*) \cap \Phi_p$ for any $J^* \in \mathcal{J}(\Phi)$.
For each variable $x\in \Phi$, if $x$ is not decided by the manipulator in $I$, then by construction $x$ is not decided by the manipulator in $I'$ and $x \in \opUPQR_\frac{1}{2}(\bm{J}_{-3},J^*)$ if and only if $x \in \opUPQR_q(\bm{J}'_{-n},J^*)$.
If $x$ is decided by the manipulator in $I$, then by construction $x$ is also decided by the manipulator in $I'$, and hence 
\[
x \in \opUPQR_\frac{1}{2}(\bm{J}_{-3},J^*)
\Leftrightarrow
x \in J^*
\Leftrightarrow
x \in \opUPQR_q(\bm{J}'_{-n},J^*).
\]
\end{proof}

We remark that the constructed equivalent instance~$I'$ in Lemma 1 is required to have
at least $n=\max\{\lceil\frac{1}{q}\rceil,\lceil\frac{1}{1-q}\rceil\}+1$
judges such that we can create variables not decided by the manipulator.
From the perspective of parameterized complexity analysis, this leaves two special cases: (1) instances with high quotas and small numbers of judges where a variable is accepted if all judges accept it and (2) instances with low quotas and small numbers of judges where a variable is accepted if at least one judge accepts it.
We discuss how to adapt our hardness reductions for these two cases in Appendix \ref{sec:app-3-2}.

\subsection{Tractable Cases of Manipulation}
\label{sec:P-case-basic-M}
We start our analysis with \Robustness and \Possible
which turn out to be linear-time solvable when the conclusions are just simple clauses.
We first show that the manipulator cannot change the outcome of a premise which is in the desired set but not in the truthful outcome due to the monotonicity of quota rules.

\begin{lemma}
\label{lem: Only gain conclusions}
For any manipulated judgment set~$J^* \in \mathcal{J}(\Phi)$, we have 
\[(\opUPQR_q(\bm{J}_{-n},J^*) \setminus \opUPQR_q(\bm{J})) \cap (J\cap \Phi_p)=\emptyset.\] 
\end{lemma} 

\begin{proof}
Assume towards a contradiction that there exists a judgment set~$J^*$ such that~
$$A=(\opUPQR_q(\bm{J}_{-n},J^*) \setminus \opUPQR_q(\bm{J})) \cap (J\cap \Phi_p) \neq \emptyset.$$
Let~$x$ be a variable such that $x \in A$ or $\neg x \in A$.
We show that $x \in A$ will lead to a contradiction. 
The other case can be proved analogously.
Suppose $x \in A$, then we have~$x \in J \subseteq J_n$ and~$x \not \in \opUPQR_q(\bm{J})$.
By definition, this implies that variable~$x$ is not decided by the manipulator.
By \cref{obs:not-decide}, the outcome of $x$ should be independent of the manipulator, which contradicts with $x \in \opUPQR_q(\bm{J}_{-n},J^*) \setminus \opUPQR_q(\bm{J})$.
\end{proof}

\begin{lemma}
\label{lem: Possible clause is in P}
\Robustness and \Possible
with conclusions being clauses are solvable in linear time.
\end{lemma} 

\begin{proof}
We first show the result for \Possible.
According to the definition of \Possible and \cref{lem: Only gain conclusions}, the question is whether there exists a manipulated judgment set~$J^* \in \mathcal{J}(\Phi)$ such that~$(\opUPQR_q(\bm{J}_{-n},J^*)  \setminus \opUPQR_q(\bm{J})) \cap (J \cap \Phi_c )\not= \emptyset$.
That is, at least one target \emph{conclusion} from the desired set which was not in the truthful outcome should be included in the manipulated outcome. 
We first compute the set~$D$ of decision variables in linear time.
Then, for every target conclusion from $J \setminus \opUPQR_q(\bm{J})$, we check one by one whether it can be satisfied
by just changing values of variables in $D$. 
The manipulation is successful if and only if at least one target conclusion from $J \setminus \opUPQR_q(\bm{J})$
is satisfiable. 
Since every conclusion is a clause (disjunction of literals)
or a negation of a clause (conjunction of literals),
checking its satisfiability can be done in time linear in the clause size.

For \Robustness,
the question is whether the manipulator can change the outcome of a formula in the desired set~$J$.
We first compute the set~$D$ of decision variables in linear time.
If there exists a variable~$x \in D$ such that~$x \in J$ or~$\neg x \in J$, then this instance can be easily manipulated as the manipulator can change the outcome of a premise in~$J$.
Otherwise, if for every variable~$x \in D$, we have that~$x \not \in J$ and~$\neg x \not \in J$, 
then a successful manipulation has to influence the outcomes of conclusions in~$J$.
That is, we need to find a manipulated judgment set~$J^*$ such that~$\opUPQR_q(\bm{J}_{-n},J^*) \cap (J \cap \Phi_c ) \not= \opUPQR_q(\bm{J}) \cap (J \cap \Phi_c )$.
For every conclusion from~$J \setminus \opUPQR_q(\bm{J})$, we check one by one whether it can be satisfied by just changing values of variables in~$D$.
For every conclusion from~$J \cap \opUPQR_q(\bm{J})$, we check one by one whether its negation can be satisfied by just changing values of variables in~$D$.
Since every conclusion is a clause (disjunction of literals)
or a negation of a clause (conjunction of literals),
checking its satisfiability can be done in time linear in the clause size.
\end{proof}
%



Next, we show that \Necessary boils down to solving a related \textsc{Satisfiability} problem.

\begin{lemma}
\label{lem: P to P}
\Necessary and \Exact with conclusions chosen from clause set~$\mathcal{C}$ \footnote{For convenience, when we say conclusions are from a clause set $\mathcal{C}$, we mean every conclusion is either a clause in $\mathcal{C}$ or the negation of a clause in $\mathcal{C}$.} can be solved by solving at most $|\Phi_c|$ instances of \textsc{$\mathcal{C}$-Sat}.
\end{lemma} 

\begin{proof}
We first show the result for \Necessary.
In \Necessary we should find a manipulated judgment set~$J^* \in \mathcal{J}(\Phi)$ such that~$\opUPQR_q(\bm{J}) \cap J \subsetneq \opUPQR_q(\bm{J}_{-n},J^*) \cap J$.
That is, the manipulated result~$\opUPQR_q(\bm{J}_{-n},J^*)$ should not only contain one more formula~$C^*$ from~$J \setminus \opUPQR_q(\bm{J})$, but also contain all formulas in 
$Q_0=\opUPQR_q(\bm{J}) \cap J$. 
According to \cref{lem: Only gain conclusions}, this formula~$C^*$ cannot be a premise, and hence, $C^*$ should be a conclusion.
Moreover, since the manipulated result should contain all formulas in $Q_0$, specifically, all premises in~$Q_0$, the manipulator is only allowed to change the values of variables from~$D^*=D \setminus \{x \mid x \in Q_0 \lor \neg x \in Q_0\}$, 
where~$D$ is the set of decision variables.
So the problem is to check whether there exists a conclusion~$C^* \in J \setminus \opUPQR_q(\bm{J})$
such that a set~$Q=(Q_0 \cap \Phi_c) \cup \{C^*\}$ of conclusions can be satisfied by just controlling the values of
variables from~$D^*$.

We can simply try all possible~$C^* \in J \setminus \opUPQR_q(\bm{J})$ and for each~$C^*$ check the set~$Q=(Q_0 \cap \Phi_c)\cup \{C^*\}$.
Each conclusion in~$Q$ is either a clause from~$\mathcal{C}$ or a negation of a clause from~$\mathcal{C}$.
To satisfy a negative clause (conjunction of literals), the values of all variables in this clause are fixed.
For all negative clauses in $Q$, we first check whether they are consistent, which can be done in linear time.
If all negative clauses in $Q$ are consistent, then we get the value for all variables in them.
Next, we need to check for each such variable whether the corresponding value can be satisfied, that is, either the value is the original value before
the manipulation or the variable is in~$D^*$.
If not, these negative clauses can not be satisfied. 
After this, we only need to check whether the remaining positive clauses in $Q$ can be satisfied.
Since all clauses in $Q$ are chosen from $\mathcal{C}$, the remaining problem forms an instance of \textsc{$\mathcal{C}$-Sat}.

For \Exact, the question is whether there exists a manipulated judgment set~$J^* \in \mathcal{J}(\Phi)$ such that~$J \subseteq \opUPQR_q( \bm{J}_{-n},J^* )$.
According to \cref{lem: Only gain conclusions}, if~$J \cap \Phi_p \not \subseteq \opUPQR_q(\bm{J})$, then for any~$J^*$, we have~$J \cap \Phi_p \not \subseteq \opUPQR_q( \bm{J}_{-n},J^* )$, which means there is no successful manipulation.
Hence in the following we can assume that~$J \cap \Phi_p  \subseteq \opUPQR_q(\bm{J})$.
Then the goal is to satisfy all conclusions in~$J \cap \Phi_c$ by changing the values of variables in~$D \setminus \{x \mid x \in J \lor \neg x \in J\}$, 
where~$D$ is the set of decision variables.
This can be reduced to an instance of \textsc{$\mathcal{C}$-Sat} in the same way as in \Necessary.
\end{proof}

As a corollary, every polynomial-time algorithm for \textsc{$\mathcal{C}$-Sat} can be adapted to these two variants of \textsc{Manipulation}.

\begin{corollary}
\label{cor: P to P}
Let $\mathcal{C}$ be a clause set. If \textsc{$\mathcal{C}$-Sat} is in P, then \Necessary and \Exact with conclusions chosen from $\mathcal{C}$ are in P.
\end{corollary} 

In the next section we show the other direction holds for all \CT sets.

\subsection{
Intractable Cases of Manipulation: 
Manipulation vs. Satisfiability}
\label{sec:hard-case-basic-M}

In this section, we give a full characterization for the computational complexity of \Necessary
by showing that \textsc{$\mathcal{C}$-Sat} and \Necessary with conclusions chosen
from $\mathcal{C}$ are actually \PvsNPequiv{}
when $\mathcal{C}$ is a \CT set (see \cref{def: CT} for the definition of \CT set).


\begin{theorem}
\label{thm: equivalence}
Let $\mathcal{C}$ be a \CT set, then \Necessary with conclusions chosen from $\mathcal{C}$ and \textsc{$\mathcal{C}$-Sat} are \PvsNPequiv{}.
\end{theorem}

In order to prove \cref{thm: equivalence}, 
we first identify the type of \CT sets $\mathcal{C}$ for which \textsc{$\mathcal{C}$-Sat} is \NP-hard. 

\begin{lemma}
\label{lem: 2 NP-hard cases}
Let $\mathcal{C}$ be a \CT set, then \textsc{$\mathcal{C}$-Sat} is \NP-hard if and only if 
 \begin{enumerate}
 \item  there is a pair of $i,j$ with $i \geq 3$ and $0<j<i$ such that $\mathcal{M}_{2}^+ \cup \mathcal{M}_{2}^- \cup \mathcal{S}_i^j \subseteq \mathcal{C}$, or
\item there is a pair of $k_1,k_2$ with $\max\{k_1,k_2\} \geq 3$ and $\min\{k_1,k_2\} \geq 2$ such that $\mathcal{M}_{k_1}^+ \cup \mathcal{M}_{k_2}^- \subseteq \mathcal{C}$.
\end{enumerate}
\end{lemma}


\begin{proof}
$\Rightarrow$ If~$\mathcal{C}$ does not contain any~$\mathcal{M}_{k_1}^+$ with $k_1 \ge 2$, then \textsc{$\mathcal{C}$-Sat} can solved in linear time as follows.
Given a formula $f$ of \textsc{$\mathcal{C}$-Sat}, we first take out all clauses in $f$ that consists of only one positive literal (clauses from $\mathcal{M}_1^+$) and set all these variables to 1.
If this makes any one of the remaining clauses in $f$ unsatisfied, then $f$ is unsatisfiable.
Otherwise, all remaining clauses contain at least one negative variable, which can be satisfied simultaneously by setting all remaining variables to 0.
Similarly, if~$\mathcal{C}$ does not contain any~$\mathcal{M}_{k_2}^-$ with $k_2 \ge 2$, then \textsc{$\mathcal{C}$-Sat} can solved in linear time.
Therefore, $\mathcal{C}$ must contain at least one $\mathcal{M}_{k_1}^+$ with $k_1 \ge 2$ and at least one $\mathcal{M}_{k_2}^-$ with $k_2 \ge 2$. 
In addition, if~$\mathcal{C}$ does not contain any clause set which contains clauses of length at least 3, then~\textsc{$\mathcal{C}$-Sat} is solvable in polynomial time since~\textsc{2-SAT} is solvable in polynomial time.
Therefore, $\mathcal{C}$ must contain some~$\mathcal{S}_i^j$ with~$i \ge 3$ and $1<j<i$ or some~$\mathcal{M}_k^+$ (or $\mathcal{M}_k^-$) for some~$k \ge 3$.

$\Leftarrow$ For the first case, we show \textsc{$(\mathcal{M}_{2}^+ \cup \mathcal{M}_{2}^- \cup \mathcal{S}_i^j)$-SAT} is \NP-hard by reducing from \textsc{$(\mathcal{M}_i^+ \cup \mathcal{M}_{2}^-)$-SAT} with~$i \ge 3$, which is \NP-hard according to \cref{lem:2-3-SAT-NPC} in Appendix~\ref{sec:app-missing}.
It suffices to show that for any clause $x_1 \lor \dots \lor x_i \in \mathcal{M}_i^+$, there is an equivalent formula in \textsc{$(\mathcal{M}_{2}^+ \cup \mathcal{M}_{2}^- \cup \mathcal{S}_i^j)$-SAT}.
This is true since 
$$x_1 \lor \dots \lor x_i \Leftrightarrow
(\neg y_1 \lor \dots \lor \neg y_j \lor x_{j+1} \lor \dots \lor x_i) 
\land (y_1 \lor x_1) \land \dots \land (y_j \lor x_j),$$
where the formula on the right side is a conjunction of clauses from $\mathcal{M}_{2}^+ \cup \mathcal{S}_i^j$.

For the second case,
since \textsc{$(\mathcal{M}_{k_1}^+ \cup \mathcal{M}_{k_2}^-)$-SAT} and \textsc{$(\mathcal{M}_{k_2}^+ \cup \mathcal{M}_{k_1}^-)$-SAT} are equivalent under linear-time reductions, we can assume $k_1 \le k_2$.
First, according to \cref{lem:2-3-SAT-NPC} in Appendix~\ref{sec:app-missing}, \textsc{$(\mathcal{M}_2^+ \cup \mathcal{M}_{k_2}^-)$-SAT} with $k_2 \geq 3$ is \NP-hard.
Then to show other cases, we construct a reduction from \textsc{$(\mathcal{M}_{k_1-1}^+ \cup \mathcal{M}_{k_2}^-)$-SAT} to \textsc{$(\mathcal{M}_{k_1}^+ \cup \mathcal{M}_{k_2}^-)$-SAT}.
It suffices to show that for any clause $X_{k_1-1}=x_1 \lor \dots \lor x_{k_1-1}$, there is an equivalent formula in \textsc{$(\mathcal{M}_{k_1}^+ \cup \mathcal{M}_{k_2}^-)$-SAT}.
This is true since 
$$X_{k_1-1} \Leftrightarrow
(X_{k_1-1} \lor y_1) \land \dots \land (X_{k_1-1} \lor y_{k_2}) \land (\neg y_1 \lor \dots \lor \neg y_{k_2}).$$
\end{proof}

Combining \cref{thm: equivalence} and \cref{lem: 2 NP-hard cases} we get a full characterization for the computational complexity of \Necessary with conclusions chosen from a \CT set $\mathcal{C}$.
As a corollary, the NP-hardness of \Necessary holds even if conclusions are clauses with length 3 or monotone clauses.

\begin{corollary}
\label{cor: manipulation for 3sat and monotone sat is NP-c}
Let $\mathcal{C}=\cup_{j=0}^3 \mathcal{S}_3^j$ or $\mathcal{C}=\cup_{k=1}^{\infty} (\mathcal{M}_k^+ \cup \mathcal{M}_k^-)$.
Then,
\Necessary with conclusions chosen from $\mathcal{C}$ is \NP-hard.
\end{corollary}


According to \cref{lem: 2 NP-hard cases}, we need to consider two cases.
We cover these two cases in the following by \cref{lem: from M(k)-SAT to M(k+1)}, \cref{lem: 3M- or 2M+ is complete}, and \cref{lem: M+ M- S is NP-c}.
The main idea for the proofs of these lemmas is as follows.
By definition, an instance of \Necessary with conclusions chosen from $\mathcal{C}$ is
a yes-instance if and only if there is one target conclusion $C^* \in J \setminus \opUPQR_q(\bm{J})$ such that $C^*$
and all formulas in $Q_0=\opUPQR_q(\bm{J}) \cap J$
can be included in the manipulated outcome simultaneously. 
Since all formulas in~$Q_0$ are in the original outcome $\opUPQR_q(\bm{J})$,
we already know that all formulas in~$Q_0$ can be satisfied simultaneously.
The question is whether it is possible to satisfy one more clause $C^* \notin Q_0$.
Therefore, \cref{thm: equivalence} implies that the additional information ``all formulas in~$Q_0$ can be satisfied simultaneously'' does not help to efficiently determine whether all conclusions in~$Q=Q_0 \cup \{C^*\}$ can be satisfied simultaneously.
Our reductions in the following lemmas will reflect this phenomenon.

We first prove a weaker version of \cref{thm: equivalence} in the following \cref{lem: from M(k)-SAT to M(k+1)}, where the clauses sets in \textsc{Satisfiability} and \textsc{Manipulation} are similar but not the same.

\begin{lemma}
\label{lem: from M(k)-SAT to M(k+1)}
If \textsc{$(\mathcal{M}_{k_1}^+ \cup \mathcal{M}_{k_2}^-)$-Sat} is \NP-hard, then \Necessary with conclusions chosen from a closely related \CT set $\mathcal{C}=\mathcal{M}_{k_1+1}^+ \cup \mathcal{M}_{k_2}^-$ is \NP-hard.
\end{lemma}

\begin{proof}
We present a po\-ly\-no\-mial-time reduction from \textsc{$(\mathcal{M}_{k_1}^+ \cup \mathcal{M}_{k_2}^-)$-Sat} to \Necessary
with conclusions chosen from $\mathcal{C}=\mathcal{M}_{k_1+1}^+ \cup \mathcal{M}_{k_2}^-$.
Given an instance 
$$f=C_1^+ \land \dots \land C_{m_1}^+ \land C_1^- \land \dots \land C_{m_2}^-$$
of \textsc{$(\mathcal{M}_{k_1}^+ \cup \mathcal{M}_{k_2}^-)$-Sat}, where $C_i^+ \in \mathcal{M}_{k_1}^+$ and $C_i^- \in \mathcal{M}_{k_2}^-$,
we construct an instance of \Necessary with~$q=\frac{1}{2}$ as follows (see also \cref{table: from M(k)-SAT to M(k+1)}).
The agenda contains all variables~$x_1, \dots, x_n$ that appear in $f$ and their negations.
In addition, we create~$y_1, \dots, y_{k_2}$ and their negations in premises.
Then we add~$C_i^+ \lor y_1$ for~$1 \le i \le m_1$, $C_i^-$ for~$1 \le i \le m_2$, $\neg  y_1 \lor \dots \lor \neg  y_{k_2}$ and their negations as conclusions.
The set of judges is~$N=\{1,2,3\}$. 
The manipulator is the third judge and his desired set~$J$ consists of all positive conclusions. 
Note that the manipulator is decisive for variables~$x_1, \dots, x_n$ and~$ y_1$. 
We now show that~$f$ is satisfiable if and only if the manipulation is feasible.

\begin{table*}
\centering
\caption{Instance of \Necessary with conclusion set~$\mathcal{C}=\mathcal{M}_{k_1+1}^+ \cup \mathcal{M}_{k_2}^-$ for the proof of \cref{lem: from M(k)-SAT to M(k+1)}.}
\label{table: from M(k)-SAT to M(k+1)}
\begin{tabular}
{p{2.5cm}llllllll p{1.5cm} p{0.8cm} p{2.5cm}}
\toprule
Judgment Set & $x_1$ & $\dots$ & $x_n$ & $y_1$ & $y_2$ & $\dots$ & $y_{k_2}$ & & $C_i^+ \lor y_1$ & $C_i^-$ & $\neg  y_1 \lor \dots \lor \neg  y_{k_2}$ \\ \midrule
$J_1$ & 1 & $\dots$ & 1 & 1 & 1 & $\dots$ & 1 & & 1 & 0 & 0\\
$J_2$ & 0 & $\dots$ & 0 & 0 & 1 & $\dots$ & 1 & & 0 & 1 & 1\\
$J_3$ & 0 & $\dots$ & 0 & 1 & 0 & $\dots$ & 0 & & 1 & 1 & 1 \\
$\opUPQR_{1/2}$ & 0 & $\dots$ & 0 & 1 & 1 & $\dots$ & 1 & $\Rightarrow$ & 1 & 1 & 0 \\
\bottomrule
\end{tabular}
\end{table*}

$\Rightarrow$ Suppose that~$f$ is satisfiable, then there is a value assignment~$x_i^*, 1 \le i \le n$ such that all clauses in~$f$ are satisfied. 
So the manipulator can set $x_i=x_i^*,1 \le i \le n$ to satisfy conclusions $C_i^+ \lor y_1$ for~$1 \le i \le m_1$ and $C_i^-$ for~$1 \le i \le m_2$.
The remaining conclusion $\neg  y_1 \lor \dots \lor \neg  y_{k_2}$ can be satisfied by setting $y_1=0$.
Thus the manipulation is feasible.  

$\Leftarrow$ Suppose that the manipulation is feasible.
Since all positive conclusions except for~$\neg  y_1 \lor \dots \lor \neg  y_{k_2}$ are already in the truthful outcome~$\opUPQR_{1/2}(\bm{J})$,
to make a successful manipulation, the manipulator has to make the manipulated outcome contain all positive conclusions.
Specifically, for conclusion~$\neg  y_1 \lor \dots \lor \neg  y_{k_2}$, since~$y_j=1,j \geq 2$ can not be changed by the manipulator, the manipulator has to set $y_1=0$.
Then, to satisfy all remaining conclusions~$C_i^+ \lor  y_1$ and $C_i^-$ is equivalent to setting values for $x_1, \dots, x_n$
to satisfy $f=C_1^+ \land \dots \land C_{m_1}^+ \land C_1^- \land \dots \land C_{m_2}^-$. 
\end{proof}



Note that in \cref{lem: from M(k)-SAT to M(k+1)} the two clause sets~$\mathcal{M}_{k_1}^+ \cup \mathcal{M}_{k_2}^-$ (in~\textsc{Satisfiability}) and~$\mathcal{M}_{k_1+1}^+ \cup \mathcal{M}_{k_2}^-$ (in \textsc{Manipulation}) are not the same.
This leaves a gap when conclusions of \Necessary are chosen from~$\mathcal{M}_2^+ \cup \mathcal{M}_3^-$ (or equivalently~$\mathcal{M}_3^+ \cup \mathcal{M}_2^-$):
We cannot adopt \cref{lem: from M(k)-SAT to M(k+1)}, since the corresponding \textsc{Satisfiability} problem is
\textsc{$(\mathcal{M}_1^+ \cup \mathcal{M}_3^-)$-SAT} (or \textsc{$(\mathcal{M}_2^+ \cup \mathcal{M}_2^-)$-SAT}), which is not \NP-hard (cf.\ \cref{lem: 2 NP-hard cases}).
Next we close this gap by giving a more involved reduction to show the \NP-hardness for the case when conclusions are chosen from~$\mathcal{M}_2^+ \cup \mathcal{M}_3^-$.

\begin{lemma}
\label{lem: 3M- or 2M+ is complete}
\Necessary with conclusions chosen from~$(\mathcal{M}_2^+ \cup \mathcal{M}_3^-)$ is \NP-hard.
\end{lemma}

\begin{table*}
\caption{Instance of \Necessary with conclusion set~$\mathcal{C}=\mathcal{M}_2^+ \cup \mathcal{M}_3^-$ for the proof of \cref{lem: 3M- or 2M+ is complete}.}
\label{table: M2M3}
\centering
\begin{tabular}
{p{2.5cm}p{0.2cm}p{0.2cm}p{0.2cm}p{0.2cm}p{0.2cm}p{0.4cm}p{1cm}p{3cm}p{1.5cm}}
\toprule
Judgment Set & $x_i$ & $y_i$ & $z_i$ & $w$ & $v$ & & $w \lor v$ & $\neg x_{i_1} \lor \neg x_{i_2} \lor \neg x_{i_3}$ & $z_{i_1} \lor z_{i_2}$  \\ \midrule
$J_1$ & 1 & 1 & 1 & 1 & 0 & & 1 & 0 & 1\\
$J_2$ & 0 & 0 & 0 & 0 & 0 & & 0 & 1 & 0\\
$J_3$ & 0 & 1 & 1 & 0 & 1 & & 1 & 1 & 1\\
$\opUPQR_{1/2}$ & 0 & 1 & 1 & 0 & 0 & $\Rightarrow$ & 0 & 1 & 1\\
\bottomrule
\end{tabular}

\begin{tabular}
{p{2.5cm}p{3cm}p{3cm}p{1.5cm}p{1.5cm}}
\toprule
Judgment Set & $\neg x_i \lor \neg y_i \lor \neg w$ & $\neg y_i \lor \neg z_i \lor \neg w$ & $x_i \lor y_i$ & $y_i \lor z_i$  \\ \midrule
$J_1$ & 0 & 0 & 1 & 1\\
$J_2$ & 1 & 1 & 0 & 0\\
$J_3$ & 1 & 1 & 1 & 1\\
$\opUPQR_{1/2}$ & 1 & 1 & 1 & 1\\
\bottomrule
\end{tabular}
\end{table*}

\begin{proof}
We present a polynomial-time reduction from \textsc{$(\mathcal{M}_2^+ \cup \mathcal{M}_3^-)$-SAT}.
Given an instance $f_1 \land f_2$ of \textsc{$(\mathcal{M}_2^+ \cup \mathcal{M}_3^-)$-SAT}, where $f_1$ is a conjunction of 
clauses of the form ``$x_{i_1} \lor x_{i_2}$'' from $\mathcal{M}_2^+$ with $x_{i_1}, x_{i_2} \in \{x_1, \dots, x_n\}$ and
$f_2$ is a conjunction of clauses of the form ``$\neg x_{i_1} \lor \neg x_{i_2} \lor \neg x_{i_3}$'' from $\mathcal{M}_3^-$
with $x_{i_1}, x_{i_2},x_{i_3} \in \{x_1, \dots, x_n\}$,
we construct an instance of \Necessary with conclusions chosen from $\mathcal{M}_2^+ \cup \mathcal{M}_3^-$ and~$q=\frac{1}{2}$ as follows (see also \cref{table: M2M3}).
The agenda contains~$x_i,y_i,z_i$ $(1 \le i \le n)$, $w$, $v$ and their negations as the premise set.
The conclusion set consists of the following conclusions and their negations.

\begin{itemize}
\item $w \lor v$.
\item $\neg x_{i_1} \lor \neg x_{i_2} \lor \neg x_{i_3}$ for every original clause~$\neg x_{i_1} \lor \neg x_{i_2} \lor \neg x_{i_3}$ in~$f_2$.
\item $z_{i_1} \lor z_{i_2}$ for every original clause~$x_{i_1} \lor x_{i_2}$ in~$f_1$.
Note that original variables~$x_i$ $(1 \le i \le n)$ are replaced by variables~$z_i$ $(1 \le i \le n)$.
\item $x_i \lor y_i$, $y_i \lor z_i$, $\neg x_i \lor \neg y_i \lor \neg w$ and~$\neg y_i \lor \neg z_i \lor \neg w$ for every~$1 \le i \le n$. 
\end{itemize}

The set of judges is~$N=\{1,2,3\}$. 
The manipulator is the third judge and his desired set~$J$ consists of all positive conclusions. 
The manipulator is decisive for all variables except for~$v$. 
We now show that~$f_1 \land f_2$ is satisfiable if and only if the manipulation is feasible.

$\Rightarrow$ Assume that~$f_1 \land f_2$ is satisfiable, then there is a value assignment~$x_i^*, 1 \le i \le n$ such that all clauses in~$f_1 \land f_2$ are satisfied. 
So the manipulator can set $x_i=z_i=x_i^*,1 \le i \le n$ to satisfy conclusions $\neg x_{i_1} \lor \neg x_{i_2} \lor \neg x_{i_3}$ and $z_{i_1} \lor z_{i_2}$.
All remaining positive conclusions can be satisfied by setting $w=1$ and $y_i=-x_i^*$ for $1 \le i \le n$.
Thus the manipulation is feasible.  
Recall that the manipulator is decisive for all variables except for~$v$.

$\Leftarrow$ Assume that the manipulation is feasible.
Since all positive conclusions, except for~$w \lor v$, are already in the truthful outcome~$\opUPQR_{1/2}(\bm{J})$,
the assumption that the manipulation is feasible means that there is a value assignment for all variables with~$v=0$ (since~$v$ is not decided by the manipulator) such that all positive conclusions can be satisfied.
Specifically, for conclusion~$w \lor v$, since~$v=0$ can not be changed by the manipulator, the manipulator has to set $w=1$.
Then we have that
\[
(\neg x_i \lor \neg y_i \lor \neg w) \land (x_i \lor y_i) \stackrel{w=1}{\Rightarrow} x_i=-y_i,
\]
\[
(\neg y_i \lor \neg z_i \lor \neg w) \land (y_i \lor z_i) \stackrel{w=1}{\Rightarrow} y_i=-z_i.
\]
This means in this value assignment~$x_i=z_i$.
Since all conclusions~$\neg x_{i_1} \lor \neg x_{i_2} \lor \neg x_{i_3}$ and~$z_{i_1} \lor z_{i_2}$ can be satisfied by this value assignment with~$x_i=z_i$,
we have that~$f_1 \land f_2$ is satisfiable.
\end{proof}



In a similar way, we can prove the result of \cref{thm: equivalence} for the first case in \cref{lem: 2 NP-hard cases}.
Before that, we show the following symmetric equivalence.

\begin{obs}
\label{obs: symmetric equivalence}
\Necessary with conclusions chosen from~
$\bigcup_{k=1}^{\ell} \mathcal{S}_{i_k}^{j_k}$
 and \Necessary with conclusions chosen from~
$\bigcup_{k=1}^{\ell} \mathcal{S}_{i_k}^{i_k-j_k}$
are equivalent under linear-time reductions.
\end{obs}

\begin{proof}
For these two problems, any instance of one problem can be transformed into an equivalent instance of the other problem by replacing every variable $x_i$ with its negation $\neg x_i$, and vice versa, and replacing the quota $q$ with $q'$ such that $\lfloor q n + 1 \rfloor=\lceil n-q'n \rceil$, where $\lfloor q n + 1 \rfloor$ is the number of judges needed for a variable $x_i$ to be included in the outcome of one problem and $\lceil n-q'n \rceil$ is the number  of judges needed for its negation $\neg x_i$ to be included in the outcome of the other problem.
If $qn$ is an integer, we set $q'=1-q-\frac{1}{n}$.
Otherwise, we set $q'=1-q$.
It is easy to verify that in both cases we have $q' \in [0,1)$ and $\lfloor q n + 1 \rfloor=\lceil n-q'n \rceil$.
\end{proof}

\begin{lemma}
\label{lem: M+ M- S is NP-c}
\Necessary with conclusions chosen from~$\mathcal{M}_2^+ \cup \mathcal{M}_2^- \cup \mathcal{S}_i^j$, where~$i \geq 3$ and~$0<j<i$, is \NP-hard.
\end{lemma}

\begin{proof}
We first prove the result for $i=3$. 
We present a polynomial-time reduction from \textsc{$(\mathcal{M}_2^+ \cup \mathcal{M}_2^- \cup \mathcal{S}_3^2)$-Sat} to \Necessary with conclusions chosen from~$\mathcal{M}_2^+ \cup \mathcal{M}_2^- \cup \mathcal{S}_3^2$.
Then, according to \cref{obs: symmetric equivalence}, \Necessary with conclusions chosen from~$\mathcal{M}_2^+ \cup \mathcal{M}_2^- \cup \mathcal{S}_3^1$ is also \NP-hard.

Given an instance~$f_1(X) \land f_2(X) \land f_3(X)$ of \textsc{$(\mathcal{M}_2^+ \cup \mathcal{M}_2^- \cup \mathcal{S}_3^2)$-Sat},
where $f_1(X)$ is a conjunction of clauses from $\mathcal{M}_2^+$, $f_2(X)$ is a conjunction of clauses from $\mathcal{M}_2^-$, $f_3(X)$ is a conjunction of clauses from $\mathcal{S}_3^2$, and all variables in~$f_1(X) \land f_2(X) \land f_3(X)$ are from $X=\{x_1,x_2,\dots,x_n\}$.
We denote by $f_1(Z)$ the formula after replacing every $x_k$ $(1\le k \le n)$ by $z_k$ in $f_1(X)$.
Since
\begin{align*}
f_1(X) \Leftrightarrow& f_1(Z) \land \left(\bigwedge_{k=1,2,\dots,n}x_k=z_k\right)\\
\Leftrightarrow& f_1(Z) \land \left(\bigwedge_{k=1,2,\dots,n}(\neg x_k \lor \neg y_k) \land (x_k \lor y_k) \land (\neg y_k \lor \neg z_k) \land (y_k \lor z_k)\right), 
\end{align*}
we construct an instance of \Necessary with~$q=\frac{1}{2}$ as follows (see also \cref{table: MMS}).
The agenda contains~$x_k,y_k,z_k$ for~$1 \le k \le n$, $w,v$ and their negations as premises.
The conclusion set consists of the following clauses and their negations.
\begin{itemize}
\item All clauses in $f_1(Z)$, $f_2(X)$ and $f_3(X)$.
\item $\neg w \land \neg v$.
\item $x_k \lor y_k$, $y_k \lor z_k$, $\neg x_k \lor \neg y_k \lor w$ and $\neg y_k \lor \neg z_k \lor w$ for every~$1 \le k \le n$.
\end{itemize}

\begin{table*}
\caption{Instance of \Necessary with conclusion set~$\mathcal{M}_2^+ \cup \mathcal{M}_2^- \cup \mathcal{S}_i^j$ for the proof of \cref{lem: M+ M- S is NP-c}.}
\label{table: MMS}
\centering
\begin{tabular}
{p{2.5cm}p{0.2cm}p{0.2cm}p{0.2cm}p{0.2cm}p{0.2cm}p{0.2cm}p{2.7cm}p{2.7cm}p{2.7cm}}
\toprule
Judgment Set & $x_k$ & $y_k$ & $z_k$ & $w$ & $v$ & & clauses in $f_1(Z)$ & clauses in $f_2(X)$ & clauses in $f_3(X)$ \\ \midrule
$J_1$ & 1 & 1 & 1 & 1 & 1 & & 1 & 0 & 1\\
$J_2$ & 0 & 0 & 0 & 0 & 1 & & 0 & 1 & 1\\
$J_3$ & 0 & 1 & 1 & 1 & 0 & & 1 & 1 & 1\\
$\opUPQR_{1/2}$ & 0 & 1 & 1 & 1 & 1 & $\Rightarrow$ & 1 & 1 & 1\\
\bottomrule
\end{tabular}
\begin{tabular}
{p{2.5cm}p{2cm}p{3cm}p{3cm}p{1.5cm}p{1.5cm}}
\toprule
Judgment Set & $\neg w \lor \neg v$ & $\neg x_k \lor \neg y_k \lor w$ & $\neg y_k \lor \neg z_k \lor w$ & $x_k \lor y_k$ & $y_k \lor z_k$ \\ \midrule
$J_1$ & 0 & 1 & 1 & 1 & 1\\
$J_2$ & 1 & 1 & 1 & 0 & 0\\
$J_3$ & 1 & 1 & 1 & 1 & 1\\
$\opUPQR_{1/2}$ & 0 & 1 & 1 & 1 & 1\\
\bottomrule
\end{tabular}
\end{table*}

Similar to the proof for \cref{lem: 3M- or 2M+ is complete}, the manipulator is the third judge who is decisive for all premises except for $v$.
The desired set contains all positive clauses. 
The truthful outcome contains all positive clauses except for $\neg w \land \neg v$ with $x_k=0, y_k=1, z_k=1, w=v=1$.
Similar to the proof for \cref{lem: 3M- or 2M+ is complete}, we now show that~$f_1(X) \land f_2(X) \land f_3(X)$ is satisfiable if and only if the manipulation is feasible.

$\Rightarrow$ Assume that~$f_1(X) \land f_2(X) \land f_3(X)$ is satisfiable, then there is a value assignment~$x_k^*, 1 \le k \le n$ such that all clauses in~$f_1(X) \land f_2(X) \land f_3(X)$ are satisfied. 
So the manipulator can set $x_k=z_k=x_k^*$ $(1 \le k \le n)$ to satisfy conclusions in $f_1(Z)$, $f_2(X)$ and $f_3(X)$.
All remaining positive conclusions can be satisfied by setting $w=0$ and $y_k=-x_k^*$ for $1 \le k \le n$.
Thus the manipulation is feasible.  
Recall that the manipulator is decisive for all variables except for~$v$.

$\Leftarrow$ Assume that the manipulation is feasible.
Since all positive conclusions, except for~$\neg w \lor \neg v$, are already in the truthful outcome~$\opUPQR_{1/2}(\bm{J})$,
the manipulated result must contain all positive conclusions.
So there is a value assignment for all variables with~$v=1$ such that all positive conclusions are satisfied.
Specifically, for conclusion~$\neg w \lor \neg v$, since~$v=1$ can not be changed by the manipulator, the manipulator has to set $w=0$.
Then we have that
\[
(\neg x_k \lor \neg y_k \lor w) \land (x_k \lor y_k) \stackrel{w=0}{\Rightarrow} x_k=-y_k,
\]
\[
(\neg y_k \lor \neg z_k \lor w) \land (y_k \lor z_k) \stackrel{w=0}{\Rightarrow} y_k=-z_k.
\]
This means in this value assignment~$x_k=z_k$ $(1 \le k \le n)$.
Since all conclusions in~$f_1(Z)$, $f_2(X)$ and $f_3(X)$ can be satisfied by this value assignment with~$x_k=z_k$ $(1 \le k \le n)$,
we have that~$f_1(X) \land f_2(X) \land f_3(X)$ is satisfiable.

When $i>3$, we can assume $j \geq 2$ according to \cref{obs: symmetric equivalence}.
We can make a reduction from \textsc{$(\mathcal{M}_2^+ \cup \mathcal{M}_2^- \cup \mathcal{S}_i^j)$-Sat} similarly as what we did in the above reduction for~$i=3$.
The only difference is that now clauses $\neg x_k \lor \neg y_k \lor w$ and $\neg y_k \lor \neg z_k \lor w$ are not allowed in conclusions since their length is $3 \not =i$.
To fix this, we just need to create some dummy variables whose value is fixed and not decided by the manipulator and add them to $\neg x_k \lor \neg y_k \lor w$ and $\neg y_k \lor \neg z_k \lor w$ such that they are in~$\mathcal{S}_i^j$.
Formally, we first create~$u_1,u_2,\dots,u_{i-3}$, where in the truthful outcome we have that~$u_1=u_2=\dots=u_{j-2}=1$ and~$u_{j-1}=\dots=u_{i-3}=0$, and they are not decided by the manipulator.
Then we add them into $\neg x_k \lor \neg y_k \lor w$ and $\neg y_k \lor \neg z_k \lor w$ in the following way:
\[
\neg x_k \lor \neg y_k \lor w \leadsto (\neg x_k \lor \neg y_k \lor \neg u_1 \lor \dots \lor \neg u_{j-2}) \lor (w \lor u_{j-1} \lor \dots \lor u_{i-3}) \in \mathcal{S}_i^j,
\]
\[
\neg y_k \lor \neg z_k \lor w \leadsto (\neg y_k \lor \neg z_k \lor \neg u_1 \lor \dots \lor \neg u_{j-2}) \lor (w \lor u_{j-1} \lor \dots \lor u_{i-3}) \in \mathcal{S}_i^j.
\]
\end{proof}

We are now ready to prove \cref{thm: equivalence}.

\begin{proof}[Proof of \cref{thm: equivalence}]
According to \cref{cor: P to P}, it suffices to show that when \textsc{$\mathcal{C}$-Sat} is \NP-hard,
then \Necessary with conclusions chosen from~$\mathcal{C}$ is \NP-hard.
According to \cref{lem: 2 NP-hard cases}, we need to consider two cases.

Case 1:
$\mathcal{M}_{2}^+ \cup \mathcal{M}_{2}^- \cup \mathcal{S}_i^j \subseteq \mathcal{C}$, where $i \geq 3$ and $0<j<i$.
This case has been proved in \cref{lem: M+ M- S is NP-c}.

Case 2: $\mathcal{M}_{k_1}^+ \cup \mathcal{M}_{k_2}^- \subseteq \mathcal{C}$, where $\max\{k_1,k_2\} \geq 3, \min\{k_1,k_2\} \geq 2$.
According to \cref{obs: symmetric equivalence}, it suffices to consider the case when $k_1 \le k_2$.
If $k_1 \geq 3$, then according to \cref{lem: 2 NP-hard cases}, \textsc{$(\mathcal{M}_{k_1-1}^+ \cup \mathcal{M}_{k_2}^-)$-Sat} is \NP-hard.
Then according to \cref{lem: from M(k)-SAT to M(k+1)}, \Necessary with conclusions chosen from~$\mathcal{M}_{k_1}^+ \cup \mathcal{M}_{k_2}^-$ is \NP-hard.
If $k_1=2$ and $k_2 \ge 4$, then according to \cref{lem: 2 NP-hard cases}, \textsc{$(\mathcal{M}_{2}^+ \cup \mathcal{M}_{k_2-1}^-)$-Sat} is \NP-hard.
Then according to \cref{lem: from M(k)-SAT to M(k+1)}, \Necessary with conclusions chosen from~$\mathcal{M}_{2}^+ \cup \mathcal{M}_{k_2}^-$ is \NP-hard.
The only remaining case is when conclusions are chosen from~$\mathcal{M}_{2}^+ \cup \mathcal{M}_{3}^-$, which is shown to be \NP-hard in \cref{lem: 3M- or 2M+ is complete}.
\end{proof}

Next, we show that the same equivalence holds for \Exact, since all reductions in \cref{lem: from M(k)-SAT to M(k+1),lem: 3M- or 2M+ is complete,lem: M+ M- S is NP-c} can also be used for \Exact. 
\begin{theorem}
\label{thm: Exact equivalent}
Let $\mathcal{C}$ be a \CT set, then 
\Exact with conclusions chosen from $\mathcal{C}$ and \textsc{$\mathcal{C}$-Sat} are \PvsNPequiv.
\end{theorem}

\begin{proof}
According to \cref{cor: P to P}, we just need to prove that when  when \textsc{$\mathcal{C}$-Sat} is \NP-hard, \Exact with conclusions clause from~$\mathcal{C}$ is also \NP-hard.
In all reductions in \cref{lem: from M(k)-SAT to M(k+1),lem: 3M- or 2M+ is complete,lem: M+ M- S is NP-c}, the desired set consists of all positive conclusions and the truthful outcome contains all but one of them, thus to achieve \Necessary is the same as to achieve \Exact, i.e., the manipulated outcome should contain all positive conclusions ($\opUPQR_q( \bm{J}_{-n},J^* )=J$). 
That means all these reductions can be directly used to prove that the corresponding \Exact is \NP-hard.
Then following the same line in the proof for \cref{thm: equivalence}, we can prove \textsc{$\mathcal{C}$-Sat} and \Exact with conclusions clause from~$\mathcal{C}$ are \PvsNPequiv.
\end{proof}

Finally, we remark that our core contribution when showing our P vs.\ NP dichotomy (\cref{thm: equivalence}) 
can be also interpreted as a pure equivalence statement about variants of \textsc{Satisfiability}.
Recall that when \textsc{$\mathcal{C}$-Sat} is NP-hard, \cref{thm: equivalence} implies that the additional information ``all conclusions included in the truthful outcome can be satisfied simultaneously'' does not help to efficiently determine whether one more conclusion can be satisfied simultaneously.
Accordingly, we introduce the following variant of \textsc{Satisfiability}.

\begin{probox}
\decprob{\textsc{Almost Satisfiable $\mathcal{C}$-Sat}} 
{A \CT set~$\mathcal{C}$ and a formula $C_1 \land \dots \land C_m$ with~$C_i \in \mathcal{C}$ $(1 \le i \le m)$  knowing that $C_1 \land \dots \land C_{m-1}$ is satisfiable.}
{Is formula $C_1 \land \dots \land C_m$ satisfiable?
}
\end{probox}

\begin{proposition}
\label{pro:almost-sat}
Let $\mathcal{C}$ be a \CT set, then  \textsc{$\mathcal{C}$-Sat} and \textsc{Almost Satisfiable $\mathcal{C}$-Sat} are equivalent under polynomial-time many-one reductions.
\end{proposition}

\begin{proof}
Note that the equivalence under polynomial-time Turing reductions between these two problems are trivial.
Here we show the equivalence under polynomial-time many-one reductions.
The reduction from \textsc{Almost Satisfiable $\mathcal{C}$-Sat} to \textsc{$\mathcal{C}$-Sat} is trivial.
For the other direction, if for a \CT set~$\mathcal{C}$, \textsc{$\mathcal{C}$-Sat} is in P, then for any instance of \textsc{$\mathcal{C}$-Sat}, we can first decide in polynomial time its satisfiability, and then reduce it to a trivial yes/no-instance of \textsc{Almost Satisfiable $\mathcal{C}$-Sat}.
The remaining case is when \textsc{$\mathcal{C}$-Sat} is \NP-hard.
For this case, we can adopt the same idea used in the proof of \cref{thm: equivalence}.
Recall that for the reductions in \cref{lem: from M(k)-SAT to M(k+1),lem: 3M- or 2M+ is complete,lem: M+ M- S is NP-c}, there exists a value assignment for all variables such that all but one positive conclusions are satisfied (in the truthful outcome), and the goal of the manipulator is to find a value assignment for decision variables such that all positive conclusions are satisfied.
The only difference 
is that in \textsc{Manipulation} we can construct a clause where some variables in this clause are decided by the manipulator, while the rest variables are not.
In other words, in \textsc{Manipulation} we can make use of constant 0 or 1.
This, however, is not easy to achieve in \textsc{$\mathcal{C}$-Sat}.
Thus we need to prove that for a \CT set~$\mathcal{C}$ such that \textsc{$\mathcal{C}$-Sat} is \NP-hard, we can use clauses in $\mathcal{C}$ to create a constant 0 and 1 (i.e., to enforce a variable to be 0 or 1).

We show that we can create a constant 1. A constant 0 can be created analogously.
According to \cref{lem: 2 NP-hard cases}, if \textsc{$\mathcal{C}$-Sat} is \NP-hard, then~$\mathcal{C}$ has to contain both $\mathcal{M}_{k_1}^+$ and $\mathcal{M}_{k_2}^-$ with $k_1,k_2 \ge 2$.
We create our target variable $x$, and a set of $(k_1-1)k_2$ variables $Y=\{y_1,y_2,\dots,y_{(k_1-1)k_2}\}$.
Then we create two groups of clauses such that to satisfy all of them, the value of~$x$ must be 1.

For the first group, we partition $Y$ into $k_1-1$ subsets, each with~$k_2$ consecutive variables (e.g., $\{y_1,y_2,\dots,y_{k_2}\}$).
For each subset, we create a clause in~$\mathcal{M}_{k_2}^-$, which is a disjunction of negations of all variables in this subset (e.g., $\neg y_1 \lor \dots \lor \neg y_{k_2}$).
To satisfy every clause, at least one variable from each subset need to be 0,
and hence at least $k_1-1$ variables from~$Y$ need to be 0.

For the second group, for any $k_1-1$ variables in~$Y$, we create a clause in~$\mathcal{M}_{k_1}^+$ which is a disjunction of $x$ and all these $k_1-1$ variables (e.g., $x \lor y_1 \dots \lor y_{k_1-1}$).
Since there are~$\binom{(k_1-1)k_2}{k_1-1}$ different choices, we create~$\binom{(k_1-1)k_2}{k_1-1}$ such clauses for the second group.
Among all these choices, there is at least one choice $y_{i_1}, \dots  ,y_{i_{k_1-1}}$ such that all of them are 0 due to clauses in the first group.
Then in the corresponded clause~$x \lor y_{i_1} \dots \lor y_{i_{k_1-1}}$, $x$ must be 1.
Therefore, to satisfy all clauses in the first and the second group, we have to set~$x=1$.
That is, we create a constant $x=1$ by adding~$k_1-1+\binom{(k_1-1)k_2}{k_1-1}$ clauses.
Note that for a fixed $\mathcal{C}$, $k_1$ and $k_2$ are constants.
\end{proof}

\section{Hamming Distance Based Manipulation}
\label{sec:HDmanip}

We now move on to \Hamming, which is the very first variant of \textsc{Manipulation}
analyzed by \citet{endriss2012complexity} for the majority threshold~$q=1/2$.
In \Hamming, the manipulator wants to make the outcome ``closer'' to his desired set and the distance between the outcome and his desired set is measured by the Hamming distance.
This models the case when each conclusion is equally important to the manipulator, so that the manipulator only cares about the number of formulas in the desired set~$J$ achieved by the collective judgment set.
The formal definition is given as follows\footnote{Similar to before, we put the threshold value~$q$ as part of the input and this has no influence on the computational complexity due to \cref{lem:any-q}.}.

\begin{probox}
\decprob{\Hamming}
{An agenda $\Phi$, a profile $\bm{J} = ( J_1 , \dots, J_n ) \in  {\mathcal{J}(\Phi)}^n$, the manipulator's desired consistent
(possibly incomplete) set $J \subseteq J_n$, and a uniform rational threshold~$q \in [0,1)$.}
{Does there exist a judgment set $J^* \in \mathcal{J}(\Phi)$ such that
$\HD(J,\opUPQR_q(\bm{J}_{-n},J^*)) < \HD(J,\opUPQR_q(\bm{J}))$?
} 
\end{probox}

Herein, the Hamming distance~$\HD(J,S)$ between the possibly incomplete desired set~$J$ and a complete collective judgment set~$S$ is the number of formulas in~$J$ which are not contained in~$S$, i.e. $\HD(J,S)=|J \setminus S|$.

Without loss of generality, in this section we assume that~$J=J_n \cap \Phi_c$,
that is, the desired set contains all conclusions from~$J_n$ but no premise: 
Every instance of \Hamming can be easily transformed into an equivalent instance with~$J=J_n \cap \Phi_c$ as follows.
If for some conclusion~$\varphi$ none of~$\varphi$ and~$\neg \varphi$ appears in~$J$,
then just delete~$\varphi$ and~$\neg \varphi$ from the agenda. 
If there is some premise~$x$ with~$x \in J$ (or $\neg x \in J$), we can remove it from~$J$, then create two clauses~$x \lor x'$ and~$\neg (x \lor x')$ in the conclusions and add~$x \lor x'$ (or $\neg (x \lor x')$) to~$J$,
where~$x'$ is a dummy variable with~$x' \not\in J_i$ for all~$1 \le i \le n$.
Note that doing so we just add positive monotone clauses with two literals into the conclusion set.


\citet{baumeister2015complexity} proved that \Hamming is \NP-hard
for positive monotone clauses.
In this section we show that this problem is \NP-hard even for positive monotone clauses of length~$\ell=3$
by reducing from a natural variant of \textsc{Vertex Cover}, which could be interesting on its own.
When the clause~length is 2, we show the problem is in P for positive monotone clauses, but \NP-hard for monotone clauses or Horn clauses.

\subsection{Condition for a Successful Manipulation}
\label{sec:HD-manipulation-condition}
In this section we give a sufficient and necessary condition for a successful manipulation when conclusions are positive monotone clauses.
First we classify all variables into the following four different classes with respect to their value in the truthful outcome~$\opUPQR_q(\bm{J})$ and the judgment set of the manipulator~$J_n$:


\begin{enumerate}
\item $P_1^1=\{x \in  \Phi_p \mid x \in J_n \land x \in \opUPQR_q(\bm{J})\}$;
\item $P_1^0=\{x \in  \Phi_p \mid x \not \in J_n \land x \in \opUPQR_q(\bm{J})\}$;
\item $P_0^0=\{x \in  \Phi_p \mid x \not \in J_n \land x \not \in \opUPQR_q(\bm{J})\}$;
\item $P_0^1=\{x \in  \Phi_p \mid x \in J_n \land x \not \in \opUPQR_q(\bm{J})\}$.
\end{enumerate} 


Next we give some definitions used in this section.

\begin{definition}
A variable~$x$ is called \emph{useful} if it is decided by the manipulator and there exists a positive conclusion~$\varphi \in J \setminus \opUPQR_q(\bm{J})$ containing~$x$.
A positive monotone clause~$\varphi$ is called a \emph{good} conclusion if~${\varphi \in J \setminus \opUPQR_q(\bm{J})}$ and is called a \emph{bad} conclusion if~$\varphi \not \in J \cup \opUPQR_q(\bm{J})$, or equivalently, $\neg \varphi \in J \cap \opUPQR_q(\bm{J})$ (recall that $J$ is complete with respect to the conclusion set).
\end{definition}

According to the definition, changing the value of a useful variable can make the outcome include at least one more desired (good) conclusion~$\varphi$ with~$\varphi \in J \setminus \opUPQR_q(\bm{J})$,
but it could also make the outcome include some undesired (bad) conclusion~$\varphi$ with~$\varphi \not \in J \cup \opUPQR_q(\bm{J})$, thus lose the desired conclusion~$\neg \varphi \in J \cap \opUPQR_q(\bm{J})$.
We observe that all useful variables are from~$P_0^0$ due to the monotonicity of quota rules.

\begin{obs}
\label{obs: useful variables in P00}
If~$x$ is a useful variable, then~$x \in P_0^0$.
\end{obs}

\begin{proof}
By definition, variables from $P_0^1 \cup P_1^0$ are not decided by the manipulator, so they are not useful.
For a variable~$x \in P_1^1$, any positive monotone clause $\varphi$ containing $x$ is already in~$J \cap \opUPQR_q(\bm{J})$, thus changing the value of~$x$ from 1 to 0 cannot make the outcome include any good conclusion from~$J \setminus \opUPQR_q(\bm{J})$.
Hence, variables from~$P_1^1$ are not useful.
Therefore, all useful variables are from~$P_0^0$.
\end{proof}


\myparagraph{Example.}
Consider the following profile $\bm{J} = ( J_1 , J_2, J_3 ) \in  {\mathcal{J}(\Phi)}^3$, where the manipulator is the third judge and its desired set $J=J_3 \cap \Phi_c$.
\begin{center}
\begin{tabular}{p{2.5cm}p{0.2cm}p{0.2cm}p{0.2cm}p{0.2cm}p{0.2cm}p{0.4cm}  p{1.5cm}p{1.5cm}p{1.5cm}p{1.5cm}}
\toprule[1pt]
Judgment Set & $x_1$ & $x_2$ & $x_3$ & $x'_3$  & $x_4$  & & $x_1 \lor x_2$ & $x_2 \lor x_3$ & $x_3 \lor x'_3$ & $x_3 \lor x_4$ \\ \midrule
$J_1$ & 1 & 0 & 1 & 1 & 1 & & 1 & 1 & 1 & 0 \\
$J_2$ & 0 & 0 & 0 & 0 & 1 & & 0 & 0 & 0 & 1 \\
$J_3$ & 1 & 1 & 0 & 0 & 0 & & 1 & 1 & 0 & 0 \\
$\opUPQR_{1/2}$ & 1 & 0 & 0 & 0 & 1 & $\Rightarrow$ & 1 & 0 & 0 & 1 \\
\bottomrule[1pt]
\end{tabular}
\end{center}
Variables~$x_1$,~$x_3$, and~$x'_3$ are decided by the manipulator, but~$x_1 \in P_1^1$ is not useful since changing its value from 1 to 0 would only exclude~$x_1 \lor x_2 \in J$ from the outcome.
Conclusion~$x_2 \lor x_3 \in J \setminus \opUPQR_q(\bm{J})$ is good, and changing~$x_3$ from 0 to 1 will make~$x_3$ and~$x_2 \lor x_3$ included in the outcome.
Conclusion~$x_3 \lor x'_3 \not \in J \cup \opUPQR_q(\bm{J})$ is bad, and changing~$x_3$ or~$x'_3$ from 0 to 1 will make~$x_3 \lor x'_3$ included in the outcome, and hence its negation~$\neg (x_3 \lor x'_3) \in J$ will be excluded from the outcome.

Now we give a sufficient and necessary condition for a successful manipulation.

\begin{lemma}
\label{lem: condition for HD manipulation}
An instance of \Hamming with all conclusions being positive monotone clauses is a yes-instance if and only if there is a set~${T \subseteq P_0^0}$ of useful variables, such that after changing their values from 0 to 1,
the number of good conclusions included in the outcome is strictly larger than the number of bad
conclusions included in the outcome:
\begin{equation*}
\begin{split}
|\{ \varphi \in J \setminus \opUPQR_q(\bm{J}) \mid T_{\varphi} \cap T \neq \emptyset \}| > 
|\{ \varphi \not \in J \cup \opUPQR_q(\bm{J}) \mid T_{\varphi} \cap T \neq \emptyset \}|,
\end{split}
\end{equation*}
where~$T_{\varphi}$ is the set of variables appearing in clause~$\varphi$.
\end{lemma}

\begin{proof}
The ``if'' direction is obvious and we prove the ``only if'' direction as follows.
According to the definition of useful variables, if an instance is a yes-instance, then the manipulator can achieve a better outcome by only changing the values of useful variables.
According to \cref{obs: useful variables in P00}, all useful variables are from $P_0^0$.
To prove this lemma, it suffices to show that we just need to consider good conclusions and bad conclusions.
If a positive monotone clause~$\varphi$ is neither good nor bad, then it must be~$\neg \varphi \in J \setminus \opUPQR_q(\bm{J})$ or~$\varphi \in J \cap \opUPQR_q(\bm{J})$.
In both cases we have that~$\varphi \in \opUPQR_q(\bm{J})$. 
Changing the values of variables in~$P_0^0$ from 0 to 1 will not change the value of~$\varphi$ ($\varphi$ is still in~$\opUPQR_q(\bm{J})$ after this change).
Therefore, we just need to consider the influence on the number of good conclusions and bad conclusions after changing the values of useful variables.
\end{proof}

%

\subsection{Positive Monotone Clauses of Length~$\ell=2$}
In this section we show \Hamming with positive monotone clauses of length~$\ell=2$ is solvable in polynomial time by a reduction to the \WMDS (\WMDSshort) problem, which can be solved in polynomial time by a reduction to the \MC problem~\cite{goldberg1984finding}.

\begin{probox}
\decprob{\WMDS}
{An undirected graph $G=(V,E)$ with nonnegative rational edge weights $w(e)$ and vertex weights $w(v)$, and a nonnegative rational number~$k$.}
{Does there exist a vertex subset $V' \subseteq V$ with~$\sum_{v \in V'}w(v)>0$ such that
$$\frac{\sum_{e \in E(G[V'])} w(e)}{\sum_{v \in V'}w(v)}>k,$$
where $G[V']$ is the subgraph induced by $V'$?
} 
\end{probox}

\begin{theorem}
\label{thm: l=2 positive monotone is in P}
\Hamming with positive monotone clauses of length~$\ell=2$ is solvable in polynomial time.
\end{theorem}

\begin{proof}
According to \cref{lem: condition for HD manipulation}, we need to find a set~$T \subseteq P_0^0$ of useful variables such that the number of good conclusions containing variables from $S$ is strictly larger than that of bad conclusions.
For any good conclusion~$\varphi \in J \setminus \opUPQR_q(\bm{J})$, since~$\varphi \in J \subseteq J_n$, there exists one variable $x$ in~$\varphi$ such that~$x \in J_n$.
Moreover, since $\varphi \not\in \opUPQR_q(\bm{J})$ we have~$x \not\in \opUPQR_q(\bm{J})$.
Thus~$\varphi$ contains at least one variable~$x \in P_0^1$, which is not decided by the manipulator.
Hence a good conclusion~$\varphi$ of length~2 contains at most one useful variable from~$P_0^0$.
However, a bad conclusion~$\varphi' \not \in J \cup \opUPQR_q(\bm{J})$ of length 2 may contain two useful variables from~$P_0^0$.
Thus, if we change the values of a set~$T \subseteq P_0^0$ of useful variables and sum up the number of included bad conclusions for each variable,
then some bad conclusions will be counted twice.

To solve this issue, we create a weighted graph~$G=(V,E)$ as follows (see also \cref{fig: weighted graph}):
First,
for every useful variable~$x \in P_0^0$, create a vertex~$v \in V$ and assign it a weight~$w(v)=n_v-p_v$, where~$n_v$ is the number of bad conclusions containing~$x$ and~$p_v$ is the number of good conclusions containing~$x$.
Thus~$w(v)$ is the increased Hamming distance when a single variable~$x$ is changed. 
Second,
for every pair of vertices~$u$ and~$v$, create an edge between them if there is a bad conclusion~$\varphi=x_u \lor x_v$, where~$x_u$ and~$x_v$ are the corresponding variables of~$u$ and~$v$.
Based on the constructed weighted graph, we can reduce an instance of \Hamming to an instance of \WMDSshort problem.

\begin{figure}[t]
\centering
    \begin{tikzpicture}[line width=0.8pt, scale=0.8] 

 	\node  (xi) at (0,4.8) {$P_0^0$};
    \node [vert] (x_1) at (0,4) {};     
    \node [vert] (x_2) at (0,3) {};  
    \node [vert] (x_3) at (0,2) {};     
    \node [vert] (x_4) at (0,1) {}; 
    \node [vert] (x_5) at (0,0) {};

    \node  (yi) at (2,4.8) {$P_0^1$};
    \node [vert] (y_1) at (2,4) {};  
    \node [vert] (y_2) at (2,3) {}; 
    \node [vert] (y_3) at (2,2) {}; 
    \node [vert] (y_4) at (2,1) {}; 
    \node [vert] (y_5) at (2,0) {};

    \draw[ACMRed] (x_1) -- (y_3);
    \draw[ACMRed] (x_2) -- (y_1);
    \draw[ACMRed] (x_2) -- (y_5);
    \draw[ACMRed] (x_3) -- (y_2);
    \draw[ACMRed] (x_4) -- (y_3);
    \draw[ACMRed] (x_5) -- (y_4);
    \draw[ACMRed] (x_5) -- (y_5);
    
    \draw[ACMDarkBlue,densely dotted] (x_1) to[out=180,in=180] (x_2);
    \draw[ACMDarkBlue,densely dotted] (x_1) to[out=180,in=180] (x_3);
    \draw[ACMDarkBlue,densely dotted] (x_1) to[out=180,in=180] (x_5);
    \draw[ACMDarkBlue,densely dotted] (x_2) to[out=180,in=180] (x_4);
    \draw[ACMDarkBlue,densely dotted] (x_3) to[out=180,in=180] (x_5);
    \draw[ACMDarkBlue,densely dotted] (x_4) to[out=180,in=180] (x_5);

    \node  at (-0.8,-0.5) {\textcolor{ACMDarkBlue}{\textbf{bad}}};
    \node  at (1,-0.5) {\textcolor{ACMRed}{good}};
    
    \begin{scope}[shift={(1,0)}]
    \node  (V) at (5,4.8) {$V$};
    \node [vert] (v_1) at (5,4) {};     
    \node [vert] (v_2) at (5,3) {};  
    \node [vert] (v_3) at (5,2) {};     
    \node [vert] (v_4) at (5,1) {}; 
    \node [vert] (v_5) at (5,0) {};    
    
    \draw[ACMDarkBlue,densely dotted] (v_1) to[out=180,in=180] (v_2);
    \draw[ACMDarkBlue,densely dotted] (v_1) to[out=180,in=180] (v_3);
    \draw[ACMDarkBlue,densely dotted] (v_1) to[out=180,in=180] (v_5);
    \draw[ACMDarkBlue,densely dotted] (v_2) to[out=180,in=180] (v_4);
    \draw[ACMDarkBlue,densely dotted] (v_3) to[out=180,in=180] (v_5);
    \draw[ACMDarkBlue,densely dotted] (v_4) to[out=180,in=180] (v_5);
    
    \node (w1) at (6,4.8) {$w$}; \node (w1) at (6,4) {$2$};
    \node (w1) at (6,3) {0};
    \node (w1) at (6,2) {1};
    \node (w1) at (6,1) {1};
    \node (w1) at (6,0) {1};
    \end{scope}
    
    \draw[->] (3,2) to (4,2);
    
    \end{tikzpicture}
    
\caption{
\small
Illustration of the constructed weighted graph in the proof of \cref{thm: l=2 positive monotone is in P}.
On the left side a vertex represents a variable from~$P_0^0$ or $P_0^1$.
A line between two vertices represents a conclusion containing the two corresponding variables.
A red solid line represents a good conclusion and a blue dotted line represents a bad conclusion.
We transform it into the vertex weighted graph on the right side, where the vertex set~$V$ corresponds to~$P_0^0$ and the weight~$w$ for a vertex~$v \in V$ is the difference between the number of bad and good conclusions that contain the corresponding variable~$x_v$.
}
\label{fig: weighted graph}
\end{figure}
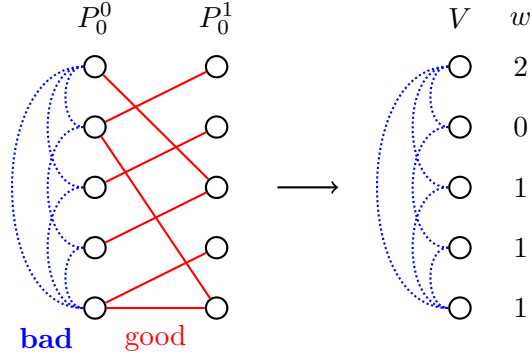

We first do the following preprocessing.
If there is a vertex~$v \in V$ with~$w(v)<0$, then changing this variable alone can strictly decrease the Hamming distance and hence the manipulation is feasible.
If there is an edge~$e=\{u,v\}$ with~$w(u)=w(v)=0$, then there is a bad conclusion $\varphi=x_u \lor x_v$ that is counted in both $w(u)$ and $w(v)$.
So changing the value of~$x_u$ and~$x_v$ can decrease the Hamming distance by 1, which means the manipulation is feasible.

In the following we can assume~$w(v) \ge 0$ for every~$v \in V$ and there is no edge~$e=\{u,v\}$ with~$w(u)=w(v)=0$.
For any vertex subset $V' \subseteq V$, changing the value of the corresponding variables can increase the Hamming distance
by $\sum_{v \in V'}{w(v)}-|E(G[V'])|$, where $G[V']$ is the subgraph induced by $V'$. 
If~$\sum_{v \in V'}{w(v)}=0$, then according to the above assumption, we have~$|E(G[V'])|=0$ and the Hamming distance has not changed.
Therefore manipulation is feasible if and only if there is a vertex subset $V'$ with~$\sum_{v \in V'}{w(v)}>0$ such that
$\sum_{v \in V'}{w(v)}-|E(G[V'])|<0$ or equivalently, 
\[\frac{|E(G[V'])|}{\sum_{v \in V'}{w(v)}}>1.\]
This is just an instance of the \WMDSshort problem where every edge has weight 1, which can be solved in polynomial time~\cite{goldberg1984finding}.
\end{proof}

\subsection{Positive Monotone Clauses of Length~$\ell=3$}

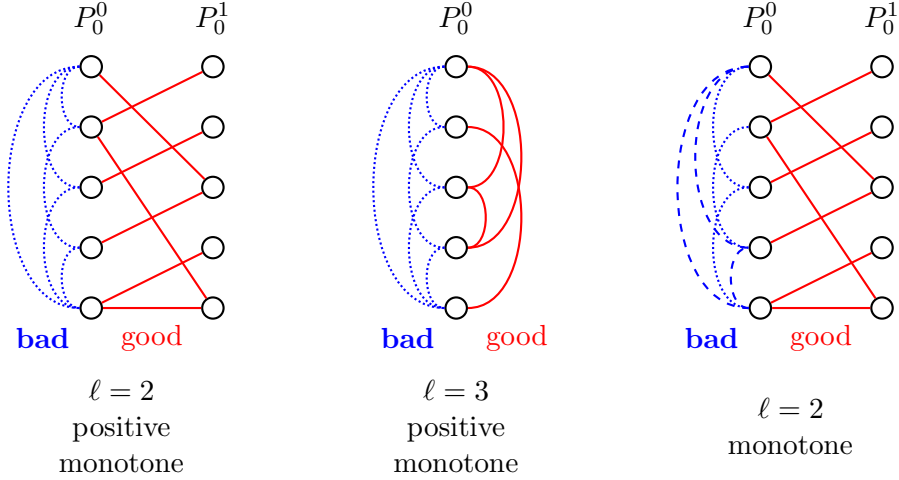
\begin{figure}[t]
\centering
    \begin{tikzpicture}[line width=0.8pt, scale=0.8] 
    
 	\node  (xi) at (0,4.8) {$P_0^0$};
    \node [vert] (x_1) at (0,4) {};     
    \node [vert] (x_2) at (0,3) {};  
    \node [vert] (x_3) at (0,2) {};     
    \node [vert] (x_4) at (0,1) {}; 
    \node [vert] (x_5) at (0,0) {};

    \node  (yi) at (2,4.8) {$P_0^1$};
    \node [vert] (y_1) at (2,4) {};  
    \node [vert] (y_2) at (2,3) {}; 
    \node [vert] (y_3) at (2,2) {}; 
    \node [vert] (y_4) at (2,1) {}; 
    \node [vert] (y_5) at (2,0) {}; 
    
    \draw[ACMRed] (x_1) -- (y_3);
    \draw[ACMRed] (x_2) -- (y_1);
    \draw[ACMRed] (x_2) -- (y_5);
    \draw[ACMRed] (x_3) -- (y_2);
    \draw[ACMRed] (x_4) -- (y_3);
    \draw[ACMRed] (x_5) -- (y_4);
    \draw[ACMRed] (x_5) -- (y_5);
    
    \draw[ACMDarkBlue,densely dotted] (x_1) to[out=180,in=180] (x_2);
    \draw[ACMDarkBlue,densely dotted] (x_1) to[out=180,in=180] (x_3);
    \draw[ACMDarkBlue,densely dotted] (x_1) to[out=180,in=180] (x_5);
    \draw[ACMDarkBlue,densely dotted] (x_2) to[out=180,in=180] (x_4);
    \draw[ACMDarkBlue,densely dotted] (x_3) to[out=180,in=180] (x_5);
    \draw[ACMDarkBlue,densely dotted] (x_4) to[out=180,in=180] (x_5);

    \node  at (-0.8,-0.5) {\textcolor{ACMDarkBlue}{\textbf{bad}}};
    \node  at (1,-0.5) {\textcolor{ACMRed}{good}};
    \node  at (0.5,-2) {\begin{tabular}{c} $\ell=2$ \\ positive \\ monotone \end{tabular}};
  	
  	\begin{scope}[shift={(1,0)}]
    \node  (V) at (5,4.8) {$P_0^0$};
    \node [vert] (v_1) at (5,4) {};     
    \node [vert] (v_2) at (5,3) {};  
    \node [vert] (v_3) at (5,2) {};     
    \node [vert] (v_4) at (5,1) {}; 
    \node [vert] (v_5) at (5,0) {};    
    
    \draw[ACMDarkBlue,densely dotted] (v_1) to[out=180,in=180] (v_2);
    \draw[ACMDarkBlue,densely dotted] (v_1) to[out=180,in=180] (v_3);
    \draw[ACMDarkBlue,densely dotted] (v_1) to[out=180,in=180] (v_5);
    \draw[ACMDarkBlue,densely dotted] (v_2) to[out=180,in=180] (v_4);
    \draw[ACMDarkBlue,densely dotted] (v_3) to[out=180,in=180] (v_5);
    \draw[ACMDarkBlue,densely dotted] (v_4) to[out=180,in=180] (v_5);
    
    \draw[ACMRed] (v_1) to[out=0,in=0] (v_3);
    \draw[ACMRed] (v_1) to[out=0,in=0] (v_4);
    \draw[ACMRed] (v_2) to[out=0,in=0] (v_5);
    \draw[ACMRed] (v_3) to[out=0,in=0] (v_4);
    
    \node  at (4.2,-0.5) {\textcolor{ACMDarkBlue}{\textbf{bad}}};
    \node  at (6,-0.5) {\textcolor{ACMRed}{good}};
   	\node  at (5,-2) {\begin{tabular}{c} $\ell=3$ \\ positive \\ monotone \end{tabular}};
  	\end{scope} 
  	   	
   	\begin{scope}[shift={(6,0)}]
	\node  (V) at (5,4.8) {$P_0^0$};
    \node [vert] (v_1) at (5,4) {};     
    \node [vert] (v_2) at (5,3) {};  
    \node [vert] (v_3) at (5,2) {};     
    \node [vert] (v_4) at (5,1) {}; 
    \node [vert] (v_5) at (5,0) {};    
    
    \node  (yi) at (7,4.8) {$P_0^1$};
    \node [vert] (z_1) at (7,4) {};  
    \node [vert] (z_2) at (7,3) {}; 
    \node [vert] (z_3) at (7,2) {}; 
    \node [vert] (z_4) at (7,1) {}; 
    \node [vert] (z_5) at (7,0) {}; 
    
    \draw[ACMDarkBlue,densely dotted] (v_1) to[out=180,in=180] (v_3);
    \draw[ACMDarkBlue,dashed] (v_1) to[out=180,in=180] (v_5);
    \draw[ACMDarkBlue,densely dotted] (v_2) to[out=180,in=180] (v_4);
    \draw[ACMDarkBlue,densely dotted] (v_3) to[out=180,in=180] (v_5);
    \draw[ACMDarkBlue,dashed] (v_1) to[out=180,in=180] (v_4);
    \draw[ACMDarkBlue,dashed] (v_5) to[out=180,in=180] (v_4);
    
    \draw[ACMRed] (v_1) -- (z_3);
    \draw[ACMRed] (v_2) -- (z_1);
    \draw[ACMRed] (v_2) -- (z_5);
    \draw[ACMRed] (v_3) -- (z_2);
    \draw[ACMRed] (v_4) -- (z_3);
    \draw[ACMRed] (v_5) -- (z_4);
    \draw[ACMRed] (v_5) -- (z_5);
    
    \node  at (4.2,-0.5) {\textcolor{ACMDarkBlue}{\textbf{bad}}};
    \node  at (6,-0.5) {\textcolor{ACMRed}{good}};
    \node  at (5.5,-2) {\begin{tabular}{c} $\ell=2$  \\ monotone \end{tabular}};
  	\end{scope}  	
   	
    \end{tikzpicture}
\caption{
\small
Comparison between different clause classes.
A vertex represents a variable from~$P_0^0 \cup P_0^1$, and only variables from~$P_0^0$ could be decided by the manipulator.
A line between two vertices represents a conclusion containing the two corresponding variables.
A line is solid or dotted if changing the value of \emph{one} of its endpoints in~$P_0^0$ will change the value of this conclusion,
while a line is dashed if changing the value of \emph{both} endpoints in~$P_0^0$ will change the value of this conclusion.
}
\label{fig: compare different clause classes}
\end{figure}

In this section we show that \Hamming with positive monotone clauses of length~$\ell = 3$ is \NP-hard. 
The main difference between~$\ell = 2$ and~$\ell=3$ is that when~$\ell=2$, every good conclusion must contain a variable from~$P_0^1$, and hence contains at most one useful variable from~$P_0^0$.
This makes it easy to count the number of included good conclusions.
When~$\ell \ge 3$, however, 
in addition to one variable from~$P_0^1$, a good conclusion can contain two useful variables from~$P_0^0$.
Hence, useful variables are not independent with respect to good conclusions.
See \cref{fig: compare different clause classes} for the comparison between~$\ell = 2$ and~$\ell=3$.
Intuitively, when~$\ell = 3$, we need to find a vertex subset of~$P_0^0$ in \cref{fig: compare different clause classes} to cover more red solid edges than blue dotted edges.
This leads us to define the following closely related graph problem.

\begin{probox}
\decprob{\textsc{Positive Vertex Cover}} 
{An undirected graph $G=(V,E^+ \cup E^-)$ with $E^+ \cap E^-= \emptyset$.}
{
Is there a vertex subset $V' \subseteq V$ which covers strictly more edges in $E^+$ than in~$E^-$? Herein, an edge is covered by a vertex subset if the vertex subset contains at least one endpoint of the edge.
}
\end{probox}
%

We show that \textsc{Positive Vertex Cover} is \NP-hard and then provide a simple reduction from \textsc{Positive Vertex Cover} to \Hamming with positive monotone clauses of length~$\ell = 3$.

\begin{lemma}
\label{lem: graph problem is NP-c}
\textsc{Positive Vertex Cover} is \NP-hard.
\end{lemma}

\begin{proof}
We construct a reduction from \textsc{Cubic Vertex Cover}, where given an undirected 3-regular graph and an integer~$k$, the task is to determine whether there exists a vertex cover of size at most~$k$.
Let $(G_0=(V_0,E_0),k)$ be an instance of \textsc{Cubic Vertex Cover}.
Denote $n=|V_0|$. Then $|E_0|=\frac{3n}{2}$.
Since a vertex cover for a cubic graph needs at least $\frac{n}{2}$ vertices to cover all $\frac{3n}{2}$ edges, we can assume $\frac{n}{2} \leq k \leq n$. 
We create an instance $G=(V,E^+ \cup E^-)$ of \textsc{Positive Vertex Cover} as follows (see also \cref{figure: graph problem is NP-c}).
First, for every original vertex~$v_i$ in~$V_0$, create a vertex~$x_i$ in~$V$, and for every edge~$\{v_i,v_j\}$ in~$E_0$, create an edge~$\{x_i,x_j\}$ in~$E^+$.
Then, create three more vertices~$y_1,y_2,y_3$ in~$V$ and create edges~$\{x_i,y_1\}$, $\{x_i,y_2\}$ and~$\{x_i,y_3\}$ for every~$1 \le i \le n$ in~$E^-$.
Add an edge $\{y_1,y_2\}$ in $E^+$.
Next, create vertices~$z_1,\dots,z_{n-p}$ in~$V$, where~$p=\frac{3n}{4}-\frac{k}{2}$, and create edges~$\{z_j,y_1\}$ and~$\{z_j,y_2\}$ for every~$1 \le j \le n-p$ in~$E^+$.
Finally, create vertices~$w_1,w_2,w_3$ in~$V$, and create edges~$\{z_j,w_1\}$, $\{z_j,w_2\}$ and~$\{z_j,w_3\}$ for every~$1 \le j \le n-p$ in~$E^-$.

%

\begin{figure}[t]

\centering

   \begin{tikzpicture}[line width=1pt, scale=1] 
    
 	\node  (xi) at (2,4.5) {$x_i$};
    \node [vert] (x_1) at (2,4) {};     
    \node [vert] (x_2) at (2,3) {};  
    \node [vert] (x_3) at (2,2) {};     
    \node [vert] (x_4) at (2,1) {}; 
    \node [vert] (x_5) at (2,0) {};    
    
    \node  (zi) at (-2,4) {$z_j$};
    \node [vert] (z_1) at (-2,3.5) {}; 
    \node [vert] (z_2) at (-2,2.5) {};  
    \node [vert] (z_3) at (-2,1.5) {}; 
    \node [vert] (z_4) at (-2,0.5) {};

    \node  (y1) at (0,3.5) {$y_1$};
    \node  (y2) at (0,1.5) {$y_2$};
    \node  (y3) at (0,0.5) {$y_3$};
    \node [vert] (y_1) at (0,3) {};  
    \node [vert] (y_2) at (0,2) {}; 
    \node [vert] (y_3) at (0,0) {};

    \draw[ACMRed] (y_1) -- (z_1);
    \draw[ACMRed] (y_1) -- (z_2);
    \draw[ACMRed] (y_1) -- (z_3);
    \draw[ACMRed] (y_1) -- (z_4);
    \draw[ACMRed] (y_2) -- (z_1);
    \draw[ACMRed] (y_2) -- (z_2);
    \draw[ACMRed] (y_2) -- (z_3);
    \draw[ACMRed] (y_2) -- (z_4);

    \draw[ACMDarkBlue,densely dotted] (y_1) -- (x_1);
    \draw[ACMDarkBlue,densely dotted] (y_1) -- (x_2);
    \draw[ACMDarkBlue,densely dotted] (y_1) -- (x_3);
    \draw[ACMDarkBlue,densely dotted] (y_1) -- (x_4);
    \draw[ACMDarkBlue,densely dotted] (y_1) -- (x_5);
    \draw[ACMDarkBlue,densely dotted] (y_2) -- (x_1);
    \draw[ACMDarkBlue,densely dotted] (y_2) -- (x_2);
    \draw[ACMDarkBlue,densely dotted] (y_2) -- (x_3);
    \draw[ACMDarkBlue,densely dotted] (y_2) -- (x_4);
    \draw[ACMDarkBlue,densely dotted] (y_2) -- (x_5);
    \draw[ACMDarkBlue,densely dotted] (y_3) -- (x_1);
    \draw[ACMDarkBlue,densely dotted] (y_3) -- (x_2);
    \draw[ACMDarkBlue,densely dotted] (y_3) -- (x_3);
    \draw[ACMDarkBlue,densely dotted] (y_3) -- (x_4);
    \draw[ACMDarkBlue,densely dotted] (y_3) -- (x_5);
    
    \node  (w1) at (-4,3.5) {$w_1$};
    \node  (w2) at (-4,2.5) {$w_2$};
    \node  (w3) at (-4,1.5) {$w_3$};
    \node [vert] (w_1) at (-4,3) {};  
    \node [vert] (w_2) at (-4,2) {}; 
    \node [vert] (w_3) at (-4,1) {}; 
    \draw[ACMDarkBlue,densely dotted] (w_1) -- (z_1);
    \draw[ACMDarkBlue,densely dotted] (w_1) -- (z_2);
    \draw[ACMDarkBlue,densely dotted] (w_1) -- (z_3);
    \draw[ACMDarkBlue,densely dotted] (w_1) -- (z_4);
    \draw[ACMDarkBlue,densely dotted] (w_2) -- (z_1);
    \draw[ACMDarkBlue,densely dotted] (w_2) -- (z_2);
    \draw[ACMDarkBlue,densely dotted] (w_2) -- (z_3);
    \draw[ACMDarkBlue,densely dotted] (w_2) -- (z_4);
    \draw[ACMDarkBlue,densely dotted] (w_3) -- (z_1);
    \draw[ACMDarkBlue,densely dotted] (w_3) -- (z_2);
    \draw[ACMDarkBlue,densely dotted] (w_3) -- (z_3);
    \draw[ACMDarkBlue,densely dotted] (w_3) -- (z_4);
    
    \path[red] (x_1) edge[bend left=90] (x_5);
    \node  (E0) at (4,2) {\textcolor{ACMRed}{$E_0$}};
    
    \draw (-4,-0.2) -- (-4,-0.7);
    \draw (-2,-0.2) -- (-2,-0.7);
    \draw (0,-0.2) -- (0,-0.7);
    \draw (2,-0.2) -- (2,-0.7);
    \draw (4,-0.2) -- (4,-0.7);
   
    \node  at (-3,-0.5) {\textcolor{ACMDarkBlue}{$E^-$}};
    \node  at (-1,-0.5) {\textcolor{ACMRed}{$E^+$}};
    \node  at (1,-0.5) {\textcolor{ACMDarkBlue}{$E^-$}};
    \node  at (3,-0.5) {\textcolor{ACMRed}{$E^+$}};
    
    \draw [dotted] [rotate=0] (0,2.5) ellipse [x radius=10pt, y radius=40pt];
    \draw [dotted] [rotate=0] (2,2.3) ellipse [x radius=15pt, y radius=75pt];
    \draw[red,line width=1mm] (y_1) -- (y_2);
    
    \end{tikzpicture}

\caption{
Illustration of the constructed instance in the proof of \cref{lem: graph problem is NP-c}.
Blue dotted edges are edges in~$E^-$ and red solid edges are edges in~$E^+$.
The manipulator just need to consider variables in $\{x_1,\dots,x_n,y_1,y_2\}$.}
\label{figure: graph problem is NP-c}
\end{figure}
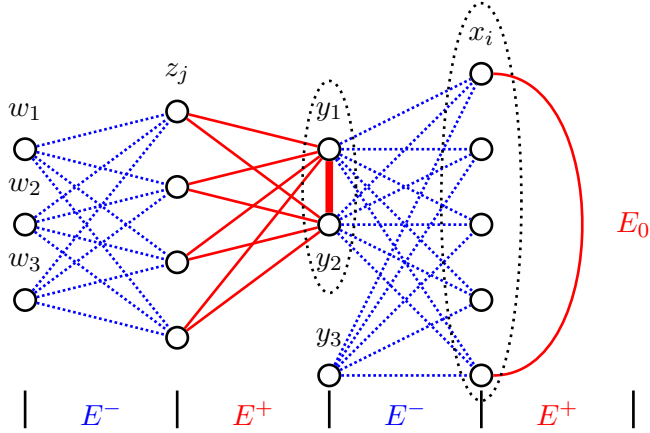


Notice that every vertex in~$\{w_1,w_2,w_3,y_3\}$ only covers edges in~$E^-$, thus it's always better to not choose them. 
Similarly, every vertex in~$\{z_1,\dots,z_{n-p}\}$ covers 3 edges in~$E^-$ and two edges in~$E^-$, thus it's always better to not choose them. 
Therefore, our choice is constrained in $\{x_1,\dots,x_n,y_1,y_2\}$.
The following lemma describes when should we choose $y_1$ or~$y_2$, which is the key point in this proof.

\begin{lemma}
\label{lem: when to choose y1y2}
If the number $k'$ of vertices chosen from~$\{x_1,\dots,x_n\}$ is less than $p$, then it is always better to not choose $y_1$ or $y_2$.
Otherwise, it is always better to choose both $y_1$ and $y_2$.
\end{lemma}

\begin{proof}
\renewcommand\qedsymbol{$\blacksquare$}
We prove the result for~$y_1$ and the result for~$y_2$ can be proved analogously.
Notice that after choosing~$k'$ vertices from~$\{x_1,\dots,x_n\}$, the number of uncovered edges incident on~$y_1$ from~$E^-$ is~$n-k'$, while the number of uncovered edges incident on~$y_1$ from~$E^+$ is~$n-p+1$.
If~$k' < p$, then~$n-k' \ge n-p+1$, and hence it is better to not choose~$y_1$.
If~$k' \ge p$, then~$n-k' < n-p+1$, and hence it is better to choose~$y_1$.
\end{proof}

Let~$V'$ be the set of vertices we will choose in the constructed instance.
Denote the number of edges in $E^+$ covered by $V'$ and the number of edges in $E^-$ covered by $V'$ by $n_1$ and $n_2$, respectively.
We show in the following that there is a vertex cover of size $k$ in $G_0$ if and only if there is a vertex subset $V' \subseteq V$ such that $n_1>n_2$.

$\Rightarrow$ Suppose that there is a vertex cover of size $k$ in $G_0$.
Let~$V'$ consists of the corresponding~$k$ vertices in~$V$ and two more vertices~$y_1$ and~$y_2$.
Then vertices in~$V'$ cover all edges in~$E^+$.
That is, all edges corresponding to~$E_0$, edges $z_jy_1$ and $z_jy_2$ for every $1 \le j \le n-p$ and $y_1y_2$.
Thus
$$n_1=\frac{3n}{2}+2(n-p)+1=2n+k+1.$$
On the other hand, in $E^-$, edges $x_iy_1$ and $x_iy_2$ for every $1 \le i \le n$ and $k$ edges from~$\{x_iy_3 \mid 1 \le i \le n\}$ are covered, so
$$n_2=2n+k<n_1.$$

$\Leftarrow$ Suppose that there is a vertex subset $V' \subseteq V$ such that $n_1>n_2$.
Let~$k'=|V' \cap \{x_1,\dots,x_n\}|$.
We first show that~$k' \ge p$.
Assume towards a contradiction that~$k' < p$.
According to \cref{lem: when to choose y1y2}, it is better to not choose~$y_1$ or~$y_2$.
Therefore, we can assume that~$V' \subseteq \{x_1,\dots,x_n\}$. 
Since every vertex in $\{x_1,\dots,x_n\}$ covers 3 edges in $E^-$, and every edge in $E^-$ is covered by at most one vertex in $\{x_1,\dots,x_n\}$, we have $n_2= 3k'$.
Since every vertex in $\{x_1,\dots,x_n\}$ covers 3 edges in $E^+$, we have $n_1 \le 3k'=n_2$, which is a contradiction.
In the following, we can assume that~$k' \ge p$.

Since~$k' \ge p$, according to \cref{lem: when to choose y1y2}, it is better to choose~$y_1$ and~$y_2$.
Thus we can assume that~$y_1 \in V'$ and~$y_2 \in V'$.
Let~$m$ be the number of edges corresponding to~$E_0$ covered by vertices in~$V'$.
Now we compute~$n_1$ and~$n_2$.
For~$n_1$, vertices in $V'$ cover edges $z_jy_1$ and $z_jy_2$ for every $1 \le j \le n-p$, $y_1y_2$ and~$m$ edges in~$E_0$, so
$$n_1=2(n-p)+1+m.$$
For~$n_2$, edges $x_iy_1$ and $x_iy_2$ for every $1 \le i \le n$ and $k'$ of edges $x_iy_3$ are covered, so
$$n_2=2n+k'.$$
Since~$n_1>n_2$, we have that~$m > k'+2p-1=\frac{3n}{2}+k'-k-1$ or
$$m \ge \frac{3n}{2}+k'-k.$$
Since~$m \le |E_0| = \frac{3n}{2}$, we have that~$k' \le k$.
If~$k'=k$, then~$m =\frac{3n}{2}$.
Thus~$V' \cap \{x_1,\dots,x_n\}$ is a vertex cover of size~$k$ for~$G_0$.
If~$k' < k$, then except for the~$k'$ edges in~~$V' \cap \{x_1,\dots,x_n\}$, we can always find~$k-k'$ more vertices to cover~$k-k'$ more edges in~$G_0$ such that~$m+k-k' \ge \frac{3n}{2}$.
Therefore, there is a vertex cover of size~$k$ for~$G_0$.
\end{proof}

Now we can show the \NP-hardness of \Hamming with positive monotone clauses of length~$\ell=3$ by a simple reduction from \PVC.
%

\begin{theorem}
\label{thm: l>=3 is in NP-c}
\Hamming with positive monotone clauses of fixed length $\ell$ $(\ge 3)$ is \NP-hard.
\end{theorem}

\begin{proof}
We prove the result for~$\ell=3$.
Other cases when~$\ell > 3$ can be shown by slightly adapting the following proof.
We present a reduction from \PVC.
Given a graph $G=(V,(E^+ \cup E^-))$ with $E^+ \cap E^- = \emptyset$, we construct an instance of \Hamming with~$q=\frac{1}{2}$ as follows (see also \cref{table: l>=3 is in NP-c}).
The agenda contains~$x_i$ for each~$v_i \in V$ ($1 \le i \le n$), $y$, $z$ and their negations as the premise set.
The conclusions set consists of clauses~$x_i \lor x_j \lor y$ for every edge~$\{v_i,v_j\}$ in $E^+$, $x_i \lor x_j \lor z$ for every edge $\{v_i,v_j\}$ in $E^-$ and their negations.
The third judge is the manipulator and the desired set~$J =J_3 \cap \Phi_c$.
Note that except for $y$ and $z$, all variables~$x_i$ are decided by the manipulator.
A successful manipulation is an action that changes the values of strictly more conclusions in~$\{x_i \lor x_j \lor y \mid \{v_i,v_j\} \in E^+\}$ than that in~$\{x_i \lor x_j \lor z \mid \{v_i,v_j\} \in E^-\}$.
Therefore, the manipulation is feasible if and only if there is a vertex subset which covers strictly more edges in $E^+$ than $E^-$. 
\end{proof}




\begin{table*}
\caption{Instance of \Hamming with positive monotone clauses of length~$\ell=3$ for the proof of \cref{thm: l>=3 is in NP-c}.}
\label{table: l>=3 is in NP-c}
\centering
\begin{tabular}
{p{2.5cm} p{0.2cm}p{0.2cm}p{0.2cm}p{0.2cm}p{0.2cm}p{0.4cm} p{2cm} p{2cm}}
\toprule[1pt]
Judgment Set & $x_1$ & $\dots$ & $x_n$ & $y$ & $z$ && $x_i \lor x_j \lor y$ &  $x_i \lor x_j \lor z$   \\ \midrule
$J_1$ & 1 & $\dots$ & 1 & 0 & 0 & & 1 & 1  \\
$J_2$ & 0 & $\dots$ & 0 & 0 & 0 & & 0 & 0  \\
$J_3$ & 0 & $\dots$ & 0 & 1 & 0 & & 1 & 0  \\
$\opUPQR_{1/2}$ & 0 & $\dots$ & 0 & 0 & 0 & $\Rightarrow$ & 0 & 0  \\
\bottomrule[1pt]
\end{tabular}
\end{table*}


\subsection{Monotone or Horn Clauses of Length~$\ell=2$}

In this section we study two remaining cases: Monotone clauses of length~$\ell=2$ and Horn clauses of length~$\ell=2$.
For these two cases, we cannot use the characterization of a successful manipulation for monotone clauses given in \cref{lem: condition for HD manipulation}.
Actually, different from positive monotone clauses of length~$\ell=2$, we will show that these two clause structures can be used to encode hard problems.


\subsubsection{Monotone Clauses of Length ${\ell=2}$}
For monotone clauses we may have both~$x_i \lor x_j$ and~$\neg x_i \lor \neg x_j$ in the conclusions.
Consider the following example, where the third judge is the manipulator and the desired set is $J=\{\neg x_1 \lor \neg x_2\}$.
\begin{center}
\begin{tabular}{p{2.5cm}p{0.2cm}p{0.2cm}p{0.4cm}p{1.8cm}}
\toprule
Judgment Set & $x_1$ & $x_2$ &  & $\neg x_1 \lor \neg x_2$  \\ \midrule
$J_1$ & 1 & 1 &  & 0  \\
$J_2$ & 0 & 0 &  & 1 \\
$J_3$ & 0 & 0 &  & 1\\
$\opUPQR_{1/2}$ & 0 & 0 & $\Rightarrow$ & 1\\
\bottomrule
\end{tabular}
\end{center}
Conclusion~$\neg x_1 \lor \neg x_2 \in J \cap \opUPQR_q(\bm{J})$ will be excluded from the outcome only when both~$x_1$ and~$x_2$ have been changed. 
Recall that for positive monotone clauses, changing one variable is enough to include a bad conclusion (see also \cref{fig: compare different clause classes} for the comparison).
Thus, for monotone clauses of length $\ell=2$ we have a new kind of ``bad'' conclusions, which can be used to encode hard problems.

\begin{theorem}
\label{thm: l=2 monotone is NP-c}
\Hamming with monotone clauses of length $\ell=2$ is \NP-hard.
\end{theorem}
\begin{proof}
We present a reduction from \textsc{Clique} on regular graphs.
Given an instance~$(G=(V,E),k,d)$ of \textsc{Clique} on regular graphs, where $d$ is the vertex degree and $k$
is the size of a desired clique,
we build an instance of \Hamming with~$q=\frac{1}{2}$ as follows (see also \cref{tab: HD monotone l=2 NP-c} for the profile).
Denote $n=|V|$.
The agenda contains~$x_i$ $(1 \le i \le n)$, $y_i$ $(1 \le i \le n)$, $x^*$, $y^*$ and their negations as the premise set.
The conclusion set consists of the following clauses and their negations. 
\begin{itemize}
\item $\neg x_i \lor \neg x_j$ $(1 \leq i < j \leq n)$.
\item $x_i \lor x_{i'}$ for each edge $\{v_i,v_{i'}\} \in E$.
\item $x_i \lor x^*$ for each $1 \le i \le n$.
\item $d+1$ copies\footnote{A very formal interpretation of our model definition with the agenda being a set (not a multiset) may not allow copies.
In this case, and later in \cref{thm: Microbriery l=2 positive monotone is NP-c}, instead of copies, we can also introduce a separate fresh $y_i$~or $y^*$~variable for each clause.
Since these variables cannot be changed, the proof does not change.}
of $x_i \lor y_i$ for each $1 \le i \le n$.
\item $n-k+1$ copies of $x^* \lor y^*$.
\end{itemize} 
There are three judges and their judgment sets are shown in \cref{tab: HD monotone l=2 NP-c}. 
The manipulator is the third judge and his desired set $J=J_3 \cap \Phi_c$.
The manipulator is decisive for all $x_i$ $(1 \le i \le n)$ and $x^*$.

\begin{table*}
\caption{ Instance of \Hamming with monotone clauses of length~$\ell=2$ for the proof of \cref{thm: l=2 monotone is NP-c}.}
\label{tab: HD monotone l=2 NP-c}
\centering
\begin{tabular}{p{2.5cm}cccccp{2cm} p{1.2cm}p{1.2cm}p{1.2cm}p{1.2cm}}
\toprule
Judgment Set & $x_i$ & $x^*$ & $y_i$ & $y^*$ &  & $\neg x_i \lor \neg x_j$ & $x_i \lor x_{i'}$ & $x_i \lor x^*$ & $x_i \lor y_i$ & $x^* \lor y^*$ \\ \midrule
$J_1$ & 1 & 1 & 0 & 0 & & 0 & 1 & 1 & 1 & 1 \\
$J_2$ & 0 & 0 & 0 & 0 & & 1 & 0 & 0 & 0 & 0\\
$J_3$ & 0 & 0 & 1 & 1 & & 1 & 0 & 0 & 1 & 1 \\
$\opUPQR_{1/2}$ & 0 & 0 & 0 & 0 & $\Rightarrow$ & 1 & 0 & 0 & 0 & 0\\
\bottomrule
\end{tabular}
\end{table*}
\medskip

The effects of changing $k'$ variables from $\{x_i \mid 1 \le i \le n\}$ are as follows.
On the one hand, $k'(d+1)$ of $x_i \lor y_i$ will be changed, which is good.
On the other hand, from the first three groups $\neg x_i \lor \neg x_j$, $x_i \lor x_{i'}$, and $x_i \lor x^*$, $p \ge k'(d+1)$ of them will be changed, which is bad.
Note that $p=k'(d+1)$ holds if the corresponding $k'$ vertices form a clique in $G$, and in this case these two effects cancel each other out. 
Now, if we continue to change $x^*$, then $n-k+1$ of $x^* \lor y^*$ will be changed, which is good, and $n-k' \le n-k$ more of $x_i \lor x^*$ will be changed, which is bad.
The key point is that if $k' \ge k$, then we have $n-k+1>n-k\ge n-k'$.
Therefore, a clique of size $k' \ge k$ in $G$ corresponds to a successful manipulation that changes the corresponding $k'$ variables of $x_i$ and $x^*$ such that overall the Hamming distance will be decreased by 1.

Now we show that there is a clique of size $k' \ge k$ in $G$ if and only if the manipulation is feasible.
The ``only if'' direction is clear from the above analysis, and we just need to show the ``if'' direction.
Suppose there exists a successful manipulation.
From the above analysis we have that $x^*$ has to be changed and the number $k'$ of changed variables from $\{x_i \mid 1 \le i \le n\}$ should satisfy $k' \ge k$.
Let $V'$ be the set of $k'$ vertices in $G$ corresponding to the $k'$ variables.
We claim that there are $k$ vertices from $V'$ that form a clique in~$G$.
To prove this claim, let us consider the induced subgraph $G[V']$.
Suppose towards a contradictions that $G[V']$ does not have a clique of size $k$, then there are at least $k'-k+1$ pairs of non-adjacent vertices in in $G[V']$.
To see this, notice that we can delete one vertex from each pair of non-adjacent vertices to get a clique.
Since $G[V']$ does not have a clique of size $k$, there should be at least $k'-k+1$ pairs of non-adjacent vertices.
Then, after the manipulation, from the first three groups $\neg x_i \lor \neg x_j$, $x_i \lor x_{i'}$, and $x_i \lor x^*$, at least 
\[k'd+(k'-k+1)+n=k'(d+1)+(n-k+1)\]
of them will be changed.
On the other hand, $k'(d+1)$ of $x_i \lor y_i$ and $n-k +1$ of $w \lor z_i$ will be changed. Thus, the Hamming distance does not change, which is a contradiction.
\end{proof}

\subsubsection{Horn Clauses of Length ${\ell=2}$}
For Horn clauses, we have conclusions of the form~$\neg x_i \lor x_j$.
This allows variables from~$P_1^1$ to be ``useful''.
To see this, consider the following example where the third judge is the manipulator and~$J=\{\neg x_1 \lor x_2\}$ is the desired set.
\begin{center}
\begin{tabular}{p{2.5cm}p{0.2cm}p{0.2cm}p{0.4cm}p{1.5cm}}
\toprule
Judgment Set & $x_1$ & $x_2$ &  & $\neg x_1 \lor x_2$  \\ \midrule
$J_1$ & 1 & 0 &  & 0 \\
$J_2$ & 0 & 0 &  & 1 \\
$J_3$ & 1 & 1 &  & 1 \\
$\opUPQR_{1/2}$ & 1 & 0 & $\Rightarrow$ & 0\\
\bottomrule
\end{tabular}
\end{center}
Changing the value of~$x_1 \in P_1^1$ from 1 to 0 can make~$\neg x_1 \lor x_2 \in J \setminus \opUPQR_q(\bm{J})$ included in the outcome.
Thus, for Horn clauses of length $\ell=2$ we have a new kind of useful variables, which allows to encode hard problems.

\begin{theorem}
\label{thm: l=2 Horn is NP-c}
\Hamming with Horn clauses of length $\ell=2$ is \NP-hard.
\end{theorem}
\begin{proof}
We adopt the reduction in the proof for \cref{thm: l=2 monotone is NP-c} with some modifications such that all clauses in the modified reduction are Horn clauses.
We copy the whole judgment profile, except for conclusions $\neg x_i \lor \neg x_j$ $(1 \leq i < j \leq n)$, which are not allowed in Horn clause.
To overcome this issue, we create a new variable~${x_i}' \in P_1^1$ for every variable~$x_i \in P_0^0$  $(1 \le i \le n)$, and add conclusions ${x_i}' \lor {x_j}'$ for all $1 \le i < j \le n$.
Notice that here ${x_i}' \lor {x_j}'$ will play the role of $\neg x_i \lor \neg x_j$ in the original reduction.
That is, changing ${x_i}'$ or ${x_j}'$ alone is not enough to change the result of ${x_i}' \lor {x_j}'$,
and changing the value of both~${x_i}'$ and~${x_j}'$ can change the result of~${x_i}' \lor {x_j}'$.

It remains to show that we can use a gadget to enforce that any successful manipulation has to either choose both of $x_i$ and ${x_i}'$, or choose none of them, for every $1 \le i \le n$.
To this end, for each pair of $x_i$ and ${x_i}'$, we add new variables $e_i^j \in P_0^0$ $(1 \le j \le N)$ and~$f_i^j \in P_0^1$ $(1 \le j \le \frac{N}{2})$, where~$N$ is a large even number whose value will be determined later.
In addition, we add clauses~$x_i \lor e_i^j$, ${x_i}' \lor e_i^j$ for all~$1 \le j \le N$, and~$x_i \lor t_j$, $\neg {x_i}' \lor t_j$ for all $1 \le j \le \frac{N}{2}$.
See \cref{table: Hamming_Horn_hard} for the judgment sets of the gadget for the pair of $x_i$ and ${x_i}'$ and see \cref{figure: l=2 Horn is NP-c} for an illustration.
Note that~$x_i$, ${x_i}'$ and all~$e_i^j$ $(1 \le j \le N)$ are decided by the manipulator.

\begin{figure}[h]
\begin{center}
   \begin{tikzpicture}[line width=1pt, scale=1] 
  \node[vert] (x) at (0,0) {};
  \node[left=0.2cm of x] {$x_i \in P_0^0$};

  \node[vert] (s1) at (3,3) {};
  \node[vert] (s2) at (3,2.5) {};
  \node (s3) at (3,2) {$\dots$};
  \node[vert] (s4) at (3,1.5) {};
  \node[vert] (s5) at (3,1) {};
  \draw[rotate=0] (3,2) ellipse [x radius=15pt, y radius=40pt];
  \node[above=0.3cm of s1] {$e_i^j \in P_0^0, 1 \le j \le N$};

  \node[vert] (x') at (6,0) {};
  \node[right=0.2cm of x'] {${x_i}' \in P_1^1$};

  \node[vert] (t1) at (3,-1) {};
  \node (t2) at (3,-1.5) {$\dots$};
  \node[vert] (t3) at (3,-2) {};
  \draw[rotate=0] (3,-1.5) ellipse [x radius=10pt, y radius=25pt];
  \node[below=0.3cm of t3] {$f_i^j \in P_0^1, 1 \le j \le \frac{N}{2}$};

  \draw (x) -- (s1);
  \draw (x) -- (s2);
  \draw (x) -- (s4);
  \draw (x) -- (s5);
  \draw (x') -- (s1);
  \draw (x') -- (s2);
  \draw (x') -- (s4);
  \draw (x') -- (s5);

  \draw (x) -- (t1);
  \draw (x) -- (t3);
  \draw (x') -- (t1);
  \draw (x') -- (t3);

  \node  at (1,2) {$x_i \lor e_i^j$};
  \node  at (5,2) {${x_i}' \lor e_i^j$};
  \node  at (1,-1.5) {$x_i \lor f_i^j$};
  \node  at (5,-1.5) {$\neg {x_i}' \lor f_i^j$};
   \end{tikzpicture}
\end{center}

\caption{Gadget for the pair of $x_i$ and ${x_i}'$.}
\label{figure: l=2 Horn is NP-c}
\end{figure}
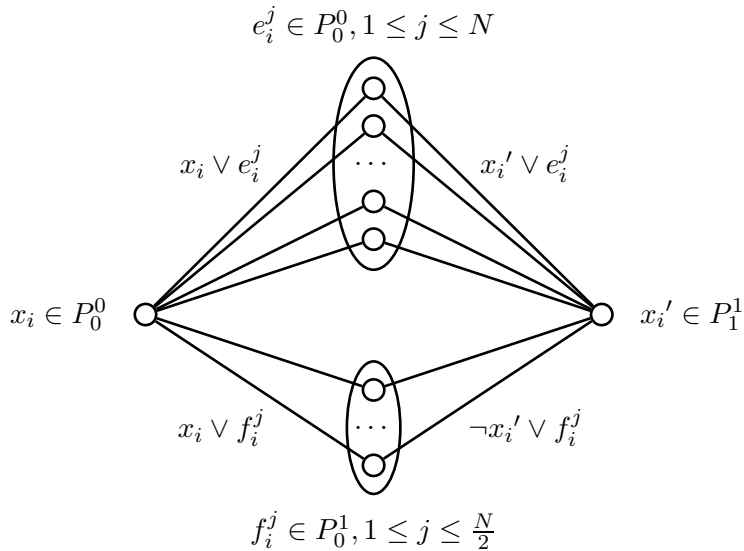

\begin{table*}
\caption{Gadget for the pair of $x_i$ and ${x_i}'$ in the proof of \cref{thm: l=2 Horn is NP-c}.}
\label{table: Hamming_Horn_hard}
\centering
\begin{tabular}
{p{2.5cm}p{0.2cm}p{0.2cm}p{0.2cm}p{0.2cm}p{0.4cm}p{1.6cm}p{1.6cm}p{1.4cm}p{1.6cm}}
\toprule
Judgment Set & $x_i$ & ${x_i}'$ & $e_i^j$ & $f_i^j$ & & $ x_i \lor e_i^j$ & $ {x_i}' \lor e_i^j$ & $x_i \lor f_i^j$ & $\neg {x_i}' \lor f_i^j$ \\ \midrule
$J_1$ & 1 & 1 & 1 & 0 && 1 & 1 & 1 & 0\\
$J_2$ & 0 & 0 & 0 & 0 && 1 & 1 & 0 & 1\\
$J_3$ & 0 & 1 & 0 & 1 && 0 & 1 & 1 & 1\\
$\opUPQR_{1/2}$ & 0 & 1 & 0 & 0 & $\Rightarrow$ & 0 & 1 & 0 & 0\\
\bottomrule
\end{tabular}
\end{table*}

Now we show the correctness of this gadget.
That is, any successful manipulation has to change both~$x_i$ and~${x_i}'$ or none of them, and for the first case, there exists a way to manipulate within the gadget such that within the gadget there is no influence on the Hamming distance.
If we change~$x_i$, ${x_i}'$ and all~$e_i^j$ $(1 \le j \le N)$, then all~$N$ conclusions from~$\{x_i \lor f_i^j, \neg {x_i}' \lor f_i^j \mid 1 \le j \le  \frac{N}{2}\}$ will be newly included into the outcome, and on the other hand, $N$ conclusions from~$\{x_i \lor e_i^j \mid 1 \le j \le N\}$ will also be newly included, while all conclusions from $\{{x_i}' \lor e_i^j \mid 1 \le j \le N\}$ remain to be included.
Therefore, the influence on the Hamming distance within the gadget is~$N-N=0$.
In other words, there exists a way to manipulate within the gadget such that we change both~$x_i$ and~${x_i}'$, and there is no influence on the Hamming distance.
However, if we just change only one of~$x_i$ and~${x_i}'$, then we get only~$\frac{N}{2}$ conclusions from~$\{x_i \lor f_i^j, \neg {x_i}' \lor f_i^j \mid 1 \le j \le  \frac{N}{2}\}$, but lose~$N$ conclusions from~$\{x_i \lor e_i^j, {x_i}' \lor e_i^j \mid 1 \le j \le N\}$ no matter how we change variables in~$\{e_i^j \mid 1 \le j \le N\}$.
We can choose the value for~$N$ to be large enough such that the difference~$\frac{N}{2}$ in this gadget can not be reverted by any choice of other variables.
\end{proof}

\section{Bribery}
\label{sec:brib}



In this section, we consider two variants of \textsc{Bribery} introduced by \citet{baumeister2015complexity}, where an external agent pays some judges to change their judgment sets so that the outcome changes towards some desired judgment set.

\begin{probox}
\decprob{\Bribery (resp. \Microbribery) }
{An agenda $\Phi$, a profile $\bm{J} = ( J_1 , \dots, J_n ) \in  {\mathcal{J}(\Phi)}^n$, a (possibly incomplete) consistent judgment 
set $J$ desired by the briber, a budget $k \in \mathbb{Z}^+$, and a uniform rational threshold~$q \in [0,1)$.}
{Is it possible to change up to $k$ individual judgment sets in $\bm{J}$ (resp. $k$ premise entries in $\bm{J}$) such that
for the resulting new profile $\bm{J'}$ it holds that $\HD(\opUPQR_q(\bm{J'}),J) < \HD(\opUPQR_q(\bm{J}),J)$?}
\end{probox}

\citet{baumeister2015complexity} proved that both \Bribery and
\Microbribery are \NP-hard when conclusions are positive monotone clauses.
However, there is no bound on the length of conclusions used in the reductions.
We study how the length of conclusions influence the computational complexity of these bribery problems.

\subsection{\Bribery}

We start with \Bribery and first consider the case when the budget~$k$ is a fixed constant.
For this case, we can guess different choices of $k$ judges to bribe.
For each choice it is easy to determine the set of variables that can be changed and then the main problem is to determine which subset of variables to change.
This is very similar to \Hamming where the manipulator needs to find a subset of decision variables to change.

Recall that in \cref{sec:HDmanip} we show that \Hamming with positive monotone clauses of length~$\ell=2$ is solvable in polynomial time via a reduction to \WMDSshort.
Using similar ideas, we show that for a fixed budget~$k$, \Bribery with the same clause set is also solvable in polynomial time.
The main difference is that for \Hamming the manipulator cannot decide the outcome of variables from $P_1^0 \cup P_0^1$ (notations from \cref{sec:HD-manipulation-condition}) and hence just need to consider variables from $P_0^0$ according to \cref{obs: useful variables in P00}, while this restriction does not hold for \Bribery.
Consequently, for \Bribery we need to consider more cases, as discussed in the following.

\begin{theorem}
\label{thm: Bribery l=2 positive monotone is XP}
\Bribery with conclusions being positive monotone clauses of length $\ell=2$ is solvable in polynomial time when the budget~$k$ is a fixed constant.
\end{theorem}

\begin{proof}
First of all, for any instance of \Bribery
 we can transform it into an equivalent instance such that~$J$ containing no premise and~$\{\varphi, \neg \varphi\} \cap J \not = \emptyset$ for every conclusion~$\varphi \in \Phi_c$, using the same method as in \cref{sec:HDmanip}.
In the following we can assume that~$J$ contains only conclusions and for any conclusion~$\varphi \in \Phi_c$ we have that~$\varphi \in J$ or~$\neg \varphi \in J$.

We can try all possible choices of bribing $k$ judges from all $n$ judges and there are $\binom{n}{k}$ different choices.
For each choice, we can find the set~$D$ of variables whose value can be changed.
For any~$x \in D$, if all positive clauses containing~$x$ are included in the desired set~$J$, then it is always better to make $x$ included in the outcome.
Otherwise, if there exists a positive clause~$\varphi$ with $\neg \varphi \in J$ that contains $x$, then we put~$x \in D$ into a set~$T_1$ if~$x \in \opUPQR_q(\bm{J})$ and put~$x \in D$ into a set~$T_0$ if~$x \not \in \opUPQR_q(\bm{J})$.
We just need to consider variables in $T_1 \cup T_0$.

We first observe that conclusions containing two variables from $T_1 \cup T_0$ are not in the desired set $J$.

\begin{lemma}
\label{lem:bribery-not-in-J}
For every conclusion~$x_i \lor x_j$ with~$x_i,x_j \in T_1 \cup T_0$, we have $x_i \lor x_j \not \in J$.
\end{lemma}

\begin{proof}
\renewcommand\qedsymbol{$\blacksquare$}
Let~$\bar J$ be any complete and consistent extension of~$J$.
For any variable~$x \in T_1 \cup T_0$, we have~$x \not \in \bar J$ since there is a positive monotone clause~$\varphi \not \in J$ that contains~$x$.
Thus, $x_i \not \in \bar J$ and $x_j \not \in \bar J$, which imply $x_i \lor x_j \not \in J$.
\end{proof}

Next, we consider our problem in the following three cases:

\begin{enumerate}
\item $T_1= \emptyset$.\\
In this case we only need to consider variables in~$T_0$.
We classify all conclusions~$x_i \lor x_j$ into three cases according to the number of variables from~$T_0$ included.
If both~$x_i$ and~$x_j$ are not in~$T_0$, then the value of~$x_i \lor x_j \in J$ is fixed and we do not need care about this conclusion.
If only one variable, say~$x_i$, is in~$T_0$, then the value of the other variable~$x_j$ is fixed and we can take it as a constant.
If~$x_j \in \opUPQR_q(\bm{J})$, then~$x_i \lor x_j$ will be included into the outcome for any~$x_i$ and hence we do not need care about this conclusion.
The interesting case is when~$x_j \not \in \opUPQR_q(\bm{J})$, where the result of~$x_i \lor x_j$ depends only on~$x_i$.
If both~$x_i$ and~$x_j$ are in~$T_0$, then we know that~$x_i \lor x_j \not \in \opUPQR_q(\bm{J})$ and~$x_i \lor x_j \not \in J$ (from \cref{lem:bribery-not-in-J}).
In this case changing any one variable will make~$x_i \lor x_j$ included into the outcome, which is bad.
To sum up, for each variable~$x \in T_0$, we need to consider conclusions~$x_i \lor x_j$ with~$x_j \not \in T_0 \cup  \opUPQR_q(\bm{J})$ or~$x_j \in T_0$.
We can assign each variable~$x_i \in T_0$ a weight to record the difference between the number of newly included desired conclusions and the number of lost desired conclusions if we change the value of~$x_i$ as follows:
\begin{align*}
w(x_i)=& |\{x_i \lor x_j \mid x_j \not \in T_0\cup  \opUPQR_q(\bm{J}) \text{ and }  x_i \lor x_j \in J\}| \\
& - |\{x_i \lor x_j \mid x_j \not \in T_0\cup  \opUPQR_q(\bm{J}) \text{ and } x_i \lor x_j \not \in J\}| \\
& - |\{x_i \lor x_j \mid x_j \in T_0\}|
\end{align*}
For any subset~$T \subseteq T_0$, let~$f(T)$ be the difference between the number of newly included desired conclusions and the number of lost desired conclusions after changing the values of variables in~$T$, then
\[
f(T)=\sum_{x_i \in T} w(x_i)+ g(T),
\]
where~$g(T)=|\{x_i \lor x_j \mid x_i \in T \text{ and } x_j \in T\}|$ is the number of lost desired conclusions counted twice in both~$w(x_i)$ and~$w(x_j)$.
Our goal is to find a subset~$T \subseteq T_0$ such that~$f(T)>0$, that is,
\[
\frac{g(T)}{-\sum_{x_i \in T} w(x_i)}>0.
\]
Using the same method as in the proof of \cref{thm: l=2 positive monotone is in P}, we can reduce this problem to \WMDSshort, which is solvable in polynomial time.
Note that in \WMDSshort we do not require non-negativity for vertex weights.

\item $T_0= \emptyset$.\\
In this case we only need to consider variables in~$T_1$.
Similar to the first case when~$T_0= \emptyset$, we can assign each variable~$x_i \in T_1$ a weight to record the difference between the number of newly included desired conclusions and the number of lost desired conclusions if we change the value of~$x_i$ as follows:
\begin{align*}
w(x_i)=&|\{x_i \lor x_j \mid x_j \not \in T_1 \cup  \opUPQR_q(\bm{J}) \text{ and }  x_i \lor x_j \not \in J\}| \\
& - |\{x_i \lor x_j \mid x_j \not \in T_1 \cup  \opUPQR_q(\bm{J}) \text{ and } x_i \lor x_j  \in J\}|.
\end{align*}
Note that here we do not count conclusions~$x_i \lor x_j$ with~$x_j \in T_1$, since for this kind of conclusions, we need to change both~$x_i$ and~$x_j$ to be 0 to change the outcome of~$x_i \lor x_j \in \opUPQR_q(\bm{J}) \setminus J$.
Instead, we count this part for any subset~$T \subseteq T_1$ as~$g(T)=|\{x_i \lor x_j \mid x_i \in T \text{ and } x_j \in T\}|$.
 Then, $f(T)$, the difference between the number of newly included desired conclusions and the number of lost desired conclusions after changing the values of variables in~$T$, can be defined as follows:
\[
f(T)=\sum_{x_i \in T} w(x_i)+ g(T).
\]
Our goal is to find a~$T$ such that~$f(T)>0$, that is,
\[
\frac{g(T)}{-\sum_{x_i \in T} w(x_i)}>0.
\]
Again, using the same method as in the proof of \cref{thm: l=2 positive monotone is in P}, we can reduce this problem to \WMDSshort, which is solvable in polynomial time.

\item $T_1 \not = \emptyset$ and $T_0 \not = \emptyset$.\\
We first consider two restricted subproblems where we only change variables from~$T_0$ or $T_1$.
These two subproblems are just the two cases considered above, which are solvable in polynomial time.
If we can succeed in one of them, then we are done.
Otherwise, there is no feasible bribery if we only change variables from~$T_0$ or $T_1$.
We claim that in this case there is no feasible bribery.

Suppose, towards a contradiction, that we can decrease the Hamming distance by $d>0$ through changing the value of variables in a set~$T \subseteq T_1 \cup T_0$, then $T \cap T_0 \neq \emptyset$ and $T \cap T_1 \neq \emptyset$.
Denote~$d_0$ the decreased Hamming distance if we only change the value of variables in $T \cap T_0$,
and~$d_1$ the decreased Hamming distance if we only change the value of variables in $T \cap T_1$.
Since both two subproblems have no solution, we have $d_0 \le 0$ and $d_1 \le 0$.
Now we consider the relation between $d$, $d_0$ and $d_1$.
To this end, we just need to consider conclusions in~$Q=\{x_i \lor x_j \mid x_i \in T \cap T_0,x_j \in T \cap T_1\}$, for which we have the following observations:
\begin{enumerate}
\item $x_i \lor x_j \in \opUPQR_q(\bm{J})$ since $x_j \in T_1$ implies $x_j \in \opUPQR_q(\bm{J})$.
\item $x_i \lor x_j \not \in J$ by $x_i,x_j \in T_0 \cup T_1$ and \cref{lem:bribery-not-in-J}.
\item Changing $x_i$ alone from 0 to 1 cannot change the value of $x_i \lor x_j$ since it is still 1.
\item Changing $x_j$ alone from 1 to 0 will make $x_i \lor x_j= 0 \lor 0=0$.
\item Changing both $x_i$ and $x_j$ cannot change the value of $x_i \lor x_j$ since it is still 1.
\end{enumerate}

From (a) and (b) we know that change the value of $x_i \lor x_j \in Q$ from 1 to 0 can decrease the Hamming distance by 1.
From (c) and (e) we know that conclusions from $Q$ has no influence on $d_0$ and $d$, while from (d) we know that these $|Q|$ conclusions are counted in $d_1$.
Thus, $d=d_0+(d_1-n_\delta) \le 0$. This is a contradiction.
\end{enumerate}
\end{proof}



Next, we consider more general clause classes with bounded length.
We present a simple linear-time reduction from \Hamming to \Bribery, which allows to transfer the hardness results for \Hamming established in \cref{sec:HDmanip} to \Bribery.

\begin{lemma}
\label{lem:manip-to-brib}
For any clause set $\mathcal{C}$, 
there is linear-time many-one reduction from \Hamming with conclusions chosen from $\mathcal{C}$ to \Bribery with conclusions chosen from $\mathcal{C}$ and the budget~$k=1$.
\end{lemma}

\begin{proof}
For an instance of \Hamming with conclusions chosen from $\mathcal{C}$,
let $\bm{J_M}$ be its profile and $J_M=J_n$ be the desired set of the manipulator.
We first compute the set $D$ of variables decided by the manipulator.
Then we partition all variables into three sets: $D$, $V^+$ and $V^-$, where $V^+$ is set of variables which are in $\opUPQR_q(\bm{J_M})$ and not decided by the manipulator, and $V^-$ is set of variables which are not in $\opUPQR_q(\bm{J_M})$ and not decided by the manipulator.

Now we construct an instance of \Bribery with conclusions chosen from $\mathcal{C}$ as follows.
The agenda remains the same and the desired set for the briber is $J_B=J_M$.
The profile $\bm{J_B}$ for the instance of \Bribery is the same as $\bm{J_M}$ for all variables in~$D$.
While for each variable $x \in V^+$, $x$ is contained in every judgment set,
and for each variable $x \in V^-$, $x$ is not contained in any judgment set.
Finally, we set~$k=1$.
See \cref{tab: Profiles for manipulation and bribery} for the two profiles for variables in $\bm{J_M}$ and $\bm{J_B}$ in an example with three judges.

\begin{table}[t]
 \caption{Profiles of variables in manipulation (left) and bribery (right).}
     \label{tab: Profiles for manipulation and bribery}
   \begin{minipage}{.5\linewidth}
     
     \centering
       \begin{tabular}{p{2.3cm}p{1.1cm}p{1.3cm}p{1.4cm}}
\toprule
Judgment Set & $x \in D$ & $x \in V^+$ & $x \in V^-$  \\ \midrule
$J_1$ & 1 & 1 &  0  \\
$J_2$ & 0 & 1 &  0 \\
$J_3$ & 0/1 & 0 &  1\\
$\opUPQR_{1/2}$ & 0/1 & 1 & 0\\ \midrule
$J_M$ & 0/1 & 0 & 1 \\
\bottomrule
\end{tabular}
   \end{minipage}%
   \begin{minipage}{.5\linewidth}
     \centering
      
\begin{tabular}{p{2.3cm}p{1.1cm}p{1.3cm}p{1.4cm}}
\toprule
Judgment Set & $x \in D$ & $x \in V^+$ &  $x \in V^-$  \\ \midrule
$J_1$ & 1 & 1 & 0 \\
$J_2$ & 0 & 1 & 0 \\
$J_3$ & 0/1 & 1 & 0 \\
$\opUPQR_{1/2}$ & 0/1 & 1 & 0\\ \midrule
$J_B$ & 0/1 & 0 & 1 \\
\bottomrule
\end{tabular}
\end{minipage} 
\end{table}

With budget $k=1$ the briber can only change the values of variables in $D$, the same as the manipulator.
In addition, we have $\opUPQR_q(\bm{J_M})=\opUPQR_q(\bm{J_B})$ and $J_M=J_B$.
Therefore the instance of \textsc{Manipulation} is a yes-instance if and only if the instance of \textsc{Bribery} is a yes-instance.
\end{proof}

Combining \cref{lem:manip-to-brib} with \cref{thm: l>=3 is in NP-c,thm: l=2 monotone is NP-c,thm: l=2 Horn is NP-c}, we get the following NP-hard results for \Bribery when conclusions have length 2 or 3.
Note that all reductions in the proofs for \cref{thm: l>=3 is in NP-c,thm: l=2 monotone is NP-c,thm: l=2 Horn is NP-c} have only three judges.

\begin{corollary}
\label{cor: Bribery l=2 Horn or monotone or l>=3 is NP-c}
\Bribery with conclusions being Horn clauses of length $\ell=2$ (resp.\ monotone clauses of length 2 or positive monotone of length $\ell \ge 3$) is \NP-hard,
even when there are only 3 judges and the budget~$k=1$.
\end{corollary}


\cref{thm: Bribery l=2 positive monotone is XP} and \cref{cor: Bribery l=2 Horn or monotone or l>=3 is NP-c} show that when the budget $k$ if a fixed constant, \Bribery and \Hamming have the same computational complexity results for many different clause classes.
Next we show that different from \Hamming, the complexity of \Bribery could also come from the choice of $k$ judges to bribe:
When the budget~$k$ is not fixed, \Bribery becomes \NP-hard even for the most basic case when conclusions are positive monotone clauses of length~$\ell=2$.
We show this result by a reduction from \OL defined as follows, which is \NP-hard and \W2-hard with respect to the budget $k$~\cite{christian2007complexity}.

\begin{probox}
\decprob{\OL}
{An $m \times n$ $(0,1)$ matrix $L$ 
and a positive integer $k$.}
{Is there a choice of $k$ rows in $L$ such that by changing the entries of these rows each column of the resulting matrix has  a strict majority of ones.}
\end{probox}

\begin{theorem}
\label{thm: Bribery l=2 positive monotone is NP-c W2}
\Bribery with conclusions being positive monotone clauses of length $\ell=2$ is \NP-hard 
and \W2-hard with respect to the budget $k$.
\end{theorem}

\begin{proof}
Let $(L,k)$ be an instance of \OL.
Without loss of generality, we can assume that each column of $L$ has a strict majority of zeros and $k \le \lfloor \frac{m}{2} \rfloor$.
We construct an instance of \Bribery with~$q=\frac{1}{2}$ as follows (see also \cref{table: Bribery l=2 positive monotone NP-c W2}).
The agenda contains variables~$x_i$ $(1 \le i \le n)$, $y_1,y_2$ and their negations as the premise set.
The conclusion set consists of~$x_i \lor y_1, x_i \lor y_2$ $(1 \le i \le n)$, $x_1 \lor x_j$ $(2 \le j \le n)$ and their negations.
There are $m$ judges and their judgment set with respect to~$x_i$ $(1 \le i \le n)$ are based on $L$, i.e., $x_i \in J_j$ if and only if $L_{i,j}=1$.
Premises~$y_1$ and~$y_2$ are not contained in any judgment set. 
The briber's desired set~$J$, which is complete, is shown in the last row of \cref{table: Bribery l=2 positive monotone NP-c W2}.
Note that since~$k \le \lfloor \frac{m}{2} \rfloor$, the briber could only change the values of variables of~$x_i$ $(1 \le i \le n)$.

\begin{table}
\caption{Instance with only positive monotone clauses of length $\ell=2$ in the proof of \cref{thm: Bribery l=2 positive monotone is NP-c W2}.}
\label{table: Bribery l=2 positive monotone NP-c W2}
\centering
\begin{tabular}{llllllllll}
\toprule
Judgment Set & $x_1$ & $\dots$ & $x_n$ & $y_1$ & $y_2$ & & $x_i \lor y_1$ & $x_i \lor y_2$ & $x_1 \lor x_j$  \\ \midrule
$J_1$ & \multicolumn{3}{c}{\multirow{3}{*}{$L$}} & 0 & 0 & & \multicolumn{3}{c}{\multirow{3}{*}{$\star$}}\\
$\dots$ & & & & $\dots$ & $\dots$ & &\\
$J_m$ & & & &  0 & 0 & &\\
$\opUPQR_{1/2}$ & 0 & $\dots$ & 0 & 0 & 0 & $\Rightarrow$ & 0 & 0 & 0 \\ \midrule
$J$ & 0 & $\dots$ & 0 & 1 & 1 & & 1 & 1 & 0 \\
\bottomrule
\end{tabular}
\end{table}


Now we show that instance~$(L,k)$ is a yes-instance of \OL if and only if the constructed instance is a yes-instance of \Bribery.

$\Rightarrow$ Assume that instance~$(L,k)$ is a yes-instance, then the briber can bribe the corresponding~$k$ judges such that the values of all~$x_i$ $(1 \le i \le n)$ become 1.
Consequently, all~$x_i \lor y_1, x_i \lor y_2$ $(1 \le i \le n)$ will be included into the outcome, and meanwhile all~$x_i$ $(1 \le i \le n)$ and~$x_i \lor x_j$ $(2 \le j \le n)$ will also be included into the outcome.
So the Hamming distance will be decreased by $2n-(2n-1)=1$ and this is successful bribery.

$\Leftarrow$ Assume that the bribery is feasible.
Let~$S \subseteq \{x_1, \dots,x_n\}$ be the set of changed variables in a successful bribery.
Our goal is to show that~$S=\{x_1, \dots,x_n\}$, which would imply that instance~$(L,k)$ is a yes-instance.
Suppose towards a contradiction that~$S \not = \{x_1, \dots,x_n\}$.
Then, on one hand, desired conclusions in~$\{x_i \lor y_1,x_i \lor y_2 \mid x_i \in S\}$ are included into the outcome and~$|\{x_i \lor y_1,x_i \lor y_2 \mid x_i \in S\}|=2|S|$.
On the other hand, all variables in~$S$ and at least~$|S|$ conclusions in~$\{x_1 \lor x_j \mid 1 \le j \le n\}$ (that is, $\{x_1 \lor x_j \mid x_j \in S\}$ if~$x_1 \not \in S$ and $\{x_1 \lor x_j \mid 2 \le j \le n\}$ if~$x_1 \in S$) will also be included into the outcome.
Thus, the Hamming distance will be increased by at least $2|S|-2|S|=0$ after this bribery, which contradicts with that the bribery is successful.
\end{proof}

\subsection{\Microbribery}

Finally, we consider \Microbribery, where the budget~$k$ is on the number of changes in premises. 
Different from \Bribery, if we restrict the budget$~k$ to be a fixed constant, then \Microbribery in general is solvable in polynomial time since we can simply try all possible $k$ premises to change.
When $k$ is not fixed, we show that \Microbribery is \NP-hard even for the most basic case when all conclusions are positive monotone clauses of length~$\ell=2$, which is the same as \Bribery (see \cref{thm: Bribery l=2 positive monotone is NP-c W2}).

\begin{theorem}
\label{thm: Microbriery l=2 positive monotone is NP-c}
\Microbribery with conclusions being positive monotone clauses of length $\ell=2$ is \NP-hard and \W1-hard with respect to budget $k$,
even if the desired set is complete.
\end{theorem}

\begin{proof}
We present a reduction from \textsc{Clique}.
Given an instance~$(G=(V,E),s)$ of \textsc{Clique},
we construct an instance of \Microbribery with three judges and~$q=\frac{1}{2}$ as follows.
Denote $n=|V|$.
The agenda contains~$x_i$ $(1 \le i \le n)$, $y_i$ $(1 \le i \le n)$, $x^*$, $y^*$ and their negations as the premise set.
The conclusion set consists of the following clauses and their negations (Without loss of generality we assume $s$ is an even number).
\begin{itemize}
\item $x_i \lor x_j$ for every edge $\{v_i,v_j\} \in E$.
\item $x_i \lor x^*$ for  each $1 \le i \le n$
\item $\frac{s}{2}$ copies of $x_i \lor y_j$ for  each $1 \le i \le n$.
\item $\frac{s}{2}-1$ copies of $x^* \lor y^*$.
\end{itemize} 
Note that the number of copies of $x^* \lor y^*$ is one smaller than the number of copies of $x_i \lor y_j$.
This will be essential to create a feasible bribery such that the Hamming distance is decreased by one.
The judgment sets of all~$2m+1$ $(m=s+1)$ judges and the desired set~$J$ of the briber are shown in \cref{table: Microbribery l=2 positive monotone NP-c}.
Finally, we set the budget of the briber to be~$k=s+1$.
Note that the briber cannot change the value of~$y_i$ or~$y^*$, since the briber needs to change~$m+1=s+2>k$ entries to change any one of them.
To change the value of~$x_i$ or $x^*$, the briber needs to change 1 entry. 
We show that there is a clique of size $s$ in $G$ if and only if there is a successful bribery with at most $k=s+1$ microbribes.

\begin{table}
\caption{Instance of \Microbribery with three judges and positive monotone clauses of length $s=2$ in the proof of \cref{thm: Microbriery l=2 positive monotone is NP-c}.}
\label{table: Microbribery l=2 positive monotone NP-c}
\centering
\begin{tabular}{lcccccclll}
\toprule
Judgment Set & $x_i$ & $x^*$ & $y_i$ & $y^*$ &  & $x_i \lor x_j$ & $x_i \lor x^*$ & $x_i \lor y_i$ & $x^* \lor y^*$ \\ \midrule
$J_i$ $(1 \le i \le m+1)$ & 1 & 1 & 0 & 0 & & 1 & 1 & 1 & 1 \\
$J_{i}$ $(m+2 \le i \le 2m+1)$ & 0 & 0 & 0 & 0 & & 0 & 0 & 0 & 0 \\
$\opUPQR_{1/2}$ & 1 & 1 & 0 & 0 & $\Rightarrow$ & 1 & 1 & 1 & 1\\ \midrule
$J$ & 0 & 0 & 1 & 1 & & 0 & 0 & 1 & 1 \\
\bottomrule
\end{tabular}
\end{table}

$\Rightarrow$ Assume there is a clique of size $s$ in $G$, then the manipulator can change the values of the corresponding $s$ variables from~$\{x_1,\dots,x_n\}$ and $x^*$. 
One one hand, $\binom{s}{2}$ of $\neg(x_i \lor x_j)$ and $s$ of $\neg(x_i \lor x^*)$ will be included in the outcome. 
So the number of newly included desired formulas in the outcome is
$$n_1=\binom{s}{2}+s=\binom{s+1}{2}.$$
One the other hand, $\frac{s}{2}s$ of $x_i \lor y_j$ and~$\frac{s}{2}-1$ of $x^* \lor y^*$ will be excluded from the outcome. 
So the number of lost desired formulas in the outcome is
$$n_2=\frac{s}{2}s+\frac{s}{2}-1=\binom{s+1}{2}-1.$$
Since $n_1>n_2$, this is a successful bribery.

$\Leftarrow$ Assume there is a successful bribery with at most $k$ microbribes. 
Denote $n_1$ the number of newly included desired formulas,
and $n_2$ the number of excluded desired formulas after a successful bribery.
Suppose $s'\le k$ variables from~$\{x_1,\dots,x_n\}$ are changed after this bribery.
Then~$p \leq \binom{s'}{2}$ of~$x_i \lor x_j$ are changed. 

If~$x^*$ is not changed, then 
\[n_1=p \leq \binom{s'}{2}.\]
On the other hand,  $\frac{s}{2}s'$ of $x_i \lor y_j$ are changed. So 
\[n_2 = \frac{s}{2}s'.\]
Since $n_1>n_2$, we have $s'-1>s=k-1$, which contradicts with $s'\le k$.

Therefore, $x^*$ has to be changed and hence $s' \le s$.
Now $p \leq \binom{s'}{2}$ of~$x_i \lor x_j$ and~$s'$ of $x_i \lor x^*$  are changed, so
\[n_1=p+s'.\]
On the other hand, $\frac{s}{2}s'$ of $x_i \lor y_j$ and~$\frac{s}{2}-1$ of $x^* \lor y^*$ are changed. So 
\[n_2 = \frac{s}{2}s'+\frac{s}{2}-1\ge\frac{s(s'+1)}{2}-1.\]
Since $n_1>n_2$, we have
\[
p >  \frac{s(s'+1)}{2}-s'-1 \Rightarrow p \ge \frac{s(s'+1)}{2}-s'.
\]
Since $s' \le s$ and $p \leq \binom{s'}{2}$, we have $s'=s$ and $p = \binom{s'}{2}$, which means the corresponding $s'=s$ vertices form a clique in $G$.
\end{proof}

Notice that in both \cref{thm: Bribery l=2 positive monotone is NP-c W2} for \Bribery and \cref{thm: Microbriery l=2 positive monotone is NP-c} for \Microbribery, the number of judges in the reductions is not bounded.
However, different from \cref{thm: Bribery l=2 positive monotone is NP-c W2}, where the complexity comes from the choice of $k$ judges to bribe, the complexity in \cref{thm: Microbriery l=2 positive monotone is NP-c} comes from the choices of variables to change.
In particular, in \cref{thm: Microbriery l=2 positive monotone is NP-c} the main role of the unbounded number of judges is to enforce that some variables ($y_i,z_i$) cannot be changed by the briber.
Then a natural question is what happens if we restrict the number of judges to be constant in \Microbribery.

\citet[Theorem 15]{baumeister2015complexity} claims that \Microbribery is \NP-hard when the number of judges is a fixed constant.
However, their proof seems to be wrong.
In their proof they present a reduction from \textsc{Dominate Set} to \Microbribery with all conclusions being positive monotone clauses, and the correctness of this reduction relies on a wrong assumption that in the constructed instance of \Microbribery the budget (corresponding to the size of the dominate set) is smaller than the number of judges (which is constant).


In the remaining part of this section, we consider \Microbribery with constant number of judges.
It turns out that the reduction in \cref{thm: Microbriery l=2 positive monotone is NP-c} can be easily adapted to this case if the desired set of the briber is incomplete.

\begin{theorem}
\label{thm: Microbriery l=2 positive monotone constant n is NP-c}
\Microbribery with conclusions being positive monotone clauses of length $\ell=2$ is \NP-hard and \W1-hard with respect to budget $k$, 
even if there are only three judges.
\end{theorem}

\begin{proof}
We present a reduction from \textsc{Clique}.
The construction is the same as the proof of \cref{thm: Microbriery l=2 positive monotone is NP-c} except that now we have only three judges and the profile is given in \cref{table: Microbribery l=2 positive monotone constant n NP-c}, where $J$ is incomplete and consistent (as it can be satisfied by setting $y^*=0$ and $y_i=0$).
We still set the budget of the briber to be~$k=s+1$.
Note that to change the value of~$x^*$ or~$x_i$ $(1 \le i \le n)$ the briber need to change one entry, while for $y^*$ or $y_i$ $(1 \le i \le n)$ the briber need to change two entries. 
We show that there is a clique of size $s$ in $G$ if and only if there is a successful bribery with at most $k=s+1$ changes.
The ``only if'' direction is the same as before and we just consider the ``if'' direction.

\begin{table}
\caption{Instance of \Microbribery with three judges and positive monotone clauses of length $s=2$ in the proof of \cref{thm: Microbriery l=2 positive monotone constant n is NP-c}.}
\label{table: Microbribery l=2 positive monotone constant n NP-c}
\centering
\begin{tabular}{lcccccclll}
\toprule
Judgment Set & $x_i$ & $x^*$ & $y_i$ & $y^*$ &  & $x_i \lor x_j$ & $x_i \lor x^*$ & $x_i \lor y_i$ & $x^* \lor y^*$ \\ \midrule
$J_1$ & 1 & 1 & 0 & 0 & & 1 & 1 & 1 & 1 \\
$J_2$ & 1 & 1 & 0 & 0 & & 1 & 1 & 1 & 1 \\
$J_3$ & 0 & 0 & 0 & 0 & & 0 & 0 & 0 & 0 \\
$\opUPQR_{1/2}$ & 1 & 1 & 0 & 0 & $\Rightarrow$ & 1 & 1 & 1 & 1\\ \midrule
$J$ &  &  &  &  & & 0 & 0 & 1 & 1 \\
\bottomrule
\end{tabular}
\end{table}

$\Leftarrow$ Assume there is a successful bribery with at most $k$ microbribes. 
Denote $n_1$ the number of newly included desired formulas,
and $n_2$ the number of excluded desired formulas after a successful bribery.
Suppose $s'\le k$ variables from~$\{x_1,\dots,x_n\}$ are changed after this bribery.
Then~$p \leq \binom{s'}{2}$ of~$x_i \lor x_j$ are changed. 
Note that $s'>0$ since changing $x^*$ alone cannot change the outcome of $x_i \lor x_j$ or $x_i \lor x^*$.

If~$x^*$ is not changed, then 
\[n_1=p \leq \binom{s'}{2}.\]
On the other hand, with the remaining budget $k-s'$ the briber can change at most $\lfloor \frac{k-s'}{2} \rfloor$ of $y_i$, then at least $\frac{s}{2}s'-\lfloor \frac{k-s'}{2} \rfloor$ of $x_i \lor y_j$ are changed. So 
\[n_2 \ge \frac{s}{2}s'-\lfloor \frac{k-s'}{2} \rfloor.\]
Since $n_1>n_2$, we have $\binom{s'}{2}>\frac{s}{2}s'-\lfloor \frac{k-s'}{2} \rfloor$.
By some computation we get $(s'-s-1)(\frac{s'-1}{2})>0$, which contradicts with $1 \le s'\le k=s+1$.

Therefore, $x^*$ has to be changed and hence $s' \le s$.
Now $p \leq \binom{s'}{2}$ of~$x_i \lor x_j$ and~$s'$ of $x_i \lor x^*$  are changed, so
\[n_1=p+s'.\]
On the other hand, with the remaining budget $k-s'-1=s-s'$ the briber can change at most $\lfloor \frac{s-s'}{2} \rfloor$ of $y_i$, then at least $\frac{s}{2}s'-\lfloor \frac{s-s'}{2} \rfloor$ of $x_i \lor y_j$ and~$\frac{s}{2}-1$ of $x^* \lor y^*$ are changed. So 
\[n_2 \ge \frac{s}{2}s'-\lfloor \frac{s-s'}{2} \rfloor+\frac{s}{2}-1\ge\frac{s'(s+1)}{2}-1.\]
Since $n_1>n_2$, we have
\[
p >  \frac{s'(s+1)}{2}-s'-1 \Rightarrow p \ge \frac{s'(s-1)}{2}.
\]
Since $p \leq \binom{s'}{2}$, we have $s=s'$ and $p = \binom{s'}{2}$, which means the corresponding $s'=s$ vertices form a clique in $G$.
\end{proof}

\cref{thm: Microbriery l=2 positive monotone is NP-c,thm: Microbriery l=2 positive monotone constant n is NP-c} show that for positive monotone clauses of length $\ell=2$, \Microbribery is \NP-hard even if the desired set is complete \emph{or} there are only three judges.
We complement these two results by showing that when the desired set is complete \emph{and} there are only constant number of judges, \Microbribery becomes solvable in polynomial time for positive monotone clauses (of unbounded length).

\begin{proposition}
\label{pro:Microbriery-postive-constant-complete-P}
\Microbribery with conclusions being positive monotone clauses is solvable in polynomial time when the number of judges is a fixed constant and the desired set is complete.
\end{proposition}

\begin{proof}
Let $m$ be the number of judges that is fixed a constant.
Without loss of generality, we can assume $k \ge m$, since otherwise we can simply try all possible $k$ premises in polynomial time.
Consequently, with budget $k$ the briber is able to change the outcome of every variable.
Let $J$ be the desired set of the briber.
Denote $J_c^+ \subseteq J$ the set of desired positive conclusions and $J_c^- \subseteq J$ the set of desired negative conclusions.
If there is a variable $x \in J$ that is not in the outcome, then changing the value of $x$ from 0 to 1 is a successful bribery since then $x$ will be included into the outcome and all conclusions containing $x$ are from $J_c^+$.
Thus, we can assume all variables in $J$ are already in the outcome, which then implies that all conclusions in $J_c^+$ are already in the outcome.
Now if there is a variable $x$ such that $\neg x \in J$ and $x$ is in the outcome, then changing the value of $x$ from 1 to 0 is a successful bribery since then $\neg x$ will be included into the outcome, all conclusions in $J_c^+$ are still in the outcome, and conclusions from $J_c^-$ that are in the original outcome are still in the outcome.
The remaining case is that all premises in $J$ are already included in the outcome, for which there is no feasible bribery, because the Hamming distance between the outcome and desired judgement set is already zero.
\end{proof}

\section{Conclusion and Discussion}
\label{sec: conclusion}
\looseness -1
This paper provides a refined picture in terms of the computational complexity of different variants of
\textsc{Manipulation} and \textsc{Bribery} in judgment aggregation.
Our results for basic variants of \textsc{Manipulation} are summarized in \cref{tabel: results of manipulation}.
\Robustness and \Possible are easy to be manipulated as long as all conclusions are clauses, 
while the computational complexity of \Necessary and \Exact with conclusions chosen from a \CT set~$\mathcal{C}$ is the same as the computational complexity of the corresponding \textsc{$\mathcal{C}$-Sat}.

The results for \Hamming, \Bribery and \Microbribery are summarized in
\cref{table: results for manipulation and bribery under HD}.
For \Hamming, we show that \NP-hardness holds even if all conclusions are positive monotone clauses
with length $\ell=3$ but that the problem becomes solvable in polynomial time when~$\ell=2$.
For monotone or Horn clauses with~$\ell=2$, the problem is also \NP-hard which is in stark contrast to
all basic variants of \textsc{Manipulation} that remain polynomial-time solvable for Horn and positive monotone clauses
of arbitrary length.
For \textsc{Bribery}, we show that both \Bribery and \Microbribery remain \NP-hard even
when all conclusions are positive monotone clauses of length~2.
Specifically, \Bribery with conclusions being positive monotone clauses of length 2 is \W2-hard with
respect to the number $k$ of judges that can be bribed, and is solvable in polynomial time when $k$ is fixed.


\begin{table}[t]
\centering
\caption{Computational complexity of basic variants of \textsc{Manipulation}.}
\label{tabel: results of manipulation}
\renewcommand{\arraystretch}{1.2}
\begin{tabular}{lll}
\toprule
\textsc{UPQR\nobreakdash-$M$\nobreakdash-Manipulation} $M=$ & \textsc{Possible} /\newline \textsc{Robustness} & \textsc{Necessary} /\newline \textsc{Exact} \\
\midrule
no restriction                     & NP-h \cite{baumeister2015complexity}              & NP-h \cite{baumeister2015complexity} \\
\CT set~$\mathcal{C}$        & P (Lem.~\ref{lem: Possible clause is in P})       & \textsc{$\mathcal{C}$-Sat} (Thm.~\ref{thm: equivalence} \& \ref{thm: Exact equivalent})  \\
clauses with length $\ell \le 3$                 & P (Lem.~\ref{lem: Possible clause is in P})   & NP-h  (Cor.~\ref{cor: manipulation for 3sat and monotone sat is NP-c}                          )                              \\
monotone clauses                   & P (Lem.~\ref{lem: Possible clause is in P})   & NP-h  (Cor.~\ref{cor: manipulation for 3sat and monotone sat is NP-c}                          )                              \\
Horn clauses                       & P (Lem.~\ref{lem: Possible clause is in P})    & P (Cor.~\ref{cor: P to P})                              \\
positive monotone clauses          & P (Lem.~\ref{lem: Possible clause is in P})    & P (Cor.~\ref{cor: P to P})                              \\
\bottomrule
\end{tabular}
\end{table}

\begin{table}[t]
\centering
\caption{Computational complexity of \Hamming, \Bribery and \Microbribery. \Bribery with unbounded $k$ is \NP-hard even for positive monotone clauses with $\ell=2$.}
\label{table: results for manipulation and bribery under HD}
\renewcommand{\arraystretch}{1.2}
\begin{tabular}{llll}
\toprule
 & \textsc{Manipulation}& \textsc{Bribery} (fixed $k$) & \textsc{Microbribery} \\
\midrule
positive monotone with $\ell=2$ & P (Thm. \ref{thm: l=2 positive monotone is in P}) & P (Thm. \ref{thm: Bribery l=2 positive monotone is XP}) & NP-h (Thm. \ref{thm: Microbriery l=2 positive monotone is NP-c})\\
positive monotone with $\ell=3$ & NP-h (Thm. \ref{thm: l>=3 is in NP-c}) & NP-h (Cor. \ref{cor: Bribery l=2 Horn or monotone or l>=3 is NP-c}) & NP-h (Thm. \ref{thm: Microbriery l=2 positive monotone is NP-c}) \\
Horn or monotone with $\ell=2$ & NP-h (Thm. \ref{thm: l=2 monotone is NP-c} \& \ref{thm: l=2 Horn is NP-c}) & NP-h (Cor. \ref{cor: Bribery l=2 Horn or monotone or l>=3 is NP-c}) & NP-h (Thm. \ref{thm: Microbriery l=2 positive monotone is NP-c})\\
positive monotone & NP-h \cite{baumeister2015complexity} & NP-h \cite{baumeister2015complexity} & NP-h \cite{baumeister2015complexity} \\
\bottomrule




\end{tabular}
\end{table}

All variants of \textsc{Manipulation} and \textsc{Bribery} we considered were known to be generally \NP-hard,
which was seen and sold as ``barrier against manipulative behavior''~\cite{baumeister2015complexity}.
The main message of this work is that several basic variants of \textsc{Manipulation} can be solved efficiently
for simple but well-motivated restrictions of conclusions (e.g.\ Horn clauses and generalizations thereof)
whereas other variants remain computationally intractable for most restrictions.
Hence, our results question whether there really is a barrier against strategic behavior in case
of realistically simple formulas.

We see our results as an important step and expect further effects decreasing the computational complexity
by considering other realistic structural properties.
Possible next steps include a systematic investigation of the parameterized complexity for both
judgment aggregation-specific parameters (e.g. ``number of judges'' or ``size of the desired set'') and
formula specific parameters (e.g. ``number of clauses'' or ``variable frequency'').
We note that considering the parameter ``number of judges'' alone, however, will not lead to tractable cases
because this parameter is fixed to three in most of our reductions.

Another direction is to extend our polynomial-time solvable results to more expressive formulas. 
In this paper, we restrict conclusions to be clauses (disjunctions of literals), under which some basic variants of \text{Manipulation} become solvable in polynomial time.
Recall that all of our results can be directly translated to the case where we restrict conclusions to be conjunctions of literals (see Section~\ref{sec:clause-restrict}).
Based on these results, one can consider more expressive restrictions, like Horn formulas (conjunctions of Horn clauses) or Krom formulas (conjunctions of clauses of length 2), to explore the boundary of tractability.

In this paper we consider the two Hamming distance based variants of \textsc{Bribery} introduced by \citet{baumeister2015complexity}.
One can also define other variants of \textsc{Bribery} (or \textsc{Microbribery}) similar to basic variants of \textsc{Manipulation}, e.g., possible or necessary, and study their computational complexity using our approach.
Note that the reduction from \Hamming to \Bribery in \cref{lem:manip-to-brib} can be directly generalized to other variants of \textsc{Manipulation} and \textsc{Bribery}, thus our hardness results for different variants of \textsc{Manipulation} under different clause restrictions can be easily adapted to the corresponding variants of \textsc{Bribery}.
On the other hand, for clause restrictions under which \textsc{Manipulation} is polynomial-time solvable, \textsc{Bribery} could still be NP-hard when the budget is not fixed since the complexity could also come from the choice of different judges to bribe, similar to the case in \cref{thm: Bribery l=2 positive monotone is NP-c W2}.
We leave this for future work.

Finally, it is interesting to apply our refined approach to \textsc{Control} in judgment aggregation based on the results of~\citet{DBLP:journals/jcss/BaumeisterEERS20} or to other judgment aggregation procedures (e.g., Kemeny procedure~\cite{DBLP:conf/atal/Haan17}).
Furthermore, it seems natural to extend the study to strategic behavior of groups of judges instead of a single judge~\cite{BNE16}. 
Note that most \NP-hard variants of \textsc{Control} studied in \citet{DBLP:journals/jcss/BaumeisterEERS20} rely on complex formulas, similar to the case in~\citet{baumeister2015complexity}.
We believe many of our \NP-hardness results can be extended to different variants of \textsc{Control}.
While for the Kemeny procedure, under which even the outcome determination is usually at least \NP-hard, it seems difficult to transfer our results there.

\section*{Acknowledgment}
We are grateful to the anonymous Information and Computation reviewers for their helpful and constructive comments.
Robert Bredereck was supported by the DFG project “AFFA” (BR 5207/1;NI 369/15).
Junjie Luo was supported by CAS-DAAD Joint Fellowship Program for Doctoral Students of UCAS, the DFG project “AFFA” (BR 5207/1;NI 369/15), and the Ministry of Education, Singapore, under its Academic Research Fund Tier 2 (MOE2019-T2-1-045).



\bibliographystyle{abbrvnat}
\bibliography{simple-JA-manipulative}  

\begin{thebibliography}{29}
\providecommand{\natexlab}[1]{#1}
\providecommand{\url}[1]{\texttt{#1}}
\expandafter\ifx\csname urlstyle\endcsname\relax
  \providecommand{\doi}[1]{doi: #1}\else
  \providecommand{\doi}{doi: \begingroup \urlstyle{rm}\Url}\fi

\bibitem[{Bartholdi~III} et~al.(1989){Bartholdi~III}, Tovey, and Trick]{BTT89a}
J.~J. {Bartholdi~III}, C.~A. Tovey, and M.~A. Trick.
\newblock The computational difficulty of manipulating an election.
\newblock \emph{{Social Choice and Welfare}}, 6\penalty0 (3):\penalty0
  227--241, 1989.

\bibitem[{Bartholdi, III} et~al.(1992){Bartholdi, III}, Tovey, and
  Trick]{BTT92}
J.~J. {Bartholdi, III}, C.~A. Tovey, and M.~A. Trick.
\newblock How hard is it to control an election?
\newblock \emph{{Mathematical and Computer Modeling}}, 16\penalty0
  (8-9):\penalty0 27--40, 1992.

\bibitem[Baumeister et~al.(2013)Baumeister, Erd{\'{e}}lyi, Erd{\'{e}}lyi, and
  Rothe]{BEER13}
D.~Baumeister, G.~Erd{\'{e}}lyi, O.~J. Erd{\'{e}}lyi, and J.~Rothe.
\newblock Computational aspects of manipulation and control in judgment
  aggregation.
\newblock In \emph{Proceedings of the 3rd International Conference on
  Algorithmic Decision Theory ({ADT}'13)}, pages 71--85. Springer, 2013.

\bibitem[Baumeister et~al.(2015)Baumeister, Erd{\'e}lyi, Erd{\'e}lyi, and
  Rothe]{baumeister2015complexity}
D.~Baumeister, G.~Erd{\'e}lyi, O.~J. Erd{\'e}lyi, and J.~Rothe.
\newblock Complexity of manipulation and bribery in judgment aggregation for
  uniform premise-based quota rules.
\newblock \emph{Mathematical Social Sciences}, 76:\penalty0 19--30, 2015.

\bibitem[Baumeister et~al.(2016)Baumeister, Erd{\'{e}}lyi, and
  Rothe]{EandC6judgement}
D.~Baumeister, G.~Erd{\'{e}}lyi, and J.~Rothe.
\newblock Judgment aggregation.
\newblock In \emph{Economics and Computation}, chapter~8, pages 361--391. 2016.

\bibitem[Baumeister et~al.(2017)Baumeister, Rothe, and
  Selker]{COMSOCtrends8judgment}
D.~Baumeister, J.~Rothe, and A.-K. Selker.
\newblock Strategic behavior in judgment aggregation.
\newblock In U.~Endriss, editor, \emph{Trends in Computational Social Choice},
  chapter~8, pages 145--168. AI Access, 2017.

\bibitem[Baumeister et~al.(2020)Baumeister, Erd{\'{e}}lyi, Erd{\'{e}}lyi,
  Rothe, and Selker]{DBLP:journals/jcss/BaumeisterEERS20}
D.~Baumeister, G.~Erd{\'{e}}lyi, O.~J. Erd{\'{e}}lyi, J.~Rothe, and A.~Selker.
\newblock Complexity of control in judgment aggregation for uniform
  premise-based quota rules.
\newblock \emph{Journal of Computer and System Sciences}, 112:\penalty0 13--33,
  2020.

\bibitem[Baumeister et~al.(2021)Baumeister, Boes, and
  Weishaupt]{DBLP:conf/atal/BaumeisterBW21}
D.~Baumeister, L.~Boes, and R.~Weishaupt.
\newblock Complexity of sequential rules in judgment aggregation.
\newblock In \emph{Proceedings of the 20th International Conference on
  Autonomous Agents and Multiagent Systems (AAMAS'21)}, pages 187--195. {ACM},
  2021.

\bibitem[Botan et~al.(2016)Botan, Novaro, and Endriss]{BNE16}
S.~Botan, A.~Novaro, and U.~Endriss.
\newblock Group manipulation in judgment aggregation.
\newblock In \emph{Proceedings of the 15th International Conference on
  Autonomous Agents and Multiagent Systems (AAMAS'16)}, pages 411--419. {ACM},
  2016.

\bibitem[Bredereck et~al.(2017)Bredereck, Faliszewski, Kaczmarczyk,
  Niedermeier, Skowron, and Talmon]{BFKNST2017}
R.~Bredereck, P.~Faliszewski, A.~Kaczmarczyk, R.~Niedermeier, P.~Skowron, and
  N.~Talmon.
\newblock Robustness among multiwinner voting rules.
\newblock In \emph{Proceedings of the 10th International Symposium on
  Algorithmic Game Theory (SAGT '17)}, volume 10504 of \emph{LNCS}, pages
  80--92. Springer, 2017.

\bibitem[Ceri et~al.(2012)Ceri, Gottlob, and Tanca]{ceri2012logic}
S.~Ceri, G.~Gottlob, and L.~Tanca.
\newblock \emph{Logic programming and databases}.
\newblock Springer Science \& Business Media, 2012.

\bibitem[Christian et~al.(2007)Christian, Fellows, Rosamond, and
  Slinko]{christian2007complexity}
R.~Christian, M.~Fellows, F.~Rosamond, and A.~Slinko.
\newblock On complexity of lobbying in multiple referenda.
\newblock \emph{Review of Economic Design}, 11\penalty0 (3):\penalty0 217--224,
  2007.

\bibitem[de~Haan(2016)]{DBLP:conf/ecai/Haan16}
R.~de~Haan.
\newblock Parameterized complexity results for the kemeny rule in judgment
  aggregation.
\newblock In \emph{Proceedings of the 22nd European Conference on Artificial
  Intelligence (ECAI'16)}, volume 285, pages 1502--1510. {IOS} Press, 2016.

\bibitem[de~Haan(2017)]{DBLP:conf/atal/Haan17}
R.~de~Haan.
\newblock Complexity results for manipulation, bribery and control of the
  kemeny judgment aggregation procedure.
\newblock In \emph{Proceedings of the 16th Conference on Autonomous Agents and
  MultiAgent Systems ({AAMAS}'17)}, pages 1151--1159. {ACM}, 2017.

\bibitem[de~Haan(2018)]{DBLP:conf/kr/Haan18}
R.~de~Haan.
\newblock Hunting for tractable languages for judgment aggregation.
\newblock In \emph{Principles of Knowledge Representation and Reasoning:
  Proceedings of the Sixteenth International Conference ({KR}'18)}, pages
  194--203. {AAAI} Press, 2018.

\bibitem[Dietrich and List(2007{\natexlab{a}})]{DL07str}
F.~Dietrich and C.~List.
\newblock Strategy-proof judgment aggregation.
\newblock \emph{Economics and Philosophy}, 23\penalty0 (3):\penalty0 269--300,
  2007{\natexlab{a}}.

\bibitem[Dietrich and List(2007{\natexlab{b}})]{dietrich2007judgment}
F.~Dietrich and C.~List.
\newblock Judgment aggregation by quota rules: Majority voting generalized.
\newblock \emph{Journal of Theoretical Politics}, 19\penalty0 (4):\penalty0
  391--424, 2007{\natexlab{b}}.

\bibitem[Elkind et~al.(2022)Elkind, Lackner, and Peters]{ELP22surv}
E.~Elkind, M.~Lackner, and D.~Peters.
\newblock Preference restrictions in computational social choice: A survey,
  2022.
\newblock URL \url{https://arxiv.org/abs/2205.09092}.

\bibitem[Endriss(2016)]{End16}
U.~Endriss.
\newblock Judgment aggregation.
\newblock In F.~Brandt, V.~Conitzer, U.~Endriss, J.~Lang, and A.~D. Procaccia,
  editors, \emph{Handbook of Computational Social Choice}, chapter~17.
  Cambridge University Press, 2016.

\bibitem[Endriss et~al.(2012)Endriss, Grandi, and
  Porello]{endriss2012complexity}
U.~Endriss, U.~Grandi, and D.~Porello.
\newblock Complexity of judgment aggregation.
\newblock \emph{Journal of Artificial Intelligence Research}, 45:\penalty0
  481--514, 2012.

\bibitem[Endriss et~al.(2015)Endriss, de~Haan, and
  Szeider]{DBLP:conf/atal/EndrissHS15}
U.~Endriss, R.~de~Haan, and S.~Szeider.
\newblock Parameterized complexity results for agenda safety in judgment
  aggregation.
\newblock In G.~Weiss, P.~Yolum, R.~H. Bordini, and E.~Elkind, editors,
  \emph{Proceedings of the 2015 International Conference on Autonomous Agents
  and Multiagent Systems (AAMAS'15)}, pages 127--136. {ACM}, 2015.

\bibitem[Endriss et~al.(2020)Endriss, de~Haan, Lang, and
  Slavkovik]{DBLP:journals/jair/EndrissHLS20}
U.~Endriss, R.~de~Haan, J.~Lang, and M.~Slavkovik.
\newblock The complexity landscape of outcome determination in judgment
  aggregation.
\newblock \emph{Journal of Artificial Intelligence Research}, 69:\penalty0
  687--731, 2020.

\bibitem[Faliszewski et~al.(2011)Faliszewski, Hemaspaandra, Hemaspaandra, and
  Rothe]{FHHR11}
P.~Faliszewski, E.~Hemaspaandra, L.~A. Hemaspaandra, and J.~Rothe.
\newblock The shield that never was: {Societies} with single-peaked preferences
  are more open to manipulation and control.
\newblock \emph{Information and Computation}, 209\penalty0 (2):\penalty0
  89--107, 2011.

\bibitem[Goldberg(1984)]{goldberg1984finding}
A.~V. Goldberg.
\newblock \emph{Finding a maximum density subgraph}.
\newblock University of California Berkeley, CA, 1984.

\bibitem[Grossi and Pigozzi(2014)]{GP14}
D.~Grossi and G.~Pigozzi.
\newblock \emph{Judgment Aggregation: A Primer}.
\newblock Morgan \& Claypool Publishers, 2014.

\bibitem[Kornhauser and Sager(1986)]{KS86}
L.~A. Kornhauser and L.~G. Sager.
\newblock Unpacking the court.
\newblock \emph{The Yale Law Journal}, 96\penalty0 (1):\penalty0 82--117, 1986.

\bibitem[List(2012)]{Lis12}
C.~List.
\newblock The theory of judgment aggregation: an introductory review.
\newblock \emph{Synthese}, 187\penalty0 (1):\penalty0 179--207, 2012.

\bibitem[List and Puppe(2009)]{LP09}
C.~List and C.~Puppe.
\newblock Judgment aggregation: A survey.
\newblock In C.~List and C.~Puppe, editors, \emph{Handbook of Rational and
  Social Choice}, chapter~19. Oxford University Press, 2009.

\bibitem[Lloyd(1987)]{lloyd87}
J.~W. Lloyd.
\newblock \emph{Foundations of Logic Programming}.
\newblock Springer-Verlag, 1987.

\end{thebibliography}

\newpage

\appendix

\section{Missing Proofs}
\label{sec:app-missing}

\begin{lemma}
\label{lem:2-3-SAT-NPC}
For any $k \ge 3$, \textsc{$(\mathcal{M}_k^+ \cup \mathcal{M}_{2}^-)$-SAT} and \textsc{$(\mathcal{M}_2^+ \cup \mathcal{M}_{k}^-)$-SAT} are \NP-hard.
\end{lemma}

\begin{proof}
Since \textsc{$(\mathcal{M}_k^+ \cup \mathcal{M}_{2}^-)$-SAT} and \textsc{$(\mathcal{M}_2^+ \cup \mathcal{M}_{k}^-)$-SAT} are equivalent under linear-time reductions, it suffices to show one of them is \NP-hard.
We present a polynomial-time reduction from \textsc{$k$-Coloring} to \textsc{$(\mathcal{M}_k^+ \cup \mathcal{M}_{2}^-)$-SAT}.
Given a graph $G=(V,E)$, we construct a formula of \textsc{$(\mathcal{M}_k^+ \cup \mathcal{M}_{2}^-)$-SAT} as follows.
For each vertex $v_i \in V$, we add vertex clauses $x_i^1 \lor \dots \lor x_i^k$ and $\neg x_i^s \lor \neg x_i^t$ for each pair of $s$ and $t$ with $1\le s<t \le k$ into the formula.
In addition, for each edge $\{v_i,v_j\} \in E$, we add edge clauses $\neg x_i^s \lor \neg x_j^s$ for every $1\le s\le k$ into the formula.

Suppose the graph $G$ is $k$-colorable, then we construct an assignment of variables by setting $x_i^s=1$ if vertex $v_i$ is colored $s$.
This assignment makes all vertex clauses satisfied since for each vertex $v_i$ exactly one  variable from $\{x_i^1,x_i^2,\dots,x_i^k\}$ is set True.
This assignment also makes all edge clauses satisfied since for each edge $\{v_i,v_j\} \in E$, $v_i$ and $v_j$ have different colors, which means the corresponding two variables $x_i^s$ and $x_j^t$ that are set True must satisfy that $s\neq t$.

Suppose there is a satisfying assignment for the formula.
For each vertex $v_i$, since $x_i^1 \lor \dots \lor x_i^k$ is satisfied, 
at least one variable from $\{x_i^1,x_i^2,\dots,x_i^k\}$ is set True.
In addition, since all vertex clauses $\neg x_i^s \lor \neg x_i^t$ with $1\le s<t \le k$ are satisfied, 
exactly one variable $x_i^t$ from $\{x_i^1,x_i^2,\dots,x_i^k\}$ is set True.
Accordingly, we color $v_i$ by $t$.
Since all edge clauses are satisfied, we have that for each edge, its two end points have different colors.
Therefore, we get a proper $k$-coloring for the graph $G$.
\end{proof}

\section{Symmetric Version of Lemma \ref{lem: 3M- or 2M+ is complete}}
\label{sec:app-3-2}

Recall that Lemma \ref{lem:any-q} shows that our hardness reductions can be adapted to work for any rational quota $q\in[0,1)$, but from the perspective of parameterized complexity analysis it leaves the following two special cases: (1) instances with high quotas and small numbers of judges where a variable is accepted if all judges accept it and (2) instances with low quotas and small numbers of judges where a variable is accepted if at least one judge accepts it.
We now discuss how to adapt our hardness reductions for case (1), and the method for case (2) is analogous.

For case (1), we can still create decision variables with $x \in J_n \cap \opUPQR_q(\bm{J})$ or $\neg x \in J_n \cap \opUPQR_q(\bm{J})$ (by setting $x\in J_i$ for every $i \in \{1,2,\dots,n-1\}$), while for non-decision variables, $x \in J_n \setminus \opUPQR_q(\bm{J})$ is possible but $x \in \opUPQR_q(\bm{J}) \setminus J_n$ is impossible.
In other words, the only difference is that, in this case, we cannot create variables that are included in the truthful outcome and cannot be changed by the manipulator.
Therefore, all of our hardness reductions that do not use this kind of non-decision variables still work for case (1).
The only exceptions are Lemma \ref{lem: from M(k)-SAT to M(k+1)} and \ref{lem: M+ M- S is NP-c}, for which we can rename each variable by its negation such that for non-decision variables only $x \in J_n \setminus \opUPQR_q(\bm{J})$ is used in the reductions. 
Note that this negation will also change the type of clauses, so after this change we actually show the symmetric versions of Lemma \ref{lem: from M(k)-SAT to M(k+1)} and \ref{lem: M+ M- S is NP-c}.
Then to finish the proof of Theorem 1, we also need to show the following symmetric version of Lemma \ref{lem: 3M- or 2M+ is complete} for the special case where a variable is accepted if all judges accept it.

\begin{lemma}
\label{lem:3M+2M-}
When $\lfloor q n + 1 \rfloor=n$,
\Necessary with conclusions chosen from~$(\mathcal{M}_3^+ \cup \mathcal{M}_2^-)$ is \NP-hard.
\end{lemma}

\begin{table*}
\caption{Instance of \Necessary with conclusion set~$\mathcal{C}=\mathcal{M}_3^+ \cup \mathcal{M}_2^-$ for the proof of \cref{lem:3M+2M-}.}
\label{table:M3M2}
\centering
\begin{tabular}
{p{2.5cm}p{0.2cm}p{0.2cm}p{0.2cm}p{0.2cm}p{0.2cm}p{0.2cm}p{0.2cm}p{0.4cm}p{1.5cm}p{2cm}p{2.5cm}}
\toprule
Judgment Set & $x_i$ & $y_i$ & $z_i$ & $w$ & $v$ & $u_1$ & $u_2$ & & $\neg w \lor \neg v$ & $v \lor u_1 \lor u_2$ & $x_{i_1} \lor x_{i_2} \lor x_{i_3}$  \\ \midrule
$J_i$ $(i\neq n)$ & 1 & 1 & 1 & 1 & 1 & 0 & 0 & & 0 & 1 & 1 \\
$J_n$ & 1 & 0 & 0 & 1 & 0 & 1 & 1 & & 1 & 1 & 1 \\
$\opUPQR_{1/2}$ & 1 & 0 & 0 & 1 & 0 & 0 & 0 & $\Rightarrow$ & 1 & 0 & 1 \\
\bottomrule
\end{tabular}

\begin{tabular}
{p{2.5cm}p{2cm}p{2cm}p{2cm}p{2cm}p{2.35cm}}
\toprule
Judgment Set & $\neg z_{i_1} \lor \neg z_{i_2}$ & $x_i \lor y_i \lor w$ & $y_i \lor z_i \lor w$ & $\neg x_i \lor \neg y_i$ & $\neg y_i \lor \neg z_i$  \\ \midrule
$J_i$ $(i\neq n)$ & 0 & 1 & 1 & 0 & 0\\
$J_n$ & 1 & 1 & 1 & 1 & 1\\
$\opUPQR_{1/2}$ & 1 & 1 & 1 & 1 & 1\\
\bottomrule
\end{tabular}
\end{table*}

\begin{proof}
We adapt the original reduction in the proof of Lemma \ref{lem: 3M- or 2M+ is complete} to prove the claimed result.
The constructed instance is shown in Table~\ref{table:M3M2}.
We first replace each variable by its negation such that all conclusions are from $\mathcal{M}_3^+ \cup \mathcal{M}_2^-$.
In addition, we make all $x_i,y_i,z_i,w,v$ decided by the manipulator by making them accepted by all other judges.
Note that since $\lfloor q n + 1 \rfloor=n$, a variable is accepted if all judges accept it.
Recall that the main idea of the original reduction for Lemma \ref{lem: 3M- or 2M+ is complete} is that since variable $v$ is not decided by the manipulator and $v$ is not included in the truthful outcome, the manipulator has to make $w$ included such that the outcome will include conclusion $w \lor v$.
For the new instance with $\lfloor q n + 1 \rfloor=n$ it is impossible to create a variable that is included in the truthful outcome but not decided by the manipulator.
However, we can achieve a similar effect in the new instance by adding two new variables $u_1,u_2$ that are not decided by the manipulator and a new conclusion $v \lor u_1 \lor u_2$ (as shown in Table~\ref{table:M3M2}).
The desired set $J$ of the manipulator consists of all positive conclusions.
Now $v \lor u_1 \lor u_2$ is the only positive conclusion in $J$ that is not included into the outcome, and to make $v \lor u_1 \lor u_2$ included, the manipulator has to make $v$ included, which then enforces the manipulator to make $\neg w$ included due to $\neg w \lor \neg v$.
Once $\neg w$ is included into the outcome, the manipulator has to solve an instance of \textsc{$(\mathcal{M}_3^+ \cup \mathcal{M}_2^-)$-SAT} (which is \NP-hard according to Lemma \ref{lem:2-3-SAT-NPC}), and the analysis is analogous to the proof of Lemma \ref{lem: 3M- or 2M+ is complete}.
\end{proof}

\end{document}